%% file: Thesis.tex
\documentclass[11pt,a4paper,twoside, openright]{book} 

\usepackage[swedish,english]{babel} 
\usepackage[latin1]{inputenc} 
\usepackage[T1]{fontenc}

\usepackage{times,amsmath,amssymb} 
\usepackage{helvet}  
\usepackage{ThesisSU} 
\ifpdf
   \usepackage[pdftex]{graphicx}
  \else
    \usepackage[dvips]{graphicx}
\fi 
\usepackage[colorlinks=true, urlcolor=blue, pagecolor=black, linkcolor=black, citecolor=black, filecolor=black, menucolor=black, pdfpagelayout=TwoColumnRight, pdfstartview=FitH, plainpages=false, pdfpagelabels]{hyperref} 

\input{BiblioInfo.tex} 

\def\mc{\mathcal}
\def\mb{\boldsymbol}
\def\mr{\mathrm}
\def\h{\widehat}
\def\tr{\mathrm{tr}\,}
\def\u{\underline}
\newcommand{\comment}[1]{}

\begin{document}


\frontmatterSU
\newpage

\pagestyle{empty}

\begin{center}{\timesEightteen Abstract}\end{center}
\vspace{30pt}
\abstracttext

\newpage

\tableofcontents

\newpage

\pagestyle{plain}\setcounter{page}{1}

\input{introduction}

\input{basic-tools}

\input{tensionless}
\input{symfromsp}

\input{nbifromstring}
\input{action}
\input{supersymmetric}

    \backmatter
\input{acknowledgements}

    \bibliographystyle{JHEP-2}
    \bibliography{References}

\input{errata}

\end{document}

%% file: BiblioInfo.tex
\newcommand{\authorSurname}{Wulff} 
\newcommand{\authorFirstName}{Linus} 
\newcommand{\dissertationTitle}{Strings, boundary fermions and coincident D-branes} 
\newcommand{\dissertationSubtitle}{}
\newcommand{\yearOfPublication}{2007} 
\newcommand{\placeOfPublication}{Stockholm} 
\newcommand{\ISBN}{91-7155-371-1 pp 1-83} 
\newcommand{\printer}{Printed in Sweden by Universitetsservice US-AB} 
\newcommand{\printingdate}{Stockholm 2007} 

\newcommand\abstracttext{%
The appearance in string theory of higher-dimensional objects known as D-branes has been a source of much of the interesting developements in the subject during the past ten years. A very interesting phenomenon occurs when several of these D-branes are made to coincide: The abelian gauge theory living on each brane is enhanced to a non-abelian gauge theory living on the stack of coincident branes. This gives rise to interesting effects like the natural appearance of non-commutative geometry. The theory governing the dynamics of these coincident branes is still poorly understood however and only hints of the underlying structure have been seen. 

This thesis focuses on an attempt to better this understanding by writing down actions for coincident branes using so-called boundary fermions, originating in considerations of open strings, instead of matrices to describe the non-abelian fields. It is shown that by gauge-fixing and by suitably quantizing these boundary fermions the non-abelian action that is known, the Myers action, can be reproduced. Furthermore it is shown that under natural assumptions, unlike the Myers action, the action formulated using boundary fermions also posseses kappa-symmetry, the criterion for being the correct supersymmetric action for coincident D-branes.

Another aspect of string theory discussed in this thesis is that of tensionless strings. These are of great interest for example because of their possible relation to higher spin gauge theories via the AdS/CFT-correspondence. The tensionless superstring in a plane wave background, arising as a particular limit of the near-horizon geometry of a stack of D3-branes, is considered and compared to the tensile case.
} 

\renewcommand{\listofpapers}
{\begin{center}{\timesEightteen List of Papers}\end{center} 
\vspace{30pt}
This thesis is based on the following papers, which are referred
to in the text by their Roman numerals. \vspace{13pt}
    \begin{romanlist}
  \item A.~Bredthauer, U.~Lindstrom, J.~Persson and L.~Wulff,
  {\it Type IIB tensionless superstrings in a pp-wave background}, {\em JHEP} {\bf 02} (2004) 051
  [\href{http://arXiv.org/abs/hep-th/0401159}{{\tt hep-th/0401159}}].
\\
  \item P.~S.~Howe, U.~Lindstrom and L.~Wulff,
  {\it Superstrings with boundary fermions}, {\em JHEP} {\bf 08} (2005) 041
  [\href{http://arXiv.org/abs/hep-th/0505067}{{\tt hep-th/0505067}}].
\\
  \item P.~S.~Howe, U.~Lindstrom and L.~Wulff,
  {\it On the covariance of the Dirac-Born-Infeld-Myers action},
  \href{http://arXiv.org/abs/hep-th/0607156}{{\tt hep-th/0607156}}.
  \end{romanlist}
\vspace{13pt} \noindent {\timesTen Reprints were made with permission from
the publisher.}}

\newcommand{\dedication}%
{\newpage
\thispagestyle{empty}
\vspace*{\stretch{3}}
\begin{flushright}
		
		{\fontfamily{pzc}\Large\selectfont
        \emph{my dedication...}}

\end{flushright}
\vspace*{\stretch{1}}} 

%% file: introduction.tex
\chapter{Introduction}
During the almost forty years that string theory has been around it has expanded into a vast subject with branches connecting to various other parts of physics and mathematics. Every month more than a hundred research papers are written on the subject. Needless to say, giving a full account of string theory will not be possible here. In this introductory chapter I will instead try to give, in a non-technical way, a feeling for some of the important underlying ideas and aspects of string theory that are relevant to the rest of the thesis. For a thorough introduction to string theory I refer to the standard text books on the subject, \cite{Polchinski:1998}, \cite{Green:1988} and \cite{Zwiebach:2004}. There are also many good review articles for example \cite{Schwarz:2000ew} and \cite{Szabo:2002ca}, just to mention two.

We start by giving some motivation for why string theory is studied. Next we give a brief review of the historical development of string theory and then discuss two important aspects in more detail: Supersymmetry and D-branes. We end with a list of some open problems and an outline of the thesis.

\section{Why string theory?}
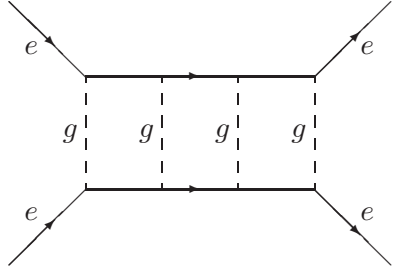
\begin{figure}[ht]
\begin{center}
\setlength{\unitlength}{1cm}
\begin{picture}(6.5,4.5)(0,0.5)
\put(1,0.5){\line(1,1){1}}
\put(1,0.5){\vector(1,1){0.6}}
\put(1.2,1.1){$e$}
\put(1,4){\line(1,-1){1}}
\put(1,4){\vector(1,-1){0.6}}
\put(1.2,3.3){$e$}
\put(2,1.5){\line(1,0){3}}
\put(3,1.5){\vector(1,0){0.5}}
\put(2,3){\line(1,0){3}}
\put(3,3){\vector(1,0){0.5}}
\put(5,1.5){\line(1,-1){1}}
\put(5,1.5){\vector(1,-1){0.6}}
\put(5.6,1.1){$e$}
\put(5,3){\line(1,1){1}}
\put(5,3){\vector(1,1){0.6}}
\put(5.6,3.3){$e$}
\put(2,1.5){\dashbox{0.15}(1,1.5)}
\put(4,1.5){\dashbox{0.15}(1,1.5)}
\put(1.7,2.2){$g$}
\put(2.7,2.2){$g$}
\put(3.7,2.2){$g$}
\put(4.7,2.2){$g$}
\end{picture}
\end{center}
\caption{Example of a process that is divergent (non-renormalizable) in a quantum field theory of gravity: Two electrons interacting gravitationally by exchanging four gravitons, the quanta of gravity.}
\label{fig:graviton-exchange}
\end{figure}

In the 20th century theories were developed that describe three of the four known forces in nature, the electromagnetic force, the weak nuclear force and the strong nuclear force, in a way that is consistent with the principles of special relativity and quantum mechanics. These theories are known as quantum field theories and the so-called standard model comprised of these three theories has been tremendously successful in accounting for all observed particle physics phenomena so far. However, this picture can not be complete since it doesn't include the fourth force, gravity. In fact, trying to reconcile Einstein's theory of gravity, known as general relativity, with quantum mechanics one encounters severe problems. Trying to formulate gravity as a quantum field theory along the lines that were so successful for the other forces gives a non-renormalizable theory, \emph{i.e.} a theory that gives infinite answers as soon as one tries to calculate something and where there is no way of removing the infinities by redefining the constants appearing in the theory (so-called renormalization). An example of a process that gives a divergent answer is given in figure \ref{fig:graviton-exchange}.

What's remarkable is that the seemingly naive idea of replacing the fundamental objects, point-like particles in quantum field theory, with extended one-dimensional objects, strings, radically changes the situation. In string theory there are no longer any arbitrary choices of what features to include; there is a single unique theory. Furthermore string theory automatically contains gravity in a quantum mechanically consistent way, and not only that; it gives a unified description of gravity together with the other kinds of forces that we observe in nature. This means that string theory has the potential of being \emph{the} theory of nature. It is important to emphasize, however, that there is, as of yet, no experimental evidence for string theory and it turns out to be extremely hard to test, at least in any direct way, due to the strings being extremely small.

\section{A brief history of string theory}
String theory was born at the end of the 1960's out of an attempt to describe the strong nuclear force, the dominant force inside the nucleus of the atom. It started with the observation by Veneziano that a particular mathematical function, known as the Euler beta function, could describe some of the properties of meson scattering that had been observed \cite{Veneziano:1968yb}. Around 1970 it was realized that these so-called dual resonance models could be thought of as describing strings. The inclusion of fermions in 1971 led to the idea of supersymmetry and the superstring \cite{Ramond:1971gb,Neveu:1971rx}. 

However, around 1973 the interest in string theory quickly faded with the invention of quantum chromodynamics (QCD) and the realization that this was the correct theory describing the strong interaction. String theory also had various problems and undesirable features; it needed more than four dimensions and predicted the existence of massless particles that were not observed in hadron experiments.

In 1974 the subject was revived and string theory as we think of it today was born. Among the massless particles predicted by string theory was a particle of spin two and it was realized by Scherk and Schwarz \cite{Scherk:1974ca} and independently by Yoneya \cite{Yoneya:1974jg} that this particle interacts like a graviton. This meant that string theory includes general relativity and this led Scherk and Schwarz to propose that it should be regarded instead as a unified theory of all interactions. As a consequence of this the typical size of the strings should not be the size of hadrons, around $10^{-15}\,m$, but rather the scale at which quantum gravity becomes important, the so-called Planck length,
\begin{equation}
l_\mr{P}\equiv\sqrt{\frac{\hbar G}{c^3}}\approx 10^{-35}\,m\,.
\end{equation}
Despite this important realization string theory remained outside of mainstream theoretical physics. One problem was that there were various different string theories but non of them resembled very closely the standard model of particle physics.

Then in 1984-85 in what has become known as the ''first superstring revolution'' the situation changed due to a number of important developments and string theory really became a legitimate part of theoretical physics. At low energy string theory is effectively described by supergravity, the supersymmetric extension of Einstein's theory of gravity. These theories were known to typically be plagued by quantum inconsistencies, or anomalies. But by 1985 it had been shown by Green and Schwarz and others that, remarkably, for the supergravity theories corresponding to five different superstring theories the anomalies were absent \cite{Alvarez-Gaume:1983ig,Green:1984sg}. The picture that emerged was that there was precisely five consistent superstring theories, all requiring ten dimensions and supersymmetry. These theories are known as type I, type IIA, type IIB, heterotic $SO(32)$ and heterotic $E_8\times E_8$. The last one in particular received a lot of interest as it could give rise to four-dimensional models, using an old idea known as compactification, that resembled the standard model.

\subsection{Compactification}
\begin{figure}[ht]
\begin{center}
\setlength{\unitlength}{1cm}
\begin{picture}(9,3)(0,1)
\put(1,3){\line(1,0){3}}
\put(1,1){\line(1,0){3}}
\qbezier(1,1)(1.5,2)(1,3)
\qbezier(1,1)(0.5,2)(1,3)
\qbezier(4,1)(4.5,2)(4,3)
\put(2.7,2){\vector(0,1){0.4}}
\put(2.7,2){\vector(0,-1){0.4}}
\put(2.7,2){\vector(1,0){0.6}}
\put(2.7,2){\vector(-1,0){0.6}}
\put(4.8,2){\vector(1,0){0.7}}
\put(6,1.8){\line(1,0){3}}
\put(6,2.2){\line(1,0){3}}
\qbezier(6,1.8)(6.1,2)(6,2.2)
\qbezier(6,1.8)(5.9,2)(6,2.2)
\qbezier(9,1.8)(9.1,2)(9,2.2)
\put(7.5,2){\vector(1,0){0.6}}
\put(7.5,2){\vector(-1,0){0.6}}
\end{picture}
\end{center}
\caption{Illustration of the idea of compactification: As the radius of the cylinder shrinks it becomes approximately one-dimensional.}
\label{fig:compactification}
\end{figure}
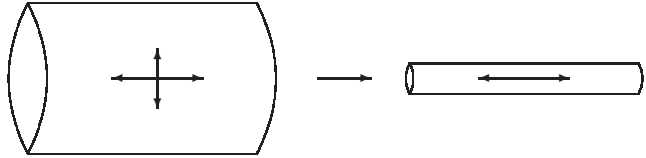

String theory requires ten dimensions, nine of space and one time, but we only observe four, three of space and one time, how can these two facts be reconciled? The answer goes back to an old idea due to Kaluza which was later elaborated on by Klein \cite{Klein:1926tv}. The idea, called Kaluza-Klein compactification, is surprisingly simple: The six extra dimensions are taken to be ''rolled up'' to a size that is smaller than can be seen in present day accelerator experiments. To illustrate the idea consider for simplicity a two-dimensional space in the shape of a cylinder, as in figure \ref{fig:compactification}. A creature living on this space can move in two independent directions, around the cylinder or along it. But, if we let the radius of the cylinder become very small the space becomes effectively one-dimensional, the dimension around the cylinder becomes so small that it can not be observed by a macroscopic observer. In string theory compactifications the compact six-dimensional space typically has a much more complicated structure but the basic idea remains the same.

\subsection{The second superstring revolution}
Around 1995 a new revolution in our understanding of string theory took place. It was discovered that the five different string theories are actually related to one another by so-called ''dualities'', implying that they are in a sense equivalent. A better way of saying it is that there is a unique underlying theory and what we have been calling different string theories should rather be thought of as perturbation expansions of the underlying theory about five different points in the space of consistent vacua. The discovery that there is indeed a unique underlying theory was reassuring, after all, why should there be five different consistent theories of nature of which only one is realized? That the theory is unique does not mean that it will uniquely predict our universe however. Indeed it seems to be the case that there are very many consistent vacuum solutions of the equations of motion for the theory, each corresponding to a possible universe, some similar to ours and others completely different.

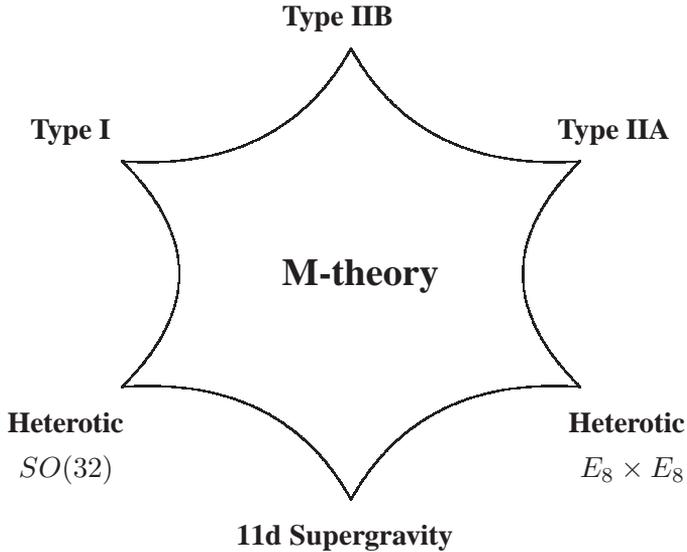
\begin{figure}[ht]
\begin{center}
\setlength{\unitlength}{3cm}
\begin{picture}(4,2.7)(2,1.7)
\qbezier(4,4)(4.3,3.45)(5,3.5)
\qbezier(5,3.5)(4.5,3)(5,2.5)
\qbezier(5,2.5)(4.3,2.55)(4,2)
\qbezier(4,2)(3.7,2.55)(3,2.5)
\qbezier(3,2.5)(3.5,3)(3,3.5)
\qbezier(3,3.5)(3.7,3.45)(4,4)
\put(3.7,2.95){\textbf{{\Large M-theory} } }
\put(3.7,4.1){\textbf{Type IIB}}
\put(4.9,3.6){\textbf{Type IIA}}
\put(4.95,2.3){\textbf{Heterotic}}
\put(5,2.1){\textbf{$E_8\times E_8$}}
\put(3.5,1.8){\textbf{11d Supergravity}}
\put(2.5,2.3){\textbf{Heterotic}}
\put(2.55,2.1){\textbf{$SO(32)$}}
\put(2.6,3.6){\textbf{Type I}}
\end{picture}
\end{center}
\caption{The five different string theories and eleven-dimensional supergravity as different limits of a single underlying theory, M-theory.}
\label{fig:Mtheoryweb}
\end{figure}

Another important discovery was that this underlying theory, often called M-theory (M could stand for Mother, Mystery, ...), also has an eleven-dimensional limit! This clarified a puzzling fact that had been known since the 1970's. It was known that in ten-dimensions there are two possible supergravity theories with maximal supersymmetry, called type IIA and type IIB supergravity. These are the low-energy limits of type IIA and type IIB superstring theory respectively. However, the highest possible dimension in which a consistent supergravity theory can exist is actually eleven dimensions and not ten. There is a single theory of eleven-dimensional supergravity which was constructed in 1978 by Cremmer, Julia and Scherk \cite{Cremmer:1978km}. The existence of this theory remained mysterious until 1995 when it was finally understood that this theory is also linked by dualities to the five ten-dimensional string theories \cite{Townsend:1995kk,Witten:1995ex}. The pattern that had emerged by the end of the second superstring revolution is illustrated in figure \ref{fig:Mtheoryweb}.

A third discovery of profound importance was that of the role played by higher-dimensional extended objects, called branes, in string theory. In particular the class of these known as Dirichlet branes, or D-branes for short. We will have more to say about these objects below and later in the thesis.

\section{Supersymmetry}
Supersymmetry is an integral part of string theory, but the reasons for supersymmetry go beyond string theory. It has been known for some time that supersymmetry could help resolve some of the outstanding problems of the standard model and particle physics. In a supersymmetric theory bosons (particles usually associated to the transmission of forces) and fermions (matter particles) come in pairs that are related to each other by supersymmetry. As this is not true for the particles that we observe supersymmetry is not present in our world. This doesn't make the idea of supersymmetry useless however. The point is that the reason why we don't observe supersymmetry could be that it is only present at higher energies, which are not available to us in present day accelerator experiments. If this is the case we say that supersymmetry is broken. The breaking of a symmetry in some energy regime is a phenomenon that occurs frequently in many areas of physics. 

In fact there are good reasons, not specific to string theory, that supersymmetry should be present down to fairly low energies, low in this case being around the so-called electroweak scale, the scale above which electromagnetism and weak interactions are unified into a single electroweak force, or in the range 100 GeV -- 1 TeV. If this is true it will probably be observed by the Large Hadron Collider (LHC) at CERN that will start operating in 2008. The three major reasons to believe in supersymmetry broken around the electroweak scale are:

\begin{itemize-indent}
\item {\em The hierarchy problem}. There is one missing piece in the standard model that has not yet been observed, the Higgs boson. It plays a crucial role in the breaking of electroweak symmetry and is responsible for explaining the masses of other elementary particles. The mass of the Higgs boson is not predicted by the standard model but if the theory is to be consistent it must be below 1 TeV. The problem is that it is unnatural for a scalar particle to be this light within the standard model because it receives contributions to its mass from so-called radiative corrections. These will typically push the mass up to a very high scale. By the inclusion of (low-energy) supersymmetry in the standard model this problem can be resolved, as supersymmetry forbids these radiative corrections and so stabilizes the mass of the Higgs.
\item {\em Grand unification}. The strength of the three forces of the standard model, the electromagnetic, weak and strong force, given by the value of their corresponding coupling constants, varies with energy. If this behavior is extrapolated to very high energies it can be seen that the strength of all three forces almost coincide at around $10^{16}$ GeV but not quite. However, if supersymmetry is included and the standard model is extended to a (minimally) supersymmetric version it can be seen that the strength of the three forces exactly converge at an energy around $2\cdot10^{16}$ GeV. If one believes that there is a unified description of the forces of the standard model at high energy their strengths should converge at this energy and this therefore suggests the presence of supersymmetry (again the argument requires low-energy supersymmetry).
\item {\em Dark matter}. From cosmological observations it is known that most of the matter in the universe is made up of stuff that we can't see with our instruments. A likely candidate for this dark matter is the heavy superpartners of ordinary particles that are present in a theory with broken supersymmetry. The masses of these would be around the scale of supersymmetry breaking. Specifically, the lightest supersymmetric particle (LSP) is stable in many models with supersymmetry and makes for an excellent dark matter candidate. Again, for this to work the mass of the LSP can not be too high, which again suggests supersymmetry broken around the weak scale.
\end{itemize-indent}

If these arguments are correct supersymmetry is likely to be discovered in the near future, and if it is not discovered radically new ideas may be needed. It should be stressed however that, although supersymmetry is a prediction of string theory, low-energy supersymmetry is not and indeed string theory could be correct without low-energy supersymmetry and low-energy supersymmetry could be present even if string theory is incorrect. In spite of this the discovery of low-energy supersymmetry would undoubtedly make us even more convinced that there is something to string theory.

\section{D-branes}
A major part of this thesis is concerned with the important higher-dimensional objects in string theory known as D-branes. To understand how they enter it is important to realize that strings can be of two different types: Open, with two ends, or closed, in the form of a small loop with no ends. D-branes can be thought of as hypersurfaces on which the open strings can end. A D$p$-brane has $p$ spatial dimensions and one time dimension and the possible values of $p$ depend on the theory; for type IIA $p$ is even and for type IIB $p$ is odd. A Dp-brane carries charge and acts as a source for a $(p+1)$-form Ramond-Ramond field present in the corresponding supergravity theory \cite{Polchinski:1995mt}. 

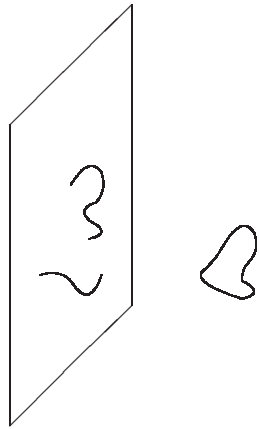
\begin{figure}[ht]
\begin{center}
\setlength{\unitlength}{0.8cm}
\begin{picture}(5,7)
\put(0.5,0){\line(0,1){5}}
\put(0.5,0){\line(1,1){2}}
\put(0.5,5){\line(1,1){2}}
\put(2.5,2){\line(0,1){5}}
\qbezier(1.5,4)(1.8,4.5)(2,4.2)
\qbezier(2,4.2)(2.1,3.9)(1.9,3.7)
\qbezier(1.9,3.7)(1.6,3.6)(1.8,3.4)
\qbezier(1.8,3.4)(2.2,3.2)(1.8,3.1)
\qbezier(1,2.5)(1.3,2.6)(1.5,2.4)
\qbezier(1.5,2.4)(1.8,1.9)(2,2.5)
\qbezier(4,3)(4.3,3.5)(4.5,3.2)
\qbezier(4.5,3.2)(4.6,2.9)(4.4,2.7)
\qbezier(4.4,2.7)(4.3,2.6)(4.3,2.4)
\qbezier(4.3,2.4)(4.7,2.2)(4.3,2.1)
\qbezier(4.3,2.1)(3.4,2.4)(3.7,2.6)
\qbezier(3.7,2.6)(3.9,2.8)(4,3)
\end{picture}
\end{center}
\caption{D-brane with open strings attached. The closed strings have no endpoints and are free to move in the whole ten-dimensional space.}
\label{fig:D-brane}
\end{figure}

D-branes are actually dynamical objects and their quantum fluctuations are given by open strings with their endpoints attached to the brane as shown in figure \ref{fig:D-brane}. These low-energy fluctuations give rise to an abelian gauge field, just like the electromagnetic field, which is confined to the surface, or worldvolume, of the brane. There are also fluctuations in the shape of the brane which give rise to a number of scalar fields on the brane, as many as the number of dimensions transverse to the brane, \emph{i.e.} $9-p$, that encode the fluctuations in shape in the transverse directions.

D-branes have played, and continue to play, a crucial role in many striking successes of string theory: Some  very important applications include the microscopic origin of black hole entropy, the so-called AdS/CFT-correspondence between string theory and gauge theories and the construction of so-called brane world scenarios as a possibility for the universe we inhabit.

\subsection{Coincident D-branes}
\begin{figure}[ht]
\begin{center}
\setlength{\unitlength}{0.9cm}
\begin{picture}(4,7)
\put(0.5,0){\line(0,1){5}}
\put(0.5,0){\line(1,1){0.4}}
\put(0.5,5){\line(1,1){2}}
\put(2.5,7){\line(0,-1){0.34}}
\put(0.9,0){\line(0,1){5}}
\put(0.9,0){\line(1,1){0.4}}
\put(0.9,5){\line(1,1){2}}
\put(2.9,7){\line(0,-1){0.34}}
\put(1.3,0){\line(0,1){5}}
\put(1.3,0){\line(1,1){0.4}}
\put(1.3,5){\line(1,1){2}}
\put(3.3,7){\line(0,-1){0.34}}
\put(1.7,0){\line(0,1){5}}
\put(1.7,0){\line(1,1){2}}
\put(1.7,5){\line(1,1){2}}
\put(3.7,7){\line(0,-1){5}}
\qbezier(2.5,4)(2.8,4.5)(3,4.2)
\qbezier(3,4.2)(3.1,3.9)(2.9,3.7)
\qbezier(2.9,3.7)(2.6,3.6)(2.8,3.4)
\qbezier(2.8,3.4)(3.2,3.2)(2.8,3.1)
\qbezier(1,2.5)(1.3,2.6)(1.5,2.4)
\qbezier(1.5,2.4)(1.8,1.9)(2,2.5)
\qbezier(2,2.5)(2,2.7)(1.8,2.8)
\qbezier(0.6,3)(0.75,2.8)(0.85,3.4)
\qbezier(1.5,5.9)(2,6.1)(1.6,5.5)
\qbezier(1.7,6.1)(1.3,6.5)(1.9,7)
\qbezier(1.9,7)(2.2,7.5)(3,7.2)
\qbezier(3,7.2)(3.3,6.9)(3,6.5)
\end{picture}
\end{center}
\caption{A stack of (almost) coincident D-branes with open strings stretching between them.}
\label{fig:coincident-branes}
\end{figure}
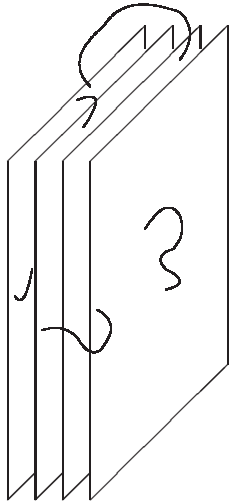

A remarkable phenomenon occurs when parallel branes of the same dimension are brought together (this is possible because the force between the branes is zero, their gravitational attraction is exactly cancelled by the repulsion due to their charge) \cite{Witten:1995im}. The phenomenon can be understood qualitatively in the following way: When there are several branes present an open string can have one endpoint attached to one brane and the other endpoint attached to another, see figure \ref{fig:coincident-branes}. 

It turns out that such a string gives rise to fluctuations that look like particles with a mass proportional to the separation of the two branes. When a number of branes become coincident these strings therefore give rise to new massless fluctuations. What actually happens is that the fields living on the branes mix together and expand into a single non-abelian (\emph{i.e.} matrix valued) field living on the stack of coincident branes. This picture holds both for the gauge field and the scalar fields describing the transverse fluctuations in shape of the brane. The fact that these scalar fields become matrices and in general no longer commute with each other means that their interpretation as fluctuations in the coordinates transverse to the brane breaks down. This gives a string theory realization of an interesting mathematical idea known as non-commutative geometry, where one considers the geometry of ''spaces'' in which the coordinates do not commute with each other. This is a very interesting subject and an active area of research in both mathematics and physics.

\section{Open problems}
We have tried to describe in this chapter some of the successes of string theory so far and give a feeling for why it is viewed by many as a promising candidate for a theory of everything. The subject is far from finished however and there are still many important questions that remain to be answered and the ultimate goal of a complete theory of nature still seems far away. We will end by listing some of the important issues that string theory will need to address if it is to reach this goal.

\begin{itemize-indent}
\item {\em What is the theory?} Although much is known about the various string theories the unique underlying theory that we have called M-theory remains to a large extent mysterious. It is not known in terms of which objects this theory is to be formulated, nor what the basic principles of the theory should be. It is likely that understanding this will require new mathematical tools and this is a question that is likely to be an important part of theoretical physics in the years to come.

\item {\em How is supersymmetry broken?} We believe that supersymmetry is present at high energies and probably even down to energies accessible to the next generation of particle accelerators but we still don't know how or why this symmetry is broken.

\item {\em Why is the cosmological constant so small?} The energy density of the vacuum, or cosmological constant, is observed to be comparable to the energy density in the form of matter in the universe. In Planck units this is a tiny number $\Lambda\sim10^{-120}$. However, supersymmetry broken at the TeV scale typically gives $\Lambda\sim10^{-60}$, which is very far off. Despite many attempts a satisfactory solution to this problem has not yet been found.

\item {\em What principle selects the vacuum?} String theory (or M-theory) seems to admit a vast number of possible vacua, and therefore different universes. It would be desireable to have a principle for choosing among these. What that principle should be is still unclear however.

\end{itemize-indent}

\section{Outline of the thesis}
The thesis is organized as follows:

\subsubsection{Chapter 2}
We give a summary of some of the mathematical machinery used in the thesis and state our conventions as well as giving a brief description of supersymmetry and superspace.

\subsubsection{Chapter 3}
This chapter presents some basic string theory and introduces the notion of the tensionless limit of the string. The quantization of the superstring in a plane wave background is described and then the corresponding thing for the tensionless string. A short discussion of the differences is given. 


\subsubsection{Chapter 4}
The superparticle is introduced and its coupling to a non-abelian gauge field using extra fermionic degrees of freedom discussed. It is shown how the requirement of kappa-symmetry forces the gauge field to be a solution of the super Yang-Mills equations of motion.

\subsubsection{Chapter 5}
The considerations of chapter 4 are applied to the case of the open superstring. This is coupled to the non-abelian gauge field on a stack of coincident D-branes using boundary fermions in a geometrical way. A generalization of the geometrical superembedding conditions is found and their consequences worked out for the case of a stack of D9-branes.


\subsubsection{Chapter 6}
Actions for the bosonic fields describing the dynamics of D-branes are discussed. The action for a single D-brane and Myers' generalization to the coincident case are reviewed. Then we present an action for the coincident case inspired by our considerations of boundary fermions and demonstrate its covariance. The proof of covariance differs from that given in the paper and is a bit more general and detailed. Finally the relation to Myers' action is discussed.


\subsubsection{Chapter 7}
The boundary fermion inspired action presented in chapter 6 is shown to possess kappa-symmetry when formulated in (type IIB) superspace, at least if two natural conditions are satisfied. A discussion of the implications of this as well as possible directions for future work is given.


%% file: basic-tools.tex
\chapter{Basic tools}
In this chapter we briefly introduce some of the mathematical tools, in particular related to supersymmetry, that we will need in the rest of the thesis. For more on supersymmetry we refer to the reviews \cite{VanProeyen:1999ni,Lykken:1996xt}. We also give the  conventions used in the rest of the thesis.

\section{Gamma matrices}
An extremely useful tool when dealing with spinors, as we will be doing a lot in this thesis, are gamma matrices. They are defined in D-dimensional spacetime as satisfying the Clifford algebra
\begin{equation}
\Gamma^a\Gamma^b+\Gamma^b\Gamma^a=2\eta^{ab}\,,
\end{equation}
where $a,b=0,1,\ldots,D-1$ and we have introduced the Minkowski metric
\begin{equation}
\eta\equiv\left(
\begin{array}{cccc}
-1     & 0 & \ldots & 0\\
 0     & 1 & \ldots & 0\\
\vdots & \vdots & \ddots &\vdots\\
0      & 0      & \ldots & 1
\end{array}
\right)\,.
\end{equation}
It is always possible to find $2^{[D/2]}\times2^{[D/2]}$ matrices $\Gamma^a$ that satisfy this relation ($[\cdot]$ denotes the integer part). Using these gamma matrices it is possible to construct the generators of Lorentz transformations acting on spinors as
\begin{equation}
\Sigma^{ab}\equiv-\frac{1}{4}\Gamma^{ab}\,.
\end{equation}
We have defined the anti-symmetrized products of gamma matrices
\begin{equation}
\Gamma^{a_1\cdots a_n}\equiv\Gamma^{[a_1}\cdots\Gamma^{a_n]}\,,
\end{equation}
where brackets denote anti-symmetrization with unit weight, so that for example $\Gamma^{[a}\Gamma^{b]}\equiv\frac{1}{2}[\Gamma^a,\Gamma^b]$ (symmetrization is denoted $(\cdots)$). It can be shown that these matrices form a basis of $2^{[D/2]}\times2^{[D/2]}$ matrices, so that in particular any matrix can be written as a linear combination of the Fierz basis elements $\{1,\Gamma^a,\Gamma^{ab},\ldots,\Gamma^{a_1\cdots a_D}\}$.

When the dimension $D$ is even the matrix
\begin{equation}
\Gamma^{(D)}\equiv(i)^{D(D-1)/2+1}\Gamma^0\Gamma^1\cdots \Gamma^{D-1}
\end{equation}
anti-commutes with all $\Gamma^a$, \emph{i.e.} 
\begin{equation}
\{\Gamma^{(D)},\Gamma^a\}\equiv\Gamma^{(D)}\Gamma^a+\Gamma^a\Gamma^{(D)}=0\,.
\end{equation}
It also has the property that it commutes with the generator of Lorentz transformation and that it squares to one. This means that the $2^{D/2}\times2^{D/2}$ representation of the gamma matrices splits up into two irreducible $2^{D/2-1}\times2^{D/2-1}$ representations called Weyl representations.

In certain dimensions it is possible to impose the condition that the gamma matrices be purely real (or imaginary) instead of complex, reducing their number of components by a factor of two. Such a representation of the gamma matrices is called a Majorana representation.

\subsection{$D=10$}
In this thesis we will be working in ten dimensions and we will need some special properties of the ten-dimensional gamma matrices. Since the dimension is even we can impose a Weyl condition on our $2^5\times 2^5=32\times 32$ gamma matrices. It turns out that in addition to this it is also consistent to impose a reality condition on the representation making it Majorana-Weyl. We can then take our gamma matrices of the form
\begin{equation}
\Gamma^a\rightarrow\left(
\begin{array}{cc}
0 & (\gamma^a)^{\alpha\beta}\\
\gamma^a_{\alpha\beta} & 0
\end{array}\right)\,,
\end{equation}
where $\gamma^a$ with indices up (down) are chiral (anti-chiral) Weyl blocks. They are real and symmetric $16\times16$ matrices. 

In this representation spinors split up accordingly as
\begin{equation}
\Theta\rightarrow\left(
\begin{array}{c}
\theta^\alpha\\
\chi_\beta
\end{array}\right)\,,
\end{equation}
where $\theta^\alpha$ ($\chi_\beta$) is a $16$-component real chiral (anti-chiral) spinor\footnote{Spinor indices will always be donted by greek letters and vector indices by lation ones.}. 

A very useful identity that the ten-dimensional gamma matrices satisfy is
\begin{equation}
\gamma^a_{(\alpha\beta}(\gamma_a)_{\gamma)\delta}=\frac{1}{3}(\gamma^a_{\alpha\beta}(\gamma_a)_{\gamma\delta}+\gamma^a_{\beta\gamma}(\gamma_a)_{\alpha\delta}
+\gamma^a_{\gamma\alpha}(\gamma_a)_{\beta\delta})=0\,.
\end{equation}
This can be proved by decomposing both sides in a Fierz basis.

\section{Supersymmetry}
Supersymmetry is an extension of the spacetime symmetries that we observe in nature; Lorentz invariance and translational invariance in space and time, the generators of which satisfy the so-called Poincar\'e algebra. In the supersymmetric version, the super Poincar\'e algebra, generators of supersymmetry $Q_\alpha$, sometimes called supercharges are added, which satisfy the anti-commutation relation\footnote{This is only the simplest possibility where one has a single supercharge.}
\begin{equation}
\{Q_\alpha,Q_\beta\}=\gamma^a_{\alpha\beta} P_a\,,
\end{equation}
where the momentum, $P_a$ is, as usual, the generator of translations. This suggests that, loosely speaking, one can think of a supersymmetry transformation as the ''square root'' of a translation. 

Because supersymmetry is an extension of the symmetries associated to spacetime, which in turn are intimately connected to the very concept of a spacetime, supersymmetry can naturally be thought of as extending spacetime to what is know as a superspace.

\subsection{Superspace}\label{sec:N=1superspace}
To define superspace one introduces in addition to the ordinary coordinates of spacetime $x^m$, where in our case $m=0,\ldots,9$, additional coordinates $\theta^{\mu}$, where in our case $\mu=1,\ldots,16$. $\theta$ forms a Majorana-Weyl spinor in ten dimensions\footnote{This is for the case of $\mc N=1$, \emph{i.e.} one supersymmetry generator $Q_\alpha$, for $\mc N=2$ one introduces two Majorana-Weyl spinors $\theta^{\mu i}$ for $i=1,2$ and so on.}. Because of the unusual property that the supercharges $Q$ anti-commute rather than commute these new coordinates are not ordinary numbers but rather so-called Grassmann numbers that satisfy
\begin{equation}
\theta^\mu\theta^\nu=-\theta^\nu\theta^\mu\,.
\end{equation}
This means that their interpretation as coordinates should not be taken too literally. Superspace (with $\mc N=1$) is then parameterized by the coordinates
\begin{equation}
z^M\equiv(x^m,\theta^\mu)
\end{equation}
and fields on superspace, or superfields, become functions $\Phi(z)=\Phi(x,\theta)$. A superfield can always be expanded in powers of $\theta$ and because of the Grassmann property of $\theta$, which implies that $(\theta^\mu)^2=0$ for any $\mu$, this expansion contains only a finite number of terms.

As for the bosonic coordinates, $x^m$, we can define derivatives with respect to $\theta^\mu$. We will call this derivative $\partial_\mu$. It acts as an ordinary derivative except that it anti-commutes with $\theta$ and with itself. It is also possible to define integrals over $\theta$ but we will not need the specific definition here.

\subsection{Differential geometry on superspace}\label{sec:diff-superspace}
Just as for an ordinary manifold we can define at each point of superspace a tangent space. This space is spanned by the basis vectors 
\begin{equation}
\partial_M=(\partial_m,\partial_\mu)\,,
\end{equation}
the derivatives with respect to the coordinates. As for an ordinary manifold there is a dual space, called the cotangent space, spanned by the one-forms
\begin{equation}
dz^M=(dx^m,d\theta^\mu)\,.
\end{equation}
This coordinate basis of tangent space is inconvenient to use however, because it is not invariant under supersymmetry transformations.

We will therefore define another basis of tangent space, invariant under supersymmetry transformations, given by $d_A=(d_a,d_\alpha)$ and related to the coordinate basis as 
\begin{equation}
d_A=E_A^{\phantom A M}\partial_M\,,
\end{equation}
where the $E_A^{\phantom A M}$ are often referred to as (super)vielbeins. We will always, as is standard practice, distinguish between between this ''preferred'' supersymmetric basis of tangent space and the coordinate basis by denoting the former with indices from the beginning of the alphabet (\emph{e.g.} $A,\beta,c$) and the latter with indices from the middle of the alphabet (\emph{e.g.} $m,N,\mu$). There is also a dual space spanned by the one-forms $e^A=(e^a,e^\alpha)$ where 
\begin{equation}
e^A=dz^M E_M^{\phantom MA}
\end{equation}
where $E_M^{\phantom MA}$ is the inverse of $E_A^{\phantom A M}$.

For flat ten-dimensional $\mc{N}=1$ superspace the supersymmetric basis of tangent space takes the form
\begin{eqnarray}
d_\alpha&=&\partial_\alpha+\frac{i}{2}(\gamma^a\theta)_\alpha\partial_a\nonumber\\
d_a&=&\partial_a
\label{eq:susy-ds}
\end{eqnarray}
and the dual (cotangent space) basis becomes
\begin{eqnarray}
e^\alpha&=&d\theta^\alpha\nonumber\\
e^a&=&dx^a-\frac{i}{2}d\theta\gamma^a\theta\,.
\label{eq:susy-es}
\end{eqnarray}
From these the components of the supervielbein $E_A^{\phantom A M}$ are easily read off.

In contrast to the case in ordinary flat space the derivatives defined in (\ref{eq:susy-ds}) do not (anti-)commute with each other. This property is encoded in the torsion
\begin{equation}
\label{eq:torsion}
T^a_{\alpha\beta}=-i(\gamma^a)_{\alpha\beta}\,,
\end{equation}
for flat $\mc N=1$ superspace. It is equal to minus the anti-commutator $\{d_\alpha,d_\beta\}$.

In analogy to ordinary space one introduces differential forms by defining a graded anti-symmetric wedge product, $\wedge$, of the elements of cotangent space (called one-forms). We have
\begin{equation}
dz^M\wedge dz^N=-(-1)^{MN}dz^N\wedge dz^M\,,
\end{equation}
where in $(-1)^{MN}$ $M$ and $N$ are thought of as even/odd (\emph{e.g.} 0/1) according to whether they are ordinary/Grassmann (we will also refer to them as bosonic/fermionic). Wedge products will always be implicit in any product of forms.

One also introduces an exterior derivative $d$ which satisfies $d^2=0$ and is taken to act from the right, \emph{e.g.}
\begin{equation}
de^a=d\left(dx^a-\frac{i}{2}d\theta\gamma^a\theta\right)=-\frac{i}{2}d\theta\gamma^a d\theta=\frac{1}{2}d\theta^\alpha d\theta^\beta T_{\alpha\beta}^a\,.
\end{equation}

An $n$-form $\omega$ is written with indices contracted in a specific order which simplifies keeping track of all the signs as
\begin{equation}
\omega=\frac{1}{n!}dz^{M_n}\cdots dz^{M_1}\omega_{M_1\cdots M_n}=\frac{1}{n!}e^{A_n}\cdots e^{A_1}\omega_{A_1\cdots A_n}\,.
\end{equation}

With $\omega$ an $n$-form and $\rho$ a $k$-form the exterior derivative satisfies
\begin{equation}
d(\omega\rho)=\omega d\rho+(-1)^k d\omega\rho\,.
\end{equation}

\section{Other conventions}
When we consider the embedding of an object in a background superspace indices corresponding to the background (sometimes called target space) will be underlined to distinguish them from worldvolume indices of the embedded object (a string or D-brane). Also indices $i,j$ always take values $1,2$.

The Pauli matrices are defined as
\begin{equation}
\sigma^1=\left(
\begin{array}{cc}
0 & 1\\
1 & 0
\end{array}\right)
\quad,\quad
\sigma^2=\left(
\begin{array}{cc}
0 & -i\\
i & 0
\end{array}\right)
\quad\mbox{and}\quad
\sigma^3=\left(
\begin{array}{cc}
1 & 0\\
0 & -1
\end{array}\right)\,.
\end{equation}

We will sometimes use the graded commutator $[A,B\}$ defined to be a commutator unless both $A$ and $B$ are fermionic in which case it is an anti-commutator.

%% file: tensionless.tex
\chapter{The tensionless string in a plane wave background} 
Although this chapter lies somewhat outside of the main focus of this thesis, which is the study of coincident D-branes it provides a good opportunity to introduce some of the basic notions and techniques of string theory in a concrete way.

We start by giving some motivation for studying tensionless strings. We then review how to obtain the action for a massless particle from the action for a massive one. The corresponding limit for the case of the bosonic string is then considered and we obtain an action for the tensionless string in Minkowski space. A brief description of the plane wave limit of $AdS_5\times S^5$ is given and we review the quantization of the superstring in this background. In section \ref{sec:tzeroppwave} we consider the tensionless string in this background and compare it to the tensile case. We end with some concluding remarks.

This chapter is based on \cite{Wulff:2004}.

\section{Motivation}
Tensionless string theory can be regarded as a high-energy limit of string theory where the masses of string states, which come in multiples of (the square root of) the string tension, are neglected. One motivation for studying tensionless strings comes from the AdS/CFT correspondence. The most studied case is the remarkable correspondence between string theory on the space $AdS_{5}\times S^{5}$ and a conformal field theory in four dimensions, $\mathcal{N}=4$ super Yang-Mills theory. Quantizing the string in $AdS_{5}\times S^{5}$ meets with difficulties however but there is a particular limit in which the space becomes a plane wave which is tractable. This limit of $AdS_{5}\times S^{5}$ gives rise to a correspondence related to the AdS/CFT correspondence, the so-called BMN correspondence between string theory on a plane wave background and a sector of $\mathcal{N}=4$ super Yang-Mills theory \cite{Berenstein:2002jq}.

Since explicit quantization of the string is possible in this background this opens up a possibility to study the field theory side from string theory (and vice versa). A subject which has become particularly interesting in connection with tensionless strings is that of massless higher spins and the relation to free super Yang-Mills, see for example \cite{Sundborg:2000wp,Lindstrom:2003mg}.

The study of tensionless strings could also shed light on the high-energy symmetries of string theory which are thought to be broken at lower energies giving rise to masses of the different string states. There is also an observation that strings appear to be tensionless in the vicinity of space-time singularities. For a discussion of these points see for example \cite{Bakas:2004jq} and references therein.

\section{The tensionless string}
\label{sec:tensionless}
A tensionless string can be thought of as the string analog of a massless particle. In this section we will introduce the action for a bosonic tensionless string in flat space using this analogy.

\subsection{The massless limit of the particle}\label{sec:m0-limit}
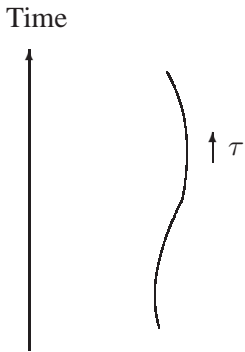
\begin{figure}[ht]
\begin{center}
\setlength{\unitlength}{1cm}
\begin{picture}(3,5)
\put(0.3,0){\vector(0,1){4}}
\put(0,4.3){\text{Time}}
\put(2.7,2.5){\vector(0,1){0.4}}
\put(2.9,2.6){$\tau$}
\qbezier(2,0.3)(1.8,1)(2.3,2)
\qbezier(2.3,2)(2.5,3)(2.1,3.7)
\end{picture}
\end{center}
\caption{The worldline of a particle parameterized by $\tau$.}
\label{fig:worldline}
\end{figure}

The equations of motion for a (scalar) particle of mass $m$ can be derived by requiring that the length of its worldline, \emph{i.e.} the curve it sweeps out in space-time, should be minimal (or extremal). The quantity that we should extremize to find the equations of motion is called the action and denoted $S$. For a particle the action is then simply proportional to the length of its worldline,
\begin{equation}
 S=-m\int ds\,,
\end{equation}
where we have introduced the constant of proportionality, $m$, which turns out to be the mass of the particle, in order to make the action dimensionless (we are working in so-called natural units in which $\hbar=c=1$ so that dimension of length is the same as inverse dimension of mass). If we parameterize the worldline of the particle with a parameter $\tau$, so that the worldline of the particle is given as $x^m(\tau)$ (see figure \ref{fig:worldline}), then the action becomes 
\begin{equation}
\label{eqn:p-action}
S=-m\int d\tau\,\sqrt{-\frac{dx^m}{d\tau}\frac{dx^n}{d\tau}\eta_{mn}}=-m\int d\tau\,\sqrt{-\dot{x}^{2}}\,,
\end{equation}
where we have put $\dot{x}^m\equiv\frac{dx^m}{d\tau}$ and $\dot x^2\equiv\dot x^m\dot x^n\eta_{mn}$ is the square of the length of the tangent vector to the worldline (in flat Minkowski space).

What about massless particles? Clearly the above action doesn't make sense when $m=0$, as it then vanishes for any choice of worldline. Nevertheless we can obtain an action for a massless particle from the action for a massive one by a neat trick that we will now describe.

We start by computing the Hamiltonian corresponding to the Lagrangian of (\ref{eqn:p-action}), 
\begin{equation}
L=-m\sqrt{-\dot{x}^{2}}\,.
\label{eq:L-particle}
\end{equation}
The conjugate momentum to $x^m$ is defined as
\begin{equation}
p_m\equiv\frac{\partial L}{\partial\dot{x}^m}=m\frac{\dot{x}_m}{\sqrt{-\dot{x}^{2}}}\,. 
\end{equation}
An important thing to note here is that the components of $p$ are not all independent, there is a constraint 
\begin{equation}
\phi\equiv p^2+m^2=m^2\frac{\dot x^2}{-\dot x^2}+m^2=0\,.
\end{equation} 
This is usually called the mass-shell condition. The reason for this constraint is that the action we started from, (\ref{eqn:p-action}), has an important symmetry: It is invariant under reparametrizations of the worldline, \emph{i.e.} choosing another parameter $\tau'=\tau'(\tau)$ to parameterize the worldline of the particle doesn't affect the action because, from (\ref{eq:L-particle}) we see that
\begin{equation}
L(\tau)\rightarrow|\partial\tau'/\partial\tau|L(\tau')\,,
\end{equation}
while
\begin{equation}
d\tau\rightarrow |\partial\tau'/\partial\tau|^{-1}d\tau'
\end{equation}
leaving $S=\int d\tau\,L$ invariant.

The canonical Hamiltonian is defined as
\begin{equation}
H_{\mr{can}}\equiv p_m\dot x^m-L=m\frac{\dot x^2}{\sqrt{-\dot x^2}}+m\sqrt{-\dot x^2}=0\,. 
\end{equation}
The vanishing of the canonical Hamiltonian is typical of reparametrization invariant systems. But because we have a constraint to take into account the canonical Hamiltonian is only defined on the ''constraint surface'', \emph{i.e.} when $p^2+m^2=0$. The full Hamiltonian is the extension of the canonical Hamiltonian off the ''constraint surface'' (as was show by Dirac \cite{Dirac:1950pj}),
\begin{equation}
H=H_{\mr{can}}+\frac{e}{2}\phi=\frac{e}{2}(p^2+m^2)\,, 
\end{equation}
where $e$ is a Lagrange multiplier enforcing the constraint $\phi=0$. Notice that $e$ must have dimension $[mass]^{-2}$.

Now that we have the Hamiltonian we can go back and write down the so-called phase space Lagrangian
\begin{equation}
\label{eqn:p-pslag}
L_{\mr{ps}}\equiv p_m\dot x^m-H=p_m \dot x^m-\frac{e}{2}(p^2+m^2)\,. 
\end{equation}
The Euler-Lagrange equations (or equations of motion) for $p_m$ read 
\begin{equation}
0=\frac{\partial L_{\mr{ps}}}{\partial p_m}-\frac{d}{d\tau}\frac{\partial L_{\mr{ps}}}{\partial\dot{p}_m}=\dot x^m-ep^m
\quad\Rightarrow\quad p_m=\frac{1}{e}\dot x_m\,. 
\end{equation}
Using this in (\ref{eqn:p-pslag}) the new action becomes 
\begin{equation}
\label{eqn:sprime}
S'=\int d\tau\,L_{\mr{ps}}=\frac{1}{2}\int d\tau\,\left(\frac{1}{e}\dot x^2-em^2\right)\,. 
\end{equation}
This is not the same action as the one we started from so what has happened? It turns out that, when $m\neq 0$, the action $S'$ is in fact equivalent to the action we started with, (\ref{eqn:p-action}). This follows if we use the Euler-Lagrange equation for $e$ to solve for $e$ and plug the result back in the action 
\begin{equation}
0=-\frac{1}{e^2}\dot{x}^2-m^2\quad\Rightarrow\quad e=\pm\frac{1}{m}\sqrt{-\dot{x}^2}\,. 
\end{equation}
Choosing the plus sign for $e$ minimizes the action $S'$ (the minus sign gives a maximum). Using this in (\ref{eqn:sprime}) we recover our original action, (\ref{eqn:p-action}), which demonstrates the equivalence.

The point of rewriting the action in this new form is that, while the original action didn't make sense for massless particles, the new action does. Taking $m=0$ in $S'$ we get an action for a massless particle
\begin{equation}
\label{eq:m0-action}
S_0'=\frac{1}{2}\int d\tau\,\frac{1}{e}\dot{x}^2\,.
\end{equation}
So we have really gained something by introducing the extra variable $e$.

Why does introducing $e$ and then taking $m\rightarrow 0$ work while taking $m\rightarrow 0$ in the action we started from didn't work? First of all $m$ is a dimensionful parameter and taking a dimensionful parameter to zero is not well defined, we can only say that \emph{compared} to another quantity of the same dimension it becomes small. This leads us to the conclusion that we should only take dimensionless combinations of quantities to zero. This is exactly what we did above. We introduced a new dimensionful parameter $e$ and took the dimensionless combination $em^2\rightarrow 0$. We will now repeat these considerations for the case of a string.

\subsection{The tensionless limit of the string}
\begin{figure}[ht]
\begin{center}
\setlength{\unitlength}{1cm}
\begin{picture}(6,5)
\put(0.3,0){\vector(0,1){4}}
\put(0,4.3){\text{Time}}
\put(4.6,2.5){\vector(0,1){0.4}}
\put(4.8,2.6){$\tau$}
\put(2.8,0.4){\vector(1,0){0.4}}
\put(2.9,0){$\sigma$}
\put(3,2.2){$\sum$}
\qbezier(2,0.3)(1.8,1)(2.3,2)
\qbezier(2.3,2)(2.5,3)(2.1,3.7)
\qbezier(4,0.3)(3.9,1.2)(4.1,2)
\qbezier(4.1,2)(4.4,3)(4,3.7)
\qbezier(2,0.5)(3.3,0.7)(4,0.4)
\qbezier(2.3,3.3)(3.3,3.7)(4.1,3.5)
\end{picture}
\end{center}
\caption{The worldsheet of an open string parameterized by $\sigma$ and $\tau$. The worldsheet of a closed string takes the form of a cylinder.}
\label{fig:worldsheet}
\end{figure}
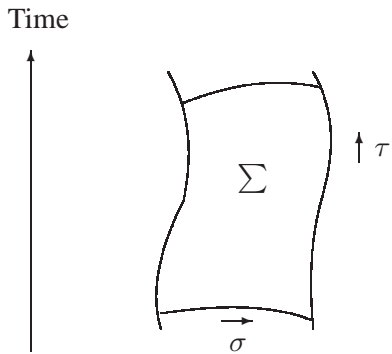

We can generalize the action for the massive particle in the last section, which was just proportional to the length of the curve the particle sweeps out in spacetime, to higher-dimensional objects by taking the action to be proportional to the corresponding ''volume'' that the object sweeps out in spacetime. For a string, which is one-dimensional, the action becomes proportional to the ''area'' of the surface it sweeps out in spacetime, its worldsheet. This describes the so-called bosonic string which contains much of the important aspects of string theory even though we need to invoke supersymmetry in order to get a completely consistent theory. We will confine ourselves to the case of the bosonic string in this section. The superstring can be handled along the same lines. The action proportional to the area of the worldsheet was first proposed independently by Nambu in 1970, \cite{Nambu:1970}, and Goto in 1971, \cite{Goto:1971ce}, and is therefore called the Nambu-Goto action. It reads
\begin{eqnarray}
\label{eqn:ng-action}
S&=&\int_{\Sigma}d^2\xi\,\mathcal{L}
=-T\int_{\Sigma}d^2\xi\,\sqrt{-\det\left(\frac{\partial x^m}{\partial\xi^i}\frac{\partial x_m}{\partial\xi^j}\right)}
\nonumber\\
&=&-T\int_{\Sigma}d^2\xi\,\sqrt{-\dot{x}^2 {x'}^2+(\dot x_m {x'}^m)^2}\,,
\end{eqnarray}
where $\xi^i=(\tau,\sigma)$ parameterizes the worldsheet $\Sigma$ (see figure \ref{fig:worldsheet}), a dot denotes derivative with respect to $\tau$ and a prime derivative with respect to $\sigma$.

The constant of proportionality, $T$, has dimension $[mass]^2$ so as to make the action dimensionless and can be (classically) interpreted as the tension of the string\footnote{This can be seen by looking at a classical solution and seeing that the energy per unit length of the string is indeed T.}. In string theory the tension is often written in terms of another parameter as $T=\frac{1}{2\pi\alpha'}$, where $\alpha'$ is known as the Regge slope (for historical reasons). As in the case of the particle this action doesn't make sense for a tensionless string. To find an action which makes sense when $T=0$ we follow the same procedure as in the particle case.

First we need the Hamiltonian of the system. The conjugate momenta (or rather momentum densities) are
\begin{equation}
p_m\equiv\frac{\partial\mathcal{L}}{\partial\dot x^m}
=T\frac{\dot x_m {x'}^2-x'_m(\dot x^m x_m')}{\sqrt{-\dot x^2 {x'}^2+(\dot x^m x_m')^2}}\,.
\end{equation}
As was the case for the particle the conjugate momenta are not all independent. We have the following two constraints
\begin{eqnarray}
p_m{x'}^m&=&0\\
p^2+T^2 x^{\prime 2}&=&0
\end{eqnarray}
and the canonical Hamiltonian (density), 
\begin{equation}
\mathcal{H}_{\mr{can}}\equiv p_m\dot x^m-\mathcal{L}\,,
\end{equation}
is again easily seen to vanish. This is a consequence of the fact that again the system is invariant under reparametrizations, $(\tau,\sigma)\rightarrow(\tau',\sigma')$ where $\tau'=\tau'(\tau,\sigma)$ and $\sigma'=\sigma'(\tau,\sigma)$ are arbitrary functions of the original parameters.

The full Hamiltonian is
\begin{equation}
\mc H=\mc H_{\mr{can}}+\lambda p_m {x'}^m+\frac{\rho}{2}(p^2+T^2 {x'}^2)\,,
\end{equation}
where $\lambda$ and $\rho$ are Lagrange multipliers enforcing the two constraints. The dimension of $\rho$ is $[mass]^{-2}$ while $\lambda$ is dimensionless.

The phase space Lagrangian becomes
\begin{equation}
\mathcal{L}_{\mr{ps}}=p_m\dot x^m-\lambda p_m {x'}^m-\frac{\rho}{2}(p^2+T^2 {x'}^2)
\end{equation}
and the Euler-Lagrange equation for $p_m$ gives
\begin{equation}
p_m=\frac{1}{\rho}\dot x_m-\frac{\lambda}{\rho} x'_m\,.
\end{equation}
Using this in the phase space Lagrangian we get the new action
\begin{eqnarray}
S'&=&\int_{\Sigma}d^2\xi\,\Big(\frac{1}{2\rho}\dot x^2-\frac{\lambda}{\rho}\dot x^m x_m'+\frac{\lambda^2-T^2\rho^2}{2\rho}{x'}^2\Big)\,.
\label{eqn:sprime2}
\end{eqnarray}
When $T\neq 0$ this action is equivalent to the Nambu-Goto action that we started with, (\ref{eqn:ng-action}). This can be easily seen by solving the Euler-Lagrange equations for $\lambda$ and $\rho$ and plugging the solutions back into (\ref{eqn:sprime2}).

The action $S'$ in (\ref{eqn:sprime2}) is usually written in another way. If we introduce the object
\begin{equation}
g^{ij}\equiv
\left(\begin{array}{cc}
-1 & \lambda\\
\lambda & T^2\rho^2-\lambda^2\\
\end{array}\right)
\end{equation}
and define 
\begin{equation}
g\equiv\det((g^{ij})^{-1})=\frac{1}{\det(g^{ij})}=-\frac{1}{T^2\rho^2}\, 
\end{equation} 
we can write the action as 
\begin{equation}
\label{eqn:polyakov}
S'=-\frac{T}{2}\int_{\Sigma}d^2\xi\,\sqrt{-g}g^{ij}\partial_i x^m\partial_j x^n\eta_{mn}\,. 
\end{equation}
This is the Brink-Di Vecchia-Howe-Deser-Zumino action, \cite{Brink:1976sc,Deser:1976rb}, or, as it is usually called, the Polyakov action \cite{Polyakov:1981rd,Polyakov:1981re}. The reason for introducing $g^{ij}$ is that under the reparametrizations considered before, $\xi^i\rightarrow\xi^{\prime i}(\xi)$,
$g^{ij}$ transforms as a (contravariant) symmetric two-tensor. It (or rather $g_{ij}\equiv(g^{ij})^{-1}$) thus has an interpretation as a metric on the worldsheet of the string. But wait a minute, we started with two parameters, $\lambda$ and $\rho$, and replaced them with a symmetric $2\times 2$ matrix $g^{ij}$ which has three independent components, what has happened? The answer is that the action (\ref{eqn:polyakov}) has an extra symmetry called Weyl invariance, it is invariant under rescaling of the worldsheet metric,
\begin{equation}
g^{ij}(\xi)\rightarrow\Lambda(\xi)g^{ij}(\xi)\,, 
\end{equation}
where $\Lambda$ is an arbitrary positive definite function. This extra symmetry can be used to remove the third component of $g^{ij}$.

When $T=0$ (we take the dimensionless combination $\rho T\rightarrow 0$) the action $S'$ in (\ref{eqn:sprime2}), which still makes sense, can no longer be written in the form (\ref{eqn:polyakov}). In this case we can instead introduce another object
\begin{equation}
V^i\equiv\frac{1}{\sqrt{\rho}}\left(\begin{array}{c}
1 \\
-\lambda\\
\end{array}\right)\,.
\end{equation}
The action for the tensionless bosonic string (in flat space) can then be written
\begin{equation}
S_0=\frac{1}{2}\int_{\Sigma}d^2\xi\,V^iV^j\partial_i x^m\partial_j x^n\eta_{mn}\,.
\end{equation}
$V^i$ turns out to transform as a vector density under reparametrizations of the worldsheet \cite{Lindstrom:1990ar,Lindstrom:1990qb}. The tensionless bosonic string, also called null-string, was first considered by Schild in \cite{Schild:1976vq} and later in \cite{Karlhede:1986wb}. A discussion similar to the one given here can be found in \cite{Isberg:1993av}.

\section{The $AdS_5\times S^5$ plane wave}\label{sec:plane-wave}
The ten-dimensional space $AdS_5\times S^5$, the product of five-dimensional anti-de Sitter space and a five-sphere, has turned out to be extremely important in string theory. It is what is known as a maximally supersymmetric space allowing $32$ supersymmetries, the maximum possible in ten-dimensions. The only other two maximally supersymmetric spaces are flat ten-dimensional Minkowski space and the plane wave space we will discuss shortly. $AdS_5\times S^5$ arises in string theory as a particular limit, zooming in on the near-horizon geometry, of a stack of $N$ coincident D3-branes. 

In 1997, motivated by a close analysis of this system, Maldacena formulated his celebrated AdS/CFT conjecture, \cite{Maldacena:1997re}, which was later made precise in the works of Witten \cite{Witten:1998qj} and Gubser, Klebanov and Polyakov \cite{Gubser:1998bc}. This conjecture says that string theory on five-dimensional anti-de Sitter space is dual to, \emph{i.e.} describes the same physics as, a four-dimensional (supersymmetric and conformal) field theory known as $\mc N=4$ super Yang-Mills theory which can be thought of as living on the stack of D3-branes. This remarkable correspondence, and similar ones, relate a theory with gravity in $d+1$ dimensions, in this case type IIB string theory in $AdS_5$, to a gauge theory (without gravity!) in $d$ dimensions. The correspondence has been tested in many ways by working out a quantity on one side of the correspondence and finding a matching quantity on the other side, but a proof of the conjecture is still lacking. AdS/CFT ideas can even been used to gain insights into QCD, the theory describing the strong interaction, by computing things in string theory (or supergravity)! This is a vast and fascinating subject but we will not have more to say about it here.

Unfortunately describing the string on $AdS_5\times S^5$ is complicated so we would like to find some way of simplifying the problem. One way of doing this is to consider the so called pp-wave limit of this space instead. The idea goes back to Penrose \cite{Penrose:1976} who showed that any space-time has a limit in which a neighborhood of a segment of null geodesic (the worldline of a massless particle) becomes a special type of space-time called a plane-parallel wave, or pp-wave. This idea was later extended by G\"{u}ven in \cite{Gueven:2000ru} to include supergravity backgrounds in ten and eleven dimensions. The limit can be roughly thought of as zooming in on the geometry seen by a massless particle moving in the space under consideration, in our case $AdS_5\times S^5$.

Taking this limit of $AdS_5\times S^5$ (along a particular null geodesic), as described in \cite{Blau:2002dy}, we get a new space with metric
\begin{equation}
\label{eq:pp-metric}
ds^2=2dx^{+}dx^{-}-\mu^2x^I x_I(dx^+)^2+dx^I dx_I\,,
\end{equation}
where the ten coordinates, $x^m$, split up into the light-cone directions $x^{\pm}\equiv\frac{1}{\sqrt{2}}(x^{9}\pm x^{0})$ and the transverse directions $x^I$ with $I=1,\ldots,8$ and $\mu=\frac{1}{R}$ where $R$ is the original radius of curvature of $AdS_5$ (which is the same as the radius of curvature of the $S^5$). Because the D3-branes carry a charge the original space, $AdS_5\times S^5$, also carries some flux of the corresponding field strength. For the case of D3-branes the field strength is a five-form field, $F^{(5)}$. The surviving components of this field strength, after taking the plane wave limit, are
\begin{equation}
\label{eq:pp-fiveform}
F^{(5)}_{+1234}=F^{(5)}_{+5678}=2\mu\,.
\end{equation}
The point of the Penrose-G\"{u}ven limit is that this space is then guaranteed to also be a solution to the (type IIB) supergravity equations and, what is more, it is guaranteed to also be maximally supersymmetric, \emph{i.e.} having 32 supersymmetries, since the space of which it is a limit has this property.

This plane wave space was first obtained in another way however, in \cite{Blau:2001ne}.

\section{The superstring in this plane wave background}
In this section we will show how to quantize the superstring in the plane wave background discussed above.

\subsection{The action}
The plane wave background (\ref{eq:pp-metric}) and (\ref{eq:pp-fiveform}) is simple enough to quantize the string in. The action for the superstring in this background was found by Metsaev in \cite{Metsaev:2001bj} and the quantization was performed by Metsaev and Tseytlin in \cite{Metsaev:2002re}.

The formulation of the superstring used is that due to Green and Schwarz in which the superstring is described as a string in a background superspace\footnote{The other formulation, called the NSR-formulation, is problematic when the background has non-trivial Ramond-Ramond fluxes, in our case of the five-form field strength $F^{(5)}$.}. The (type IIB) superspace has the usual ten bosonic coordinates $x^m$ for $m=0,\ldots,9$, but in addition to this it has two fermionic ''coordinates'' $\theta^{1\alpha}$ and $\theta^{2\alpha}$ for $\alpha=1,\ldots,16$ which are Majorana-Weyl spinors of $SO(9,1)$. 

In the Green-Schwarz formulation of the superstring there is an important fermionic symmetry related to supersymmetry known as kappa-symmetry (we will have more to say about this symmetry in the coming chapters). This symmetry can be removed (or gauge-fixed) by making a specific choice of gauge as follows
\begin{equation}
\gamma^+\theta^1=\gamma^+\theta^2=0\,,
\end{equation}
sometimes called fermionic light-cone gauge. With this choice of gauge the action simplifies and it takes the form
\begin{equation}
\label{eqn:metsaev-action}
S=-\frac{T}{2}\int_{\Sigma}d^{2}\xi\,\left(\sqrt{-g}g^{ij}h_{ij}-2i\varepsilon^{ij}\partial_ix^+\left(\theta^1\gamma^-\partial_j\theta^1-\theta^2\gamma^-\partial_j\theta^2\right)\right)\,,
\end{equation}
where we have defined the induced metric on the worldsheet\footnote{Here we denote it by $h$ to avoid confusion with the independent worldsheet metric $g$.}
\begin{eqnarray}
h_{ij}&=&2\partial_{(i}x^+\partial_{j)}x^- -\left(\mu^2x^Ix_I+2\mu\theta^1\gamma^-\Pi\theta^2\right)\partial_ix^+\partial_jx^+ 
\nonumber\\
&&{}+\partial_{(i}x^+\left(\theta^1\gamma^-\partial_{j)}\theta^1 + \theta^2\gamma^-\partial_{j)}\theta^2\right)+\partial_ix^I\partial_jx_I\,.
\label{eq:h_ij}
\end{eqnarray}
We have defined the product of gamma matrices $\Pi\equiv\gamma^{1234}$ which arises from terms involving $F^{(5)}_{+\mu_1\cdots\mu_4}\gamma^{\mu_1\cdots\mu_4}$.

\subsection{The light-cone gauge}
The action (\ref{eqn:metsaev-action}) is invariant under reparametrizations of the worldsheet,
$\xi^i\rightarrow\xi'^i(\xi)$, and Weyl transformations, $g_{ij}\rightarrow \Lambda g_{ij}$, as discussed previously for the bosonic string. These symmetries allow us to choose a specific form of the worldsheet metric $g_{ij}$ that allows us to quantize the theory. A convenient gauge choice is to take
\begin{equation}
g_{ij}=\eta_{ij}
\equiv\left(
\begin{array}{cc}
-1 & 0\\
0 & 1
\end{array}\right)\,.
\end{equation}
But we still have to satisfy the equations of motion (\emph{i.e.} the Euler-Lagrange equations) of the worldsheet metric $g_{ij}$ which, after choosing this gauge, become constraints that have to be imposed on the system. These are known as the Virasoro constraints and amount to the vanishing of the worldsheet energy-momentum tensor,
\begin{equation}
\label{eqn:virasoro}
0=T_{ij}\equiv-\frac{2}{T\sqrt{-g}}\frac{\delta S}{\delta g^{ij}}=h_{ij}-\frac{1}{2}g_{ij}g^{kl}h_{kl}\,,
\end{equation}
where we have used the relation $\delta g=-g g_{ij}\,\delta g^{ij}$ when varying $g^{ij}$ in the action (remember that
$g\equiv\det{g_{ij}}$).

Even after choosing $g_{ij}=\eta_{ij}$ it turns out that there is still some freedom left in the choice of parametrization of the worldsheet. The reparametrizations
\begin{eqnarray}
\xi^{+}&\rightarrow & f(\xi^{+})\nonumber\\
\xi^{-}&\rightarrow & h(\xi^{-})\,,
\end{eqnarray}
where $\xi^{\pm}\equiv\tau\pm\sigma$ and $f$ and $h$ are arbitrary functions, combined with the Weyl transformation
\begin{equation}
g_{ij}\rightarrow f'h'g_{ij}\,,
\end{equation}
where $f'$ and $h'$ denotes the derivatives of $f$ and $h$, leaves the form of $g_{ij}=\eta_{ij}$ unchanged. 

In order to write an action that contains only the physical degrees of freedom we should also fix this remaining gauge symmetry. It turns out that this can be done by choosing the so-called light-cone gauge
\begin{equation}
\label{eq:lc-condition}
x^+=\frac{p^+}{T}\tau\,.
\end{equation}
With this choice $p^{+}$ becomes the momentum canonically conjugate to $x^{-}$.

This gauge is very important because it lets us solve the Virasoro conditions, (\ref{eqn:virasoro}), explicitly. In light-cone gauge they read
\begin{eqnarray}
p^{+}x^{-\prime}+ip^{+}(\theta^1\gamma^{-}\theta^{1\prime}
+\theta^2\gamma^-\theta^{2\prime})+T\dot{x}^{I}x_{I}' &=&0\label{eq:lc-virasoro1}\\
\nonumber\\
2p^{+}\dot{x}^{-}+2ip^{+}(\theta^1\gamma^-\dot{\theta}^1+\theta^2\gamma^-\dot{\theta}^2)
-4imp^{+}\theta^1\gamma^-\Pi\theta^2 &&
\nonumber\\
{}-m^{2}Tx_{I}^{2}+T\dot{x}^{I}\dot{x}_{I}+Tx^{I\prime}x_{I}'&=&0\,,
\label{eq:lc-virasoro2}
\end{eqnarray}
where we have introduced the dimensionless quantity $m\equiv\mu p^{+}/T$. These conditions can be solved
for $x^{-}$ (up to an integration constant) in terms of the $x^I$ and $\theta^i$. We have now fixed the
reparametrization invariance of the action completely and in doing so we have eliminated $x^{+}$ and $x^{-}$ and
we are left with an action for the transverse coordinates $x^{I}$ only\footnote{The term $p^{+}\dot{x}^{-}$ has been omitted
since it doesn't contribute to the equations of motion, it is however needed in checking the supersymmetry of the action.}
\begin{eqnarray}
S_{\mr{lc}}&=&-\frac{T}{2}\int_{\Sigma}d^2\xi\,\Big(\partial_ix^{I}\partial^ix_{I}+m^2x^I x_I
\nonumber\\
&&{}-2i\frac{p^+}{T}\left(\theta^1\gamma^-\partial_+\theta^1 + \theta^2\gamma^-\partial_-\theta^2\right)
+4im\frac{p^+}{T}\theta^1\gamma^-\Pi\theta^2\Big)\,,
\nonumber\\
\label{eq:lc-action}
\end{eqnarray}
where we have defined $\partial_{\pm}\equiv\partial/\partial\tau\pm\partial/\partial\sigma$. This action is now quadratic in the fields and can therefore be easily quantized as we will now describe.

\subsection{Quantization and the spectrum}
The equation of motion for $x^I$ simply becomes the Klein-Gordon equation
\begin{equation}
\left(-\frac{\partial^2}{\partial\tau^2}+\frac{\partial^2}{\partial\sigma^2}\right)x^I-m^2x^I=0\,. 
\end{equation}
We will consider the case of a closed string. If we take $\sigma\in[0,1]$ the appropriate boundary condition is then
\begin{equation}
x^I(\tau,\sigma+1)=x^I(\tau,\sigma)\,.
\end{equation}
The solution to the Klein-Gordon equation subject to this condition is 
\begin{eqnarray}
x^I(\tau,\sigma)&=&x_0^I\cos(m\tau)+\frac{p_0^I}{mT}\sin(m\tau)
\nonumber\\
&&{}+i\sum_{n\neq 0}\frac{1}{\omega_{n}}e^{-i\omega_{n}\tau}\left( \alpha_{n}^{1I}e^{i2\pi n\sigma}+\alpha_{n}^{2I}e^{-i2\pi n\sigma}\right)\,,
\end{eqnarray}
where $\omega_{n}\equiv\mr{sign}(n)\sqrt{(2\pi n)^{2}+m^2}$. The momentum conjugate to $x^I$ is
\begin{equation}
p_I\equiv\frac{\partial\mathcal{L}}{\partial\dot{x}^I}=T\dot{x}_I\,.
\end{equation}
The canonical equal-time commutation relation,
\begin{equation}
[p_I(\tau,\sigma),x^J(\tau,\sigma')]=-i\delta_I^J\delta(\sigma-\sigma')\,,
\end{equation}
gives the commutation relations for the modes
\begin{equation}
\label{eq:alfa-commutator}
[p_0^I,x_0^J]=-i\delta^{IJ}\quad,\quad
[\alpha_m^{iI},\alpha_n^{jJ}]=\frac{1}{2}\frac{\omega_m}{T}\delta_{n+m,0}\delta^{IJ}\delta^{ij}\,,
\end{equation}
for $n,m=\pm1,\pm2,\ldots$.

Now let's consider the fermions. The equations of motion are
\begin{equation}
\partial_+\theta^1-m\Pi\theta^2=0
\end{equation}
and
\begin{equation}
\partial_-\theta^2+m\Pi\theta^1=0\,.
\end{equation}
These are to be supplemented by the closed string boundary condition 
\begin{equation}
{\theta^i(\sigma+1,\tau)} =\theta^i(\sigma,\tau)\,.
\end{equation}
The solutions are
\begin{eqnarray}
\theta^1(\tau,\sigma)&=&\theta_0^1\cos(m\tau)+\Pi\theta_0^2\sin(m\tau)
\nonumber\\
&&{}+i\sum_{n\neq 0}c_n e^{-i\omega_n\tau}\left(\theta_n^1e^{i2\pi n\sigma}+i\frac{\omega_n-2\pi n}{m}\Pi\theta_n^2e^{-i2\pi n\sigma}\right)
\nonumber\\
\end{eqnarray}
and
\begin{eqnarray}
\theta^2(\tau,\sigma)&=&\theta_{0}^{2}\cos(m\tau)-\Pi\theta_{0}^{1}\sin(m\tau)
\nonumber\\
&&{}+i\sum_{n\neq 0}c_{n}e^{-i\omega_{n}\tau}\left(\theta_{n}^{2}e^{i2\pi n\sigma}-i\frac{\omega_{n}-2\pi n}{m}\Pi\theta_{n}^{1}e^{-i2\pi n\sigma}\right)\,,
\nonumber\\
\end{eqnarray}
where $c_{n}=1/\sqrt{1+(\omega_{n}-2\pi n)^{2}/m^{2}}$. The momenta conjugate to $\theta^{1\alpha}$ and $\theta^{2\alpha}$ are
\begin{equation}
p^i_\alpha\equiv\frac{\partial\mathcal{L}}{\partial\dot{\theta}^{\alpha i}}=-ip^+(\theta^i\gamma^-)_\alpha\,.
\end{equation}
The fact that the momenta depend only on the $\theta^i$ themselves and not on the $\dot{\theta}^i$ means that this is a constraint. When
there is a constraint we must use Poisson-Dirac brackets, which take the constraint into account, and doing this we find the anti-commutation relations for the modes
\begin{equation}
\label{eq:theta-commutator}
\{\theta_m^{\alpha i},\theta_n^{\beta j}\}=\frac{1}{4p^+}(\gamma^+)^{\alpha\beta}\delta_{n+m,0}\delta^{ij}\quad n,m=0,\pm1,\pm2,\ldots\,.
\end{equation}
The $\gamma^+$ on the right is due to the fermionic light-cone condition $\gamma^+\theta^i=0$.

We must not forget to also impose closed string boundary conditions on the remaining coordinates $x^{\pm}$. $x^{+}$ doesn't depend on
$\sigma$ so it is okay but for $x^{-}$ we get, by using the first of the Virasoro conditions, (\ref{eq:lc-virasoro1}), and the solutions
for $x^{I}$ and $\theta^{i}$
\begin{eqnarray}
0&=&x^{-}(\tau,1)-x^{-}(\tau,0)=\int_{0}^{1}d\sigma x^{-\prime}
\nonumber\\
&=&-\int_{0}^{1}d\sigma\Big(i(\theta^1\gamma^-\theta^{1\prime}+\theta^2\gamma^-\theta^{2\prime})+\frac{T}{p^{+}}\dot{x}^{I}x_{I}'\Big)
\nonumber\\
&=&\sum_{n\neq 0}n\left(\frac{T}{p^{+}\omega_{n}}\alpha^{1I}_{-n}\alpha^{1}_{In}+\theta_{-n}^1\gamma^{-}\theta_n^2\right)-(1\leftrightarrow 2)\,.
\end{eqnarray}
This extra condition can be written
\begin{equation}
\label{eqn:class-levelmatching}
N^1=N^2\qquad\mathrm{where}\qquad
N^i=\sum_{n\neq 0}n\left(\frac{T}{\omega_n}\alpha^{iI}_{-n}\alpha^i_{In}+p^{+}\theta_{-n}^i\gamma^{-}\theta_n^i\right)\,.
\end{equation}
(Note that there is no sum on $i$ in the above expression). The light-cone Hamiltonian density is defined as
\begin{eqnarray}
\mathcal{H}_{\mr{lc}}&\equiv&\dot{\theta}^{1\alpha}p_{\alpha}^{1}+\dot{\theta}^{2\alpha}p_{\alpha}^{2}+\dot{x}^{I}p_{I}-\mathcal{L}_{\mr{lc}}
\nonumber\\
&=&\frac{1}{2T}(p^{I}p_{I}+T^{2}x^{I\prime}x_{I}'+m^{2}T^{2}x^{I}x_{I})
\nonumber\\
&&{}-ip^{+}\left(\theta^1\gamma^{-}\theta^{1\prime}-\theta^2\gamma^{-}\theta^{2\prime}\right)
+2imp^{+}\theta^1\gamma^{-}\Pi\theta^2\,.
\end{eqnarray}
Using the mode expansions for the fields the total light-cone Hamiltonian becomes
\begin{eqnarray}
H&\equiv&\int_0^1d\sigma\,\mathcal{H}_{\mr{lc}}=\frac{1}{2T}p_{0I}p_0^{I}+\frac{1}{2}m^{2}Tx_0^Ix_{0I}
+2imp^{+}\theta_0^1\gamma^-\Pi\theta_0^2
\nonumber\\
&&{}+2\sum_{n=1}^\infty\sum_{i=1}^2\left(T\alpha_{-n}^{iI}\alpha_{In}^i+p^{+}\omega_n\theta^i_{-n}\gamma^-\theta_n^i\right)\,,
\end{eqnarray}
where the expression has been put in normal-ordered form (there is no normal-ordering ambiguity here since each term coming from reordering the $\alpha$'s is exactly cancelled by a corresponding one coming from reordering the $\theta$'s).

We can introduce the following dimensionless creation and annihilation operators
\begin{eqnarray}
a_0^I=\sqrt{\frac{T}{2m}}(\frac{p_0^I}{T}-imx_0^I)&,&a_{0}^{\dagger I}=\sqrt{\frac{T}{2m}}(\frac{p_0^I}{T}+imx_0^I)\nonumber\\
a_n^{iI}=\sqrt{\frac{2T}{\omega_{n}}}\alpha_{n}^{iI}&,&a_{n}^{\dagger iI}=\sqrt{\frac{2T}{\omega_{n}}}\alpha_{-n}^{iI}\quad n=1,2,\ldots\nonumber\\
\eta_0=\sqrt{\frac{p^{+}}{2}}(\theta^1_0-i\theta^2_0)&,&\eta_0^\dagger=\sqrt{\frac{p^{+}}{2}}(\theta^1_0+i\theta^2_0)\nonumber\\
\eta_{n}^i=\sqrt{2p^{+}}\theta_{n}^i&,& \eta_{n}^{\dagger i}=\sqrt{2p^{+}}\bar{\theta}_{-n}^i\quad n=1,2,\ldots
\end{eqnarray}
in terms of which the commutation relations (\ref{eq:alfa-commutator}) and (\ref{eq:theta-commutator}) become
\begin{eqnarray}
[a_0^I,a_0^{\dagger J}]=\delta^{IJ} &,&
[a_{m}^{iI},a_{n}^{\dagger jJ}]=\delta_{mn}\delta^{IJ}\delta^{ij}
\\
\{\eta_0^\alpha,\eta_0^{\dagger\beta}\}=\frac{1}{4}(\gamma^+)^{\alpha\beta} &,&
\{\eta_m^{i\alpha},\eta_n^{\dagger j\beta}\}=\frac{1}{2}(\gamma^+)^{\alpha\beta}\delta_{mn}\delta^{ij}
\end{eqnarray}
and the Hamiltonian becomes
\begin{eqnarray}
H&=&4m+ma_0^{\dagger I}a_{0I}+2m\eta^{\dagger}_0\gamma^-\Pi\eta_0
\nonumber\\
&&{}+\sum_{n=1}^{\infty}\sqrt{(2\pi n)^{2}+m^{2}}\,\sum_{i=1}^2\left(a_{n}^{\dagger iI}a_{In}^i+\eta^{\dagger i}_{n}\gamma^{-}\eta_{n}^i\right)\,.
\label{eq:hamiltonian}
\end{eqnarray}
The $4m$ comes from normal-ordering the bosonic zero-modes, the reordering of the fermionic zero-modes gives no contribution because of the relation $\mr{tr}(\gamma^{+}\gamma^{-}\Pi)=0$. Recalling that $m=\mu p^+/T$ we see that the energies of the states depend on two scales, the string tension $T$ and the inverse radius of curvature of the original $AdS_{5}\times S^{5}$ solution $\mu$ (the actual space-time energy is given by $\frac{T}{p^{+}}H$). We can also see that the fermionic zero-modes break the rotation symmetry down from $SO(8)$ to $SO(4)\times SO(4)\times\mathbb{Z}_{2}$ because of the appearance of $\Pi$. This is because, as is easily seen from (\ref{eq:pp-metric}) and (\ref{eq:pp-fiveform}), although the metric of the plane wave has $SO(8)$ symmetry the five-form field strength $F^{(5)}$ does not.

The vacuum state is defined to be the state annihilated by all the annihilation operators
\begin{equation}
a_0^I|0\rangle=0,\quad a_n^{iI}|0\rangle=0,\quad\eta_0^\alpha|0\rangle=0,\quad\eta_n^{i\alpha}|0\rangle=0\quad (n=1,2,\ldots)\,.
\end{equation}
The Fock space is then built up by acting on the vacuum state with a product of the creation operators $a_0^{\dagger I}$,
$a_n^{\dagger iI}$, $\eta_0^{\dagger\alpha}$ and $\eta_n^{\dagger i\alpha}$. The subspace of physical states is obtained
by imposing the condition
\begin{equation}
(N^1-N^2)|\Phi_{phys}\rangle=0,\quad\mathrm{where}\quad N^i=\sum_{n=1}^{\infty}n\left(a^{\dagger iI}_{n}a^i_{In}+\eta_{n}^{\dagger i}\gamma^{-}\eta_n^i\right)
\end{equation}
and we have again normal-ordered the classical expression (\ref{eqn:class-levelmatching}). As for the Hamiltonian there is no
normal-ordering ambiguity. This condition means that a general state in the Fock space
\begin{equation}
a^{\dagger1 I_{1}}_{n_{1}}\ldots a^{\dagger1 I_{N}}_{n_{N}}
a^{\dagger2 J_{1}}_{m_{1}}\ldots a^{\dagger2 J_{M}}_{m_{M}}
\eta^{\dagger1 \alpha_{1}}_{p_{1}}\ldots \eta^{\dagger1 \alpha_{P}}_{p_{P}}
\eta^{\dagger2 \beta_{1}}_{q_{1}}\ldots \eta^{\dagger2 \beta_{Q}}_{q_{Q}}
\underbrace{a_{0}^{\dagger}\ldots}_{K}
\underbrace{\eta_{0}^{\dagger}\ldots}_{L}
|0\rangle
\end{equation}
is physical if and only if
\begin{equation}
\sum_{i=1}^{N}n_{i}+\sum_{i=1}^{P}p_{i}=\sum_{i=1}^{M}m_{i}+\sum_{i=1}^{Q}q_{i}\,.
\end{equation}
This is known as the level-matching condition.

The (physical) states built up by acting on the vacuum state with the zero-mode operators $a_{0}^{\dagger I}$ and 
$\eta_{0}^{\dagger\alpha}$ have energies set by $\mu$, the inverse curvature radius of the original space, and they
can be shown to give the supergravity spectrum in the $AdS_{5}\times S^{5}$ plane wave background, see \cite{Metsaev:2002re}. There the generators of the super Poincar\'{e} group are also computed and it is shown that they give a realization of the correct supersymmetry algebra. This is quite important to check because we fixed the
light-cone gauge in the classical theory and then quantized the system, but there is no guarantee that the gauge fixing conditions can be maintained in the quantum theory in general. Any problems in a treatment where the super Poincar\'{e} symmetry is not explicitly maintained should show up as an anomaly in the algebra of the generators of this group in the quantum theory. For example this is how, in light-cone gauge quantization of the superstring in flat space, one discovers the condition that space-time must be ten-dimensional. Here there is no anomaly since we have already put the string in a consistent supergravity background.

\section{The tensionless superstring in the plane wave}\label{sec:tzeroppwave}
In this section we show how to describe the tensionless superstring in the
$AdS_5\times S^5$ plane wave background, a Penrose-G\"{u}ven limit of the
near-horizon geometry of $N$ D3-branes, described in section
\ref{sec:plane-wave}. This is done in \cite{Bredthauer:2004kv}.

\subsection{The tensionless limit of the action}
The action for the superstring in the $AdS_5\times S^5$ plane wave background given in (\ref{eqn:metsaev-action}) in the previous section is
\begin{equation}
S=\int d^2\xi\,\left(-\frac{T}{2}\sqrt{-g}g^{ij}h_{ij}+iT\varepsilon^{ij}\partial_ix^{+}
\left(\theta^1\gamma^-\partial_j\theta^1-\theta^2\gamma^-\partial_j\theta^2\right)\right)\,,
\label{eqn:ksymfixedL}
\end{equation}
where the induced metric $h_{ij}$ is given in (\ref{eq:h_ij}).

To be able to take the tensionless limit we must first write the action in Nambu-Goto form by eliminating $g^{ij}$. Then we can follow
the same steps as in section \ref{sec:tensionless} where we took the tensionless limit in flat Minkowski space. The Euler-Lagrange equation for
$g^{ij}$ was given in (\ref{eqn:virasoro}) and it reads
\begin{equation}
\frac{1}{2}g_{ij}g^{kl}h_{kl}=h_{ij}\,.
\end{equation}
Taking the square-root of minus the determinant of this equation we get
\begin{equation}
\frac{1}{2}\sqrt{-g}g^{ij}h_{ij}=\sqrt{-\det{h}}\,,
\end{equation}
and if we plug this back in the action (\ref{eqn:ksymfixedL}) we get the Nambu-Goto form of the action
\begin{equation}
S_{\mr{NG}}=-T\int_{\Sigma}d^2\xi\,\Big(\sqrt{-\det{h}}
-i\varepsilon^{ij}\partial_{i}x^+\left(\theta^1\gamma^-\partial_j\theta^1-\theta^2\gamma^-\partial_j\theta^2\right)\Big)\,.
\end{equation}
We can now take the tensionless limit of this action in an exactly analogous way to the case of the bosonic string in flat space described in section \ref{sec:tensionless}. As described there the idea is to construct the phase-space action and then take $T\rightarrow0$ \cite{Lindstrom:1990ar,Lindstrom:1990qb} (this process is described in detail in \cite{Wulff:2004}). Doing this we end up with an action for the tensionless superstring on the $AdS_5\times S^5$ plane wave background which takes the form
\begin{eqnarray}
S_0&=&\frac{1}{2}\int d^2\xi\,V^iV^j\Big(2\partial_ix^{+}\left(\partial_jx^{-}+i\left(\theta^1\gamma^{-}\partial_j\theta^1+\theta^2\gamma^{-}\partial_j\theta^2\right)\right)
\nonumber\\
&&{}-\left(\mu^{2}x^{I}x_{I}+4i\mu\theta^1\gamma^{-}\Pi\theta^2\right)\partial_ix^{+}\partial_jx^{+}+\partial_ix^{I}\partial_jx_{I}\Big)\,.
\nonumber\\
\label{eq:action-with-v}
\end{eqnarray}

\subsection{Light-cone gauge}
The action (\ref{eq:action-with-v}) is, just as in the tensile case, invariant under reparametrizations of the worldsheet. Under these $V^i$ transforms as a vector density (see \cite{Isberg:1993av})
\begin{eqnarray}
\delta x^m&=& \epsilon^i\partial_ix^m\nonumber\\
\delta\theta^{\alpha i}&=&\epsilon^j\partial_j\theta^{\alpha i}\nonumber\\
\delta V^i&=&-V^j\partial_j\epsilon^i+\epsilon^j\partial_jV^i+\frac{1}{2}\partial_j\epsilon^jV^i\,.
\end{eqnarray}
And in analogy to the tensile case one can show that this symmetry allows us to fix the gauge $V^i=(v,0)$ with $v$ a constant of dimension $[mass]$.

As in the tensile case there is still a residual symmetry left after fixing this form of $V^i$. It corresponds to the conformal symmetry considered earlier. The symmetry transformations in this case are
\begin{eqnarray}
\tau\rightarrow\tilde{\tau}&=&f'(\sigma)\tau+g(\sigma)\nonumber\\
\sigma\rightarrow\tilde{\sigma}&=&f(\sigma)\,,
\end{eqnarray}
with $f$ and $g$ arbitrary functions of $\sigma$.

This allows us to go to the light-cone gauge in much the same way as in the tensile case and choose
\begin{equation}
x^+=\frac{p^+}{v^2}\tau\,.
\end{equation}
The completely gauge-fixed action for the tensionless string in this plane wave background then becomes
\begin{eqnarray}
S_{0,\mr{lc}}&=&\int_{\Sigma}d^2\xi\,\Big(\frac{v^2}{2}(\dot{x}^I\dot{x}_I-m_0^2x^Ix_I)+ip^+\left(\theta^1\gamma^{-}\dot{\theta}^1
+\theta^2\gamma^-\dot{\theta}^2\right)
\nonumber\\
&&\qquad\qquad{}-2im_0p^+\theta^1\gamma^-\Pi\theta^2\Big)\,,
\label{eq:T0-lc-action}
\end{eqnarray}
where we have set $m_0\equiv\frac{\mu p^+}{v^2}$.

The above action differs from the corresponding one for a tensile string, (\ref{eq:lc-action}), in that it contains no
$\sigma$-derivatives. This makes the $\sigma$-dependence of the fields more or less arbitrary and makes the tensionless string quite different from the tensile one. In a way it behaves more like a (continuous) collection of massless particles \cite{Karlhede:1986wb}.

The analog of the Virasoro conditions, (\ref{eq:lc-virasoro1}) and (\ref{eq:lc-virasoro2}), coming from varying $V^i$ in (\ref{eq:action-with-v}), become
\begin{eqnarray}
2p^+\dot{x}^{-}+v^2\dot{x}^I\dot{x}_I-v^2m^2x^Ix_I\qquad\qquad\qquad\qquad&&
\nonumber\\
{}+2ip^{+}\left(\theta^1\gamma^-\dot{\theta}^1+\theta^2\gamma^-\dot{\theta}^2-2m\theta^1\gamma^-\Pi\theta^2\right)&=&0
\label{eq:T0-lc-virasoro1}\\
\nonumber\\
p^{+}x^{-\prime}+v^2x_I'\dot{x}^I+ip^{+}\left(\theta^1\gamma^-\theta^{1\prime}+\theta^2\gamma^-\theta^{2\prime}\right)&=&0\,.\label{eq:T0-lc-virasoro2}
\end{eqnarray}

\subsection{Comparison to the tensile case}
We could now follow the same route as in the tensile case and quantize the
tensionless string on this background, this procedure is described in \cite{Bredthauer:2004kv}. Here we will take a short-cut and observe that the tensionless case can be obtained directly as a limit of the tensile one. The appropriate limit turns out to be
\begin{equation}
T\rightarrow0\qquad\mbox{with}\qquad\frac{T}{\tau}=v^2=\mbox{fixed}\,.
\end{equation}
The need for rescaling $\tau$ comes from the fact that it is related to the spacetime coordinate $x^+$ through the light-cone gauge condition (\ref{eq:lc-condition}). Indeed, it's not hard to see that taking this limit in (\ref{eq:lc-action}), (\ref{eq:lc-virasoro1}) and (\ref{eq:lc-virasoro2}) produces the corresponding equations for the tensionless case, (\ref{eq:T0-lc-action}), (\ref{eq:T0-lc-virasoro1}) and (\ref{eq:T0-lc-virasoro2}).

We can now obtain the Hamiltonian for the tensionless case directly by taking this limit in (\ref{eq:hamiltonian}) and remembering that the actual space-time energy, which has to remain finite in this limit, is $T/p^+$ times the Hamiltonian. The Hamiltonian for the tensionless string then becomes
\begin{equation}
\label{eq:T0-hamiltonian}
H_0=4m_0+m_0a_0^{\dagger I}a_{0I}+2m_0\eta_{0}^{\dagger1}\gamma^{-}\Pi\eta_{0}^{2}
+m_0\sum_{n=1}^{\infty}\sum_{i=1}^2\left(a_n^{\dagger iI}a_{n}^{iI}+\eta^{\dagger i}_{n}\gamma^{-}\eta_n^i\right)\,.
\end{equation}
Because there is no dependence on $n$ in the oscillator sum there will be an infinite degeneracy in the spectrum due to the $a_n^{\dagger iI}$ and $\eta_n^{\alpha i}$ contributing the same energy irrespective of $n$. This is of course in sharp contrast to the tensile case.

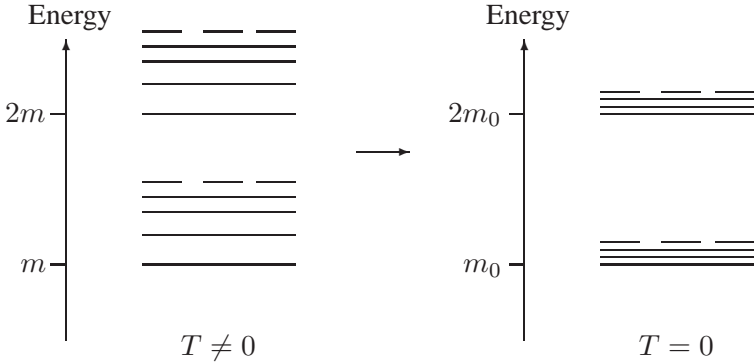
\begin{figure}
\setlength{\unitlength}{1cm}
\centering
\begin{picture}(12,6)
\put(1,0.5){\vector(0,1){4}}
\put(0.5,4.7){Energy}
\put(0.8,1.5){\line(1,0){0.2}}
\put(0.4,1.4){$m$}
\put(2,1.5){\line(1,0){2}}
\put(2,1.9){\line(1,0){2}}
\put(2,2.2){\line(1,0){2}}
\put(2,2.4){\line(1,0){2}}
\put(2,2.6){\line(1,0){0.5}}
\put(2.8,2.6){\line(1,0){0.5}}
\put(3.5,2.6){\line(1,0){0.5}}
\put(0.8,3.5){\line(1,0){0.2}}
\put(0.2,3.4){$2m$}
\put(2,3.5){\line(1,0){2}}
\put(2,3.9){\line(1,0){2}}
\put(2,4.2){\line(1,0){2}}
\put(2,4.4){\line(1,0){2}}
\put(2,4.6){\line(1,0){0.5}}
\put(2.8,4.6){\line(1,0){0.5}}
\put(3.5,4.6){\line(1,0){0.5}}
\put(2.5,0.3){$T\neq 0$}
\put(4.8,3){\vector(1,0){0.7}}
\put(7,0.5){\vector(0,1){4}}
\put(6.5,4.7){Energy}
\put(6.8,1.5){\line(1,0){0.2}}
\put(6.2,1.4){$m_0$}
\put(8,1.5){\line(1,0){2}}
\put(8,1.6){\line(1,0){2}}
\put(8,1.7){\line(1,0){2}}
\put(8,1.8){\line(1,0){0.5}}
\put(8.8,1.8){\line(1,0){0.5}}
\put(9.5,1.8){\line(1,0){0.5}}
\put(6.8,3.5){\line(1,0){0.2}}
\put(6,3.4){$2m_0$}
\put(8,3.5){\line(1,0){2}}
\put(8,3.6){\line(1,0){2}}
\put(8,3.7){\line(1,0){2}}
\put(8,3.8){\line(1,0){0.5}}
\put(8.8,3.8){\line(1,0){0.5}}
\put(9.5,3.8){\line(1,0){0.5}}
\put(8.5,0.3){$T=0$}
\end{picture}
\caption{Schematic picture showing how the energy levels become degenerate in the tensionless limit.}
\label{fig:spectra}
\end{figure}

The states in the Fock space are built from the vacuum in the same way as in the tensile case and the subspace of physical states is obtained by imposing the condition 
\begin{equation}
(N^{1}-N^{2})|\Phi_{phys}\rangle=0,\quad\mathrm{where}\quad N^i=\sum_{n=1}^{\infty}n\left(a^{\dagger iI}_{n}a^i_{In}+\eta_{n}^{\dagger i}\gamma^{-}\eta_n^i\right)\,.
\end{equation}
This level-matching condition is exactly the same as in the tensile case. The physical states will therefore be of exactly the same form, although the spectra are quite different. An illustration of how the spectrum of the tensile case, (\ref{eq:hamiltonian}), degenerates in the tensionless limit, (\ref{eq:T0-hamiltonian}), is given in figure \ref{fig:spectra}. 

Although we have not described it here the generators of the super Poincar\'{e} group can also be constructed in this case and the algebra verified. They again agree with the limit described above of the tensile ones.

\section{Conclusions}
In section \ref{sec:tzeroppwave} we obtained the quantized tensionless
superstring in the $AdS_{5}\times S^{5}$ plane wave background as a limit of the
quantized tensile string. As is shown in \cite{Bredthauer:2004kv} the same result is obtained if one instead quantizes the classical action for the tensionless string directly. The fact that the two procedures give the same result means that the following diagram commutes
\setlength{\unitlength}{0.8cm}
\begin{center}
\begin{picture}(12,5)
\put(3,4){\circle*{0.1}}
\put(3,1){\circle*{0.1}}
\put(9,4){\circle*{0.1}}
\put(9,1){\circle*{0.1}}
\put(3,3.8){\vector(0,-1){2.6}}
\put(9,3.8){\vector(0,-1){2.6}}
\put(3.2,4){\vector(1,0){5.6}}
\put(3.2,1){\vector(1,0){5.6}}
\put(1.2,2.5){$\mathbf{T\rightarrow 0}$}
\put(4.8,4.5){\textbf{Quantization}}
\put(1.6,4.5){Classical}
\put(1.3,4.2){tensile string}
\put(7.8,0.5){Quantized}
\put(7.3,0.1){tensionless string}
\end{picture}
\end{center}
This property, that taking the tensionless limit ''commutes'' with quantization, is not in general obvious. One important condition is that the background should be a solution to all orders in $\alpha'$ ($\propto\frac{1}{T}$), otherwise the limit $T\rightarrow 0$ is in some sense uncontrollable. The plane wave background we've considered here indeed fulfills this criterion and so does flat Minkowski space. There is however an important difference between the tensionless limit in flat space and in the $AdS_5\times S^5$ plane wave background. In flat Minkowski space there is no other energy scale than that set by the tension, whereas in the plane wave there is another energy scale set by $\mu=\frac{1}{R}$ where $R$ is the radius of curvature of the original $AdS_5\times S^5$ solution that arose as the near-horizon geometry of N coincident D3-branes. Since $\mu$ and $T$ appear in the combination $m=\frac{\mu p^{+}}{T}$ it seems that we can think of the tensionless limit instead as $\mu\rightarrow\infty$ or equivalently $R\rightarrow 0$,
the infinite curvature limit. This is a hint that there could be a problem with taking the tensionless limit in flat space, since flat space is recovered in the plane wave solution by taking $\mu\rightarrow 0$, in some sense the opposite to the tensionless limit. The tensionless limit in flat space indeed seems less well-behaved than in the plane wave background. In \cite{Metsaev:2002re} it is even suggested that the parameter $\mu$ can be viewed as a ''regularization'' introduced in order to define a non-trivial tensionless limit of the superstring in flat space.

An important physical question is whether it is really sensible to actually take the tension of a string to be zero since the tension is what holds the string together in the first place. Taking the tension to zero might cause the string to break up into smaller pieces so that the correct objects to study in this limit are no longer strings. This is an interesting idea in view of the BMN-limit which states that the operators in the super Yang-Mills theory correspond in some sense to discretized strings that are built up of a finite number of partons. This suggests looking at so-called string bit models which were first proposed in the 1970's as models of strings of partons. This picture seems also to be connected to the study of massless higher spins and the correspondence to free super Yang-Mills theory. In that setting there seems to be a critical string tension of order $1/R^2$, which is small but non-zero, \cite{Sezgin:2002rt}.

Despite all the important reasons to understand tensionless strings the subject is still not very well developed and some new ideas seem to be needed, maybe the idea of massless higher spin theories can give us a better understanding of tensionless string theory, the two certainly seem to be related.

We will leave the subject of tensionless strings here and turn to other considerations.

%% file: symfromsp.tex
\chapter{The superparticle and super Yang-Mills theory}
As a warm-up before discussing the string case we will in this chapter consider the simpler case of a superparticle coupled to a non-abelian gauge field in order to introduce some of the ideas needed later. We give a description of the superparticle (in flat superspace) and introduce the important notion of kappa-symmetry. We then couple the superparticle to a non-abelian gauge field. The coupling is achieved by introducing additional fermionic degrees of freedom, called boundary fermions in the case of the string, that are associated to the particle. The gauge field then becomes a function of these fermions instead of a matrix which makes it easier to handle. It is shown how kappa-symmetry of the superparticle forces the gauge field to be a solution to the super Yang-Mills equations. In order to pass back from a description in terms of functions of the boundary fermions to a description in terms of matrices the fermions must be quantized. This is done in the standard way by replacing Poisson-brackets by commutators.

\section{The superparticle action}
The action for an ordinary massless (scalar) particle in flat Minkowski space was obtained in section \ref{sec:m0-limit} of the last chapter as a limit of the massive one. It was shown to be given by
\begin{equation}
\label{eq:S_p}
S_{\mr p}=\frac{1}{2}\int d\tau\,e^{-1}\dot{x}^2\,,
\end{equation}
where a dot denotes the $\tau$-derivative and $\dot{x}^2\equiv\dot{x}^m\dot{x}^n\eta_{mn}$. The equations of motion, arrived at by varying $x$ and $e$ in the above action and setting the variation to zero, are
\begin{equation}
\partial_\tau(e^{-1}\dot{x}^m)=0\qquad\mbox{and}\qquad\dot{x}^2=0\,.
\end{equation}
The first equation says that the particle moves in a straight line and the second that its velocity is light-like, as is appropriate for a massless particle.

We will now generalize this to an action for a massless superparticle in flat ten-dimensional ($\mc N=1$) superspace described in sections \ref{sec:N=1superspace} and \ref{sec:diff-superspace}. The simplest generalization is to replace the tangent to the particle trajectory, $\dot{x}^m$, with the corresponding supersymmetric invariant, 
\begin{equation}
E_\tau^{\phantom\tau a}\equiv \dot{z}^M E_M^{\phantom{M}a}=\dot{x}^a-\frac{i}{2}\dot{\theta}\gamma^a\theta\,,
\end{equation}
the pull-back of the vielbein to the worldline of the particle. The action for this so-called Brink-Schwarz superparticle (\cite{Brink:1981nb}) is then
\begin{equation}
\label{eq:bs-action}
S_{\mr{BS}}=\frac{1}{2}\int d\tau\, e^{-1}E_\tau^2
\end{equation}
with $E_\tau^2\equiv E_\tau^{\phantom\tau a} E_\tau^{\phantom\tau b}\eta_{ab}$. This action is invariant under reparametrizations of the worldline
just as for the ordinary massless particle and since it's written in terms of a supersymmetric invariant it is also (manifestly) invariant under supersymmetry transformations.

What is surprising is that, as noted by Siegel in \cite{Siegel:1983hh}, this action is also invariant under another (local) symmetry, called kappa-symmetry, with fermionic transformation parameter $\kappa_\alpha(\tau)$. The transformations simplify if we write them in terms of 
\begin{equation}
\delta z^A\equiv\delta z^M E_M^{\phantom{M}A}\,. 
\end{equation}
The distinguishing property of kappa-symmetry is then that $\delta z^a=0$. The fermionic superspace coordinates transform as
\begin{eqnarray}
\delta z^\alpha &=&iE_\tau^{\phantom{\tau}a}\left(\gamma_a\kappa\right)^\alpha
\end{eqnarray}
and the transformation of $e$ can be determined from the requirement of kappa-invariance of the action as follows. The transformation of the pull-back of the
vielbein becomes
\begin{eqnarray}
\delta E_\tau^{\phantom\tau a}&=&\frac{d}{d\tau}\delta z^M E_M^{\phantom{M}a}+\dot{z}^M\delta E_M^{\phantom{M}a}
\nonumber\\
&=&-\delta z^M\frac{d}{d\tau}E_M^{\phantom{M}a}+\dot{z}^M\delta E_M^{\phantom{M}a}
=i\delta\theta\gamma^a\dot{\theta}\,,
\label{eq:deltaEtau}
\end{eqnarray}
so that
\begin{equation}
\label{eq:deltaEtau2}
\delta E_\tau^2=2iE_\tau^{\phantom\tau a}\delta\theta\gamma_a\dot{\theta}=-2E_\tau^2\kappa\dot{\theta}\,.
\end{equation}
Thus we see that the action (\ref{eq:bs-action}) is invariant if we take
\begin{equation}
\label{eq:delta-e1}
\delta e=-2e\kappa\dot{\theta}\,.
\end{equation}

Why is this extra symmetry present for the superparticle? We can get a clue by looking at the equations of motion coming from the action 
(\ref{eq:bs-action}), they are
\begin{eqnarray}
E_\tau^2=0\quad,\quad\partial_\tau\left(e^{-1}E_{\tau}^{\phantom{\tau}a}\right)=0\quad\mbox{and}\quad E_\tau^{\phantom{\tau}a}(\gamma_a\dot{\theta})_\alpha=0\,.
\end{eqnarray}
Now consider the matrix $E_\tau^{\phantom{\tau}a}\gamma_a$ which acts on $\dot{\theta}$. Because of the first equation above we have
\begin{equation}
E_\tau^{\phantom{\tau}a}\left(\gamma_a\right)_{\alpha\beta}E_\tau^{\phantom{\tau}b}\gamma_b^{\beta\gamma}=E_\tau^2=0\,.
\end{equation}
In particular this means that $(\gamma_aE_\tau^{\phantom\tau a})_{\alpha\beta}$ commutes with $(\gamma_aE_\tau^{\phantom\tau a})^{\alpha\beta}$ so that they can be simultaneously diagonalized. The above equation says that their product is zero which then implies that in general half the eigenvalues of $E_\tau^{\phantom{\tau}a}\gamma_a$ (with spinor indices up or down) are zero. Because this matrix multiplies $\theta$ in its equation of motion this means that half of the components of $\theta$ are actually decoupled from the theory! The effect of kappa-symmetry is to remove these components. This means that kappa-symmetry is related to supersymmetry because a supersymmetric theory must have an equal number of fermionic and bosonic degrees of freedom and, as it turns out, kappa-symmetry ensures this.

When we couple the superparticle, and, later on, the superstring, to background fields it will turn out that the requirement of kappa-symmetry of the action will mean that the background fields have to satisfy their equations of motion, \emph{i.e.} they must be on-shell. This means that kappa-symmetry gives us a way to determine the equations satisfied by the background fields, which can be very useful in cases when they are not known.

There is a deeper geometrical reason for kappa-symmetry related to supersymmetry. We will have more to say about this in the next chapter when we consider the case of the superstring. Here it will be enough to take the naive approach and not worry about the origin of this symmetry.

\section{Interaction with a non-abelian gauge field}
We now wish to see what happens when we let the particle carry a charge, so that it can interact with a gauge field. Consider first the case of a scalar particle interacting with an abelian (\emph{i.e.} $U(1)$) gauge field, $A_m(x)$, as in electrodynamics. It is easy to see that the correct equation of motion, the Lorentz force law, is obtained if we add to the free particle action (\ref{eq:S_p}) the interaction piece
\begin{equation}
\int d\tau\,\dot{x}^m A_m=\int A\,,
\end{equation}
which is just the integral along the worldline of the particle of the component of $A$ tangent to the worldline. In the final step we've written it in a geometric way as the integral of the pull-back of the gauge field one-form $A=dx^m A_m$. It is now clear how to generalize this to the case of the superparticle. The gauge (super)field is now $A=dz^M A_M$ and the interaction is again given by the integral of the pull-back of this to the worldline,
\begin{equation}
\int A=\int d\tau\,\dot{z}^M A_M=\int d\tau\,E_\tau^{\phantom\tau A} A_A\,.
\end{equation}

The case of a non-abelian gauge field is more complicated, however. It doesn't make sense to replace $A$ with a matrix in the above equations. It
turns out that the correct thing to do is to introduce the interaction in the path-integral for the particle in the following way
\begin{equation}
\label{eq:path-integral}
Z=\int\mc{D}x\,e^{iS_{\mr p}}W_C\,,
\end{equation}
where $W_C$ is known as the Wilson loop. It is given by the expression
\begin{equation}
W_C\equiv\tr\mc{P}e^{i\int_{C}\mb{A}}\,,
\end{equation}
where $C$, the worldline of the particle, is a closed loop and we have written the gauge field one-form as $\mb A$ to emphasize that it is matrix valued,
\begin{equation}
\mb A_m=A_m^r\mb T^r\,,
\end{equation}
where $\mb T^r$ are the generators of the gauge group. (We will write matrices related to the gauge group in boldface throughout this chapter to avoid confusion.) The symbol $\mc P$ denotes path-ordering, which is defined as
\begin{equation}
\mc P(\mb A_m(\tau)\mb A_n(\tau'))=\left\{
\begin{array}{cc}
\mb A_m(\tau)\mb A_n(\tau')& \tau>\tau'\\
\mb A_n(\tau')\mb A_m(\tau)& \tau<\tau'
\end{array}
\right.\,.
\end{equation}
The Wilson loop is very important because it is a gauge-invariant observable. Indeed, under a gauge transformation
\begin{equation}
\mb A_m\rightarrow\mb g\mb A_m\mb g^{-1}-i\partial_m\mb g\mb g^{-1}\,,
\end{equation}
with $\mb g$ an element of the gauge group, we have
\begin{equation}
\mc{P}e^{i\int_{C}\mb{A}}\rightarrow\mb g\mc{P}e^{i\int_{C}\mb{A}}\mb g^{-1}
\end{equation}
and we see that because of the cyclicity of the trace the Wilson loop is invariant. The generalization to the case of the superparticle is straightforward.

The problem with (\ref{eq:path-integral}) is that the integrand is not of the form $e^{iS}$, which means that there is no classical action for the system. This complicates things because it means that we have to work with the whole path-integral at once and although this can be done it is a little bit awkward. Luckily there is a way in which we can define an action for this system. The trick is to introduce some extra degrees of freedom for the particle. This idea was proposed in 1986 for the case of the open string, which we will consider in the next chapter, by Marcus and Sagnotti in \cite{Marcus:1986cm} although similar ideas had been used previously in the case of particles, \emph{e.g.}
\cite{Samuel:1978iy,Ishida:1979bc,Barducci:1980xk}.

Consider introducing $2q$ additional fermionic degrees of freedom for the particle with $q$ an integer. We will call them $\eta^{\h\mu}$ for $\h\mu=1,\ldots,2q$ and take
their free action to be
\begin{equation}
\label{eq:S_eta}
S_\eta=\int d\tau\,\frac{i}{4}\dot{\eta}^{\h\mu}\eta^{\h\nu}\delta_{\h\mu\h\nu}\,.
\end{equation}
Canonical quantization then gives the equal-time commutation relations\footnote{Because of the constraint that the conjugate momentum to $\eta$
is proportional to $\eta$ itself we must use the Dirac bracket when quantizing the system. This makes the commutator differ by a factor of
two from the naive case.}
\begin{equation}
\{\eta^{\h\mu},\eta^{\h\nu}\}=2\delta^{\h\mu\h\nu}\,.
\end{equation}
This means that as quantum mechanical operators these fermions represent a Clifford algebra, \emph{i.e.} they are just gamma-matrices in
$2q$ dimensions, $\mb\gamma^{\h\mu}$. Since the Dirac representation has dimension $2^{D/2}$ in $D$ dimensions (for even $D$) these gamma-matrices are $2^q\times2^q$ matrices. In fact any $2^q\times 2^q$ matrix can be expanded in a basis of anti-symmetrized products of these gamma matrices. We will consider only products of an even number of gamma matrices in the expansion\footnote{Terms with an odd number of gamma matrices turn out to give the wrong statistics.}. The Dirac representation then splits up into two $2^{q-1}$-dimensional Weyl representations. A gauge field $\mb A_m$ with gauge group $U(2^{q-1})\times U(2^{q-1})$ or a subgroup thereof can then be expanded as
\begin{equation}
\mb A_m=(A_m)_0+(A_m)_{\h\mu\h\nu}\mb\gamma^{\h\mu\h\nu}+(A_m)_{\h\mu\h\nu\h\rho\h\sigma}\mb\gamma^{\h\mu\h\nu\h\rho\h\sigma}+\ldots\,.
\end{equation}
We can therefore, if we want, think of $\mb A_m$ not as a matrix, but instead as the quantum operator
\begin{equation}
A_m(x,\eta)=(A_m)_0+(A_m)_{\h\mu\h\nu}\eta^{\h\mu}\eta^{\h\nu}
+(A_m)_{\h\mu\h\nu\h\rho\h\sigma}\eta^{\h\mu}\eta^{\h\nu}\eta^{\h\rho}\eta^{\h\sigma}+\ldots\,.
\end{equation}
This suggests taking
\begin{equation}
\label{eq:S_int}
S_{\mr{int}}=\int d\tau\,\dot{x}^m A_{m}(x,\eta)
\end{equation}
as the interaction. The path-integral for the particle extended with these fermions then becomes
\begin{equation}
\int\mc Dx\mc D\eta\,e^{i\left(S_{\mr p}+S_\eta+S_{\mr{int}}\right)}\,.
\end{equation}

To show that this is indeed correct we use the general relation between the Hamiltonian and Lagrangian formulation (in one dimension)
\begin{equation}
\label{eq:H-L-relation}
\tr Te^{-i\int dt\,H(\phi,t)}=\int\mc{D}\phi\,e^{iS[\phi]}\,,
\end{equation}
where $T$ stands for time-ordering, $S[\phi]=\int dt\,L(\phi)$ and $\phi$ is some collection of fields. Consider now this relation for the $\eta$ field. Taking the Lagrangian to be
\begin{equation}
L=\frac{i}{4}\dot{\eta}^{\h\mu}\eta^{\h\nu}\delta_{\h\mu\h\nu}+\dot{x}^m A_m(x,\eta)
\end{equation}
the Hamiltonian is
\begin{equation}
H=-\dot{x}^m A_m(x,\eta)\,.
\end{equation}
Thus the general relation (\ref{eq:H-L-relation}) for $\phi=\eta$ implies that\footnote{This can also be shown in a direct way by making a perturbative expansion of the $\eta$ path-integral, using that the propagator is essentially a step function and noting that this produces the path-ordering (see \cite{Marcus:1986cm}).}
\begin{eqnarray}
\int\mc D\eta\,e^{i\left(S_\eta+S_{\mr{int}}\right)}=\tr\mc P e^{i\int_C d\tau\,\dot{x}^m\mb A_m(x)}=W_C\,,
\end{eqnarray}
as time-ordering is the same as path-ordering in this case. Using this relation to express the Wilson loop in (\ref{eq:path-integral}) as a path-integral over the $\eta$ variables we see that we can define a classical action also in the non-abelian case which now includes the additional fermionic degrees of freedom. It is given by the sum of (\ref{eq:S_p}), (\ref{eq:S_eta}) and (\ref{eq:S_int}),
\begin{equation}
S[x,\eta]=S_{\mr p}+S_\eta+S_{\mr{int}}\,.
\end{equation}
However, the $\eta$ variables do not really have a good classical interpretation and it is clear from our discussion above that to pass to the usual matrix description we should quantize the $\eta$s, \emph{i.e.} replace them by gamma-matrices. Of course, ideally we should quantize the whole system including also the coordinates $x$, but this is of course much more difficult and we will not consider it in this thesis.

It is now easy to extend all this to the superparticle case. The path-integral becomes
\begin{equation}
\int\mc Dx\mc D\theta\mc D\eta\,e^{iS}\,,
\end{equation}
with
\begin{equation}
\label{eq:S-full}
S=S_{\mr{BS}}+S_\eta+\int d\tau\,E_\tau^{\phantom\tau A}A_A(z,\eta)\,.
\end{equation}
Varying this action we find the equations of motion
\begin{eqnarray}
\dot{\eta}^{\h\mu}&=&-(-1)^A2iE_\tau^{\phantom\tau A}\partial^{\h\mu}A_A\nonumber\\
E_\tau^2&=&0\nonumber\\
\partial_\tau\left(e^{-1}E_{\tau a}\right)&=&E_\tau^{\phantom{\tau}A}(d_a A_A-d_A A_a)-\dot{\eta}^{\h\mu}\partial_{\h\mu}A_a\nonumber\\
iE_\tau^{\phantom{\tau}a}(\gamma_a\dot{\theta})_\mu&=&
\partial_\tau\left(e^{-1}E_{\tau a}\right)E_\mu^{\phantom{\mu}a}-(-1)^A E_\tau^{\phantom{\tau}A}\partial_\mu A_A
+E_{\mu}^{\phantom{\mu}A}\dot{\eta}^{\h\mu}\partial_{\h\mu}A_A
\nonumber\\
&&{}+E_{\mu}^{\phantom{\mu}A}E_{\tau}^{\phantom{\tau}B}d_B A_A-i(\gamma^a\dot{\theta})_\mu A_a\,,
\end{eqnarray}
which, after simplification, become
\begin{eqnarray}
\dot{\eta}^{\h\mu}&=&-(-1)^A2iE_\tau^{\phantom{\tau}A}\partial^{\h\mu}A_A\nonumber\\
E_\tau^2&=&0\nonumber\\
\partial_\tau\left(e^{-1}E_{\tau a}\right)&=&-E_\tau^{\phantom{\tau}A}F_{Aa}\nonumber\\
iE_\tau^{\phantom{\tau}a}(\gamma_a\dot{\theta})_\alpha&=&E_{\tau}^{\phantom{\tau}A}F_{A\alpha}\,.
\end{eqnarray}
Here we have introduced the super field strength $F$ with components
\begin{equation}
\label{eq:F-components}
F_{AB}=d_A A_B-(-1)^{AB}d_B A_A-(-1)^A2i\partial^{\h\mu}A_A\partial_{\h\mu}A_B+T_{AB}^C A_C\,,
\end{equation}
where the supersymmetric derivatives are defined in (\ref{eq:susy-ds}) and the torsion of superspace is given in (\ref{eq:torsion}). When we quantize the $\eta$s and replace them by gamma matrices $A$ becomes a gauge field with gauge group $U(2^{q-1})\times U(2^{q-1})$ as we have discussed, but what about the term with $\eta$-derivatives?  In fact, for two functions of $\eta$, $f$ and $g$, the expression
\begin{equation}
(f,g)\equiv-2i\delta^{\h\mu\h\nu}f\overleftarrow{\partial}_{\h\mu}\partial_{\h\nu}g
\end{equation}
is simply the Poisson-Dirac bracket of $f$ and $g$. Canonical quantization then says that we should replace this by $-i$ times the commutator of $f$ and $g$ which become matrices when we take $\eta\rightarrow\mb\gamma$. Therefore, when we quantize the $\eta$ variables in this way we find that
\begin{equation}
F_{AB}(\eta)\rightarrow \mb F_{AB}=d_A\mb A_B-(-1)^{AB}d_B\mb A_A-i[\mb A_A,\mb A_B]+T_{AB}^C\mb A_C\,.
\end{equation}
$\mb F_{AB}$ is indeed the non-abelian super field strength which transforms in a covariant manner
\begin{equation}
\mb F_{AB}\rightarrow\mb g\mb F_{AB}\mb g^{-1}
\end{equation}
under the non-abelian gauge transformation
\begin{equation}
\mb A_B\rightarrow\mb g\mb A_B\mb g^{-1}-id_B\mb g\mb g^{-1}\,,
\end{equation}
with $\mb g\in U(2^{q-1})\times U(2^{q-1})$ since we started with $2q$ fermions.

We can express $F$ defined in (\ref{eq:F-components}) more succinctly in form language in terms of the two-form $F\equiv\frac{1}{2}e^Be^AF_{AB}$ as
\begin{equation}
\label{eq:F-def}
F=dA-i\partial^{\h\mu}A\partial_{\h\mu}A\,,
\end{equation}
where we've used the fact that $T^A=de^A$. Here it is important that, since in our conventions $d$ acts from the right, we should also take $i\partial^{\h\mu}A\partial_{\h\mu}$ to act from the right in order to be
consistent.

\subsection{Kappa-symmetry}
Kappa-symmetry must be present also in the case when the superparticle is coupled to a gauge field for the theory to be consistent. As we shall see this imposes constraints on the gauge field $A$ and in fact it has to be a solution of the super Yang-Mills equations for the superparticle to be kappa-symmetric\footnote{Kappa-symmetry of the superparticle coupled to a gauge field using fermions is considered in \cite{Sezgin:1993xg} for the case of an $SO(2q)$ gauge group.}. First we have to find the kappa-transformations of the fields in our action. We will take the transformations of the superspace coordinates $x$ and $\theta$ to be as before, \emph{i.e.}
\begin{equation}
\delta z^a=0\qquad\mbox{and}\qquad\delta z^\alpha =iE_\tau^{\phantom{\tau}a}\left(\gamma_a\kappa\right)^\alpha\,,
\end{equation}
and the transformations of $e$ and $\eta$ to be determined from the condition of kappa-symmetry.

The kappa-variation of the action (\ref{eq:S-full}) is then, making use of the expression for the variation of the vielbein (\ref{eq:deltaEtau}) and (\ref{eq:deltaEtau2}),
\begin{eqnarray}
\delta_\kappa S&=&\int d\tau\,\Big(-\frac{1}{2}e^{-2}\delta_\kappa e E_\tau^2-e^{-1}E_\tau^2\kappa\dot{\theta}
+\frac{i}{2}\dot{\eta}^{\h\mu}\delta_\kappa\eta_{\h\mu}+i\delta_\kappa\theta\gamma^a\dot{\theta}A_a
\nonumber\\
&&\qquad\qquad {}-\delta_\kappa\theta^\alpha\dot{A}_\alpha+E_\tau^{\phantom{\tau}A}\delta_\kappa A_A\Big)
\nonumber\\
&=&\int d\tau\,\Big(-\frac{1}{2}e^{-2}\delta_\kappa e E_\tau^2
-e^{-1}E_\tau^2\kappa\dot{\theta}+\frac{i}{2}\dot{\eta}^{\h\mu}\delta_\kappa\eta_{\h\mu}+i\delta_\kappa\theta\gamma^a\dot{\theta}A_a
\nonumber\\
&&\qquad\quad {}-\delta_\kappa\theta^\alpha\dot{\eta}^{\h\mu}\partial_{\h\mu}A_\alpha-\delta_\kappa\theta^\alpha E_\tau^{\phantom{\tau}A}d_A A_\alpha
+E_\tau^{\phantom{\tau}A}\delta_\kappa\eta^{\h\mu}\partial_{\h\mu}A_A
\nonumber\\
&&\qquad\qquad {}+E_\tau^{\phantom{\tau}A}\delta_\kappa\theta^\alpha d_\alpha A_A\Big)\,.
\end{eqnarray}
If we now take
\begin{equation}
\delta_\kappa\eta^{\h\mu}=2i\partial^{\h\mu}A_\alpha\delta_\kappa\theta^\alpha
\end{equation}
we see that the two terms involving $\dot\eta$ cancel and we are left with
\begin{equation}
\delta_\kappa S=\int d\tau\,\left(-\frac{1}{2}e^{-2}\delta_\kappa e E_\tau^2
-e^{-1}E_\tau^2\kappa\dot{\theta}+E_\tau^{\phantom{\tau}A}\delta_\kappa\theta^\alpha F_{\alpha A}\right)\,.
\label{eq:kappa}
\end{equation}
If we took $e$ to transform as in (\ref{eq:delta-e1}) the first two terms would cancel and we would be left with the condition
\begin{equation}
F_{A\alpha}=0\,.
\end{equation}
This condition turns out to be too strong however, as the Bianchi identities would then imply that the entire super field strength vanishes, 
$F_{AB}=0$.

The solution is to modify the kappa-variation of $e$. If we look at the last term in (\ref{eq:kappa})
\begin{equation}
E_\tau^{\phantom{\tau}A}\delta_\kappa\theta^\alpha F_{\alpha A}
=iE_\tau^{\phantom{\tau}a}E_\tau^{\phantom{\tau}b}(\kappa\gamma_b)^\alpha F_{\alpha a}
+i\dot{\theta}^\beta E_\tau^{\phantom{\tau}b}(\kappa\gamma_b)^\alpha F_{\alpha\beta}\,,
\end{equation}
we see that the first term has a chance to be cancelled by a term coming from the variation of $e$ as it is quadratic in $E_\tau^{\phantom\tau a}$. Let's therefore
try to impose only the condition
\begin{equation}
\label{eq:Falphabeta=0}
F_{\alpha\beta}(x,\theta,\eta)=0
\end{equation}
and see where that leads us.

\subsubsection{Bianchi identities}
We now have to consider the Bianchi identities for $F$. In the abelian case the Bianchi identity is $dF=0$, which is simply a consequence of
the definition $F\equiv dA$ and the fact that the exterior derivative squares to zero, $d^2=0$. In our case we find instead, using the definition of
our non-abelian field strength (\ref{eq:F-def}), that
\begin{equation}
dF=2i\partial^{\h\mu}A\partial_{\h\mu}dA=2i\partial^{\h\mu}A\partial_{\h\mu}F
-4\partial^{\h\nu}A\partial^{\h\mu}A\partial_{\h\mu}\partial_{\h\nu}A\,.
\end{equation}
The last term is zero because it is symmetric in $\h\mu$ and $\h\nu$ and we are left with
\begin{equation}
DF=0\,,
\end{equation}
where we have defined the covariant exterior derivative as
\begin{equation}
\label{eq:D-def}
D\equiv d-2i\partial^{\h\mu}A\partial_{\h\mu}\,.
\end{equation}
The components of the covariant derivative are given by
\begin{equation}
\label{eq:D-components}
D_A\equiv d_A-(-1)^A2i\partial^{\h\mu}A_A\partial_{\h\mu}\,.
\end{equation}
This new exterior derivative obeys
\begin{equation}
D^2=-2i\partial^{\h\mu}F\partial_{\h\mu}\,,
\end{equation}
or in components,
\begin{equation}
\label{eq:Dcommutator}
[D_A,D_B\}+T^C_{AB} D_C=-(-1)^{A+B}2i\partial^{\h\mu}F_{AB}\partial_{\h\mu}\,.
\end{equation}
In components of the Bianchi identity $DF=0$ reads
\begin{eqnarray}
3!D_{(\alpha}F_{\beta\gamma)}+3!T^a_{(\alpha\beta}F_{|a|\gamma)}&=&0\label{eq:bianchi1}\\
D_{a}F_{\alpha\beta}+2D_{(\alpha}F_{\beta)a}+T^b_{\alpha\beta}F_{ba}&=&0\label{eq:bianchi2}\\
2D_{[a}F_{b]\alpha}+D_{\alpha}F_{ab}&=&0\label{eq:bianchi3}\\
3!D_{[a}F_{bc]}&=&0\,,\label{eq:bianchi4}
\end{eqnarray}
where we've used the fact that the only non-zero component of the the torsion in flat superspace is $T^a_{\alpha\beta}$. Imposing our constraint from kappa-symmetry of the superparticle action, $F_{\alpha\beta}=0$, the first equation becomes
\begin{equation}
0=3\gamma^a_{(\alpha\beta}F_{|a|\gamma)}=\gamma^a_{\alpha\beta}F_{a\gamma}
+\gamma^a_{\gamma\alpha}F_{a\beta}+\gamma^a_{\beta\gamma}F_{a\alpha}\,.
\label{eq:bianchi1-2}
\end{equation}
Contracting this equation with $(\gamma_b)^{\alpha\beta}$ gives
\begin{equation}
0=16F_{b\gamma}+2(\gamma^a\gamma_b)_\gamma^{\phantom{\gamma}\beta}F_{a\beta}
=20F_{b\gamma}-2(\gamma_b)_{\gamma\alpha}(\gamma^a)^{\alpha\beta}F_{a\beta}\,.
\end{equation}
If we now define the spinor $\psi^\alpha\equiv-\frac{i}{10}(\gamma^a)^{\alpha\beta}F_{a\beta}$ we get
\begin{equation}
\label{eq:Fbgamma}
F_{b\gamma}=i(\gamma_b\psi)_\gamma\,.
\end{equation}
This is in fact also a solution to (\ref{eq:bianchi1-2}) by virtue of the identity
\begin{equation}
(\gamma^a)_{(\alpha\beta}(\gamma_a)_{\gamma)\delta}=0
\end{equation}
which holds for gamma-matrices in ten dimensions.

Thus we get for the kappa-variation in (\ref{eq:kappa})
\begin{eqnarray}
\delta_\kappa S&=&\int d\tau\,\left(-\frac{1}{2}e^{-2}\delta_\kappa e E_\tau^2-e^{-1}E_\tau^2\kappa\dot{\theta}
+E_\tau^{\phantom{\tau}a}E_\tau^{\phantom{\tau}b}\kappa\gamma_b\gamma_a\psi\right)
\nonumber\\
&=&\int\,d\tau E_\tau^2\left(-\frac{1}{2}e^{-2}\delta_\kappa e-e^{-1}\kappa\dot{\theta}+\kappa\psi\right)\,,
\end{eqnarray}
so that if we take
\begin{equation}
\delta_\kappa e=-2e\kappa_\alpha\left(\dot{\theta}^\alpha-e\psi^\alpha\right)
\end{equation}
the action for the superparticle coupled to a non-abelian gauge field is indeed kappa-symmetric provided only that $F_{\alpha\beta}=0$.

\subsection{Super Yang-Mills theory}
Let's now look at the remaining Bianchi identities. Equation (\ref{eq:bianchi2}) tells us that
\begin{equation}
\label{eq:bianchi2-1}
2i(\gamma_a D_{(\alpha}\psi)_{\beta)}=-i\gamma_{\alpha\beta}^b F_{ba}\,.
\end{equation}
Expanding $D_\alpha\psi^\beta$ in a Fierz basis of anti-symmetric products of gamma-matrices as
\begin{equation}
D_\alpha\psi^\beta=\psi_0\delta_\alpha^\beta+\psi_{ab}(\gamma^{ab})_\alpha^{\phantom\alpha\beta}\,,
\end{equation}
where higher terms in the expansion are excluded by the fact that the right-hand-side of (\ref{eq:bianchi2-1}) is proportional to $\gamma$, we get
\begin{eqnarray}
-i\gamma_{\alpha\beta}^b F_{ba}
&=&i(\gamma_a)_{\alpha\beta}\psi_0+2i\psi_{bc}(\gamma^{bc}\gamma_a)_{(\alpha\beta)}
\nonumber\\
&=&i(\gamma_a)_{\alpha\beta}\psi_0+4i(\gamma^b)_{\alpha\beta}\psi_{ba}\,.
\end{eqnarray}
Multiplying with $\gamma^a$ we see that $\psi_0=0$ and $\psi_{ab}=-\frac{1}{4}F_{ab}$, so that
\begin{equation}
\label{eq:Dalphapai}
D_\alpha\psi^\beta=-\frac{1}{4}F_{ab}(\gamma^{ab})_\alpha^{\phantom\alpha\beta}\,,
\end{equation}
which indeed solves (\ref{eq:bianchi2}).

The third Bianchi identity, (\ref{eq:bianchi3}), gives
\begin{equation}
\label{eq:bianchi3-2}
2i(\gamma_{[a}D_{b]}\psi)_\alpha+D_\alpha F_{ab}=0\,.
\end{equation}
Multiplying with $(\gamma^a\gamma^b)_\beta^{\phantom\beta\alpha}$ we get
\begin{eqnarray}
0&=&i(\gamma^a\gamma^b\gamma_a D_b\psi)_\beta-10i(\gamma^a D_a\psi)_\beta
+D_\alpha F_{ab}(\gamma^{ab})_\beta^{\phantom\beta\alpha}
\nonumber\\
&=&-18i(\gamma^{a}D_a\psi)_\beta+4D_\alpha D_\beta\psi^\alpha
\end{eqnarray}
where we have used (\ref{eq:Dalphapai}), which also implies that $D_\alpha\psi^\alpha=0$. Using this fact together with
\begin{equation}
\{D_\alpha,D_\beta\}=i\gamma^a_{\alpha\beta}D_a-2i\partial^{\h\mu}F_{\alpha\beta}\partial_{\h\mu}
\end{equation}
coming from (\ref{eq:Dcommutator}), where the last term is zero, we get
\begin{equation}
i\gamma^a D_a\psi=0\,.
\end{equation}
This says that $\psi$ satisfies the ten-dimensional Dirac equation.

Multiplying (\ref{eq:bianchi3-2}) with $\gamma^a$ instead we get
\begin{eqnarray}
0&=&10iD_{b}\psi^\beta-i(\gamma^a\gamma_b D_a\psi)^\beta+(\gamma^a)^{\beta\alpha} D_\alpha F_{ab}
\nonumber\\
&=&8iD_{b}\psi^\beta+(\gamma^a)^{\beta\alpha}D_\alpha F_{ab}\,.
\end{eqnarray}
Acting with $D_\beta$ on this and using
\begin{equation}
[D_\alpha, D_b]=2i\partial^{\h\mu}F_{\alpha b}\partial_{\h\mu}=2(\gamma_b\partial^{\h\mu}\psi)_\alpha\partial_{\h\mu}\,,
\end{equation}
coming from (\ref{eq:Dcommutator}), we obtain the equation
\begin{equation}
D^aF_{ab}=-2\partial^{\h\mu}\psi\gamma_b\partial_{\h\mu}\psi\,.
\end{equation}

Finally it is easy to check that the last Bianchi identity is already satisfied, since by definition
\begin{equation}
\label{eq:Fab}
F_{ab}=\partial_a A_b-\partial_b A_a-2i\partial^{\h\mu}A_a\partial_{\h\mu}A_b\,,
\end{equation}
which solves (\ref{eq:bianchi4}).

To summarize we have found that, after solving the Bianchi identities for the super field strength $F$, subject to the condition $F_{\alpha\beta}=0$ coming from the requirement of kappa-symmetry, everything can be expressed in terms of the superfields $A_a$ and $\psi^\alpha$ obeying the equations
\begin{eqnarray}
D^aF_{ab}&=&-2\partial^{\h\mu}\psi\gamma_b\partial_{\h\mu}\psi\\
i\gamma^a D_a\psi&=&0\\
D_\alpha F_{ab}&=&-2i(\gamma_{[a}D_{b]}\psi)_\alpha\\
D_\alpha\psi^\beta&=&-\frac{1}{4}F_{ab}(\gamma^{ab})_\alpha^{\phantom\alpha\beta}\,.
\end{eqnarray}
It is in fact not hard to see that it is enough to impose the first two equations for the lowest component in the $\theta$-expansion of
$A_a$ and $\psi^\alpha$. This is because the higher components in the expansion are then determined by the last two equations in terms of 
lower ones.

Finally, making the transition to the matrix description by replacing the $\eta$s with gamma-matrices and replacing the Poisson-Dirac bracket with $-i$ times the (graded) commutator as we have suggested, which amounts to replacing
\begin{equation}
2f\overleftarrow\partial^{\h\mu}\partial_{\h\mu}g\longrightarrow [\mb f,\mb g\}\,,
\end{equation}
for any functions of $\eta$ $f$ and $g$, we get the equations 
\begin{eqnarray}
\mb D^a\mb F_{ab}&=&2\mb\psi\gamma_b\mb\psi\label{eq:eom-A}\\
i\gamma^a\mb D_a\mb\psi&=&0\label{eq:eom-psi}\,,
\end{eqnarray}
where we've used the fact that the right-hand-side of the first equation is proportional to the anti-commutator. The Yang-Mills covariant derivative and field strength are given by
\begin{eqnarray}
\mb D_a\mb\psi&=&\partial_a\mb\psi-i[\mb A_a,\mb\psi]
\nonumber\\
\mb F_{ab}&=&\partial_a\mb A_b-\partial_b\mb A_a-i[\mb A_a,\mb A_b]\,,
\end{eqnarray}
as follows from (\ref{eq:D-components}) and (\ref{eq:Fab}). In the above $\mb A_a$ and $\mb\psi$ refer to the lowest ($\theta=0$) components in the expansion of the corresponding superfields.

It's not hard to see that the equations of motion for $\mb A_a$ and $\mb\psi$, equations (\ref{eq:eom-A}) and (\ref{eq:eom-psi}), can be derived from the action\footnote{Factors of the coupling constant \emph{etc} are absorbed in the definition of the fields and therefore don't appear explicitly here.}
\begin{equation}
\label{eq:SYM}
S=\int d^{10}x\,\tr\left(\frac{1}{4}\mb F_{ab}\mb F^{ab}-i\mb\psi\gamma^a\mb D_a\mb\psi\right)\,.
\end{equation}
This is the action for $\mc N=1$ super Yang-Mills theory in ten dimensions \cite{Brink:1976bc}. $\mc N=1$ refers to the fact that it is invariant under one global supersymmetry with parameter $\epsilon$ which is a 16 component spinor. The supersymmetry transformations are given by
\begin{eqnarray}
\delta\mb A_a&=&i\epsilon\gamma_a\mb\psi\nonumber\\
\delta\mb\psi^\alpha&=&\frac{1}{4}\mb F_{ab}\left(\gamma^{ab}\epsilon\right)^\alpha\,.
\end{eqnarray}

This theory is very important, especially its reduction to four dimensions, $\mc N=4$ super Yang-Mills theory. As we have mentioned before this theory is conjectured, based on string theory considerations, to be dual to supergravity (or string theory) in five dimensional anti-de Sitter space. Because the theory is conformal and has a high degree of supersymmetry it is much easier to handle than its non-supersymmetric and non-conformal cousin QCD, the theory of the strong nuclear force. The AdS/CFT-correspondence opens up the possibility to learn things about QCD from a study of $\mc N=4$ super Yang-Mills and its supergravity (or string) dual. This is currently a very active area of research.

From these preliminary considerations of the superparticle coupled to a Yang-Mills field we now turn to the analogous case of the open superstring coupled to a non-abelian gauge field. Kappa-symmetry will in this case turn out to require that the gauge field satisfy equations that are non-linear, ''stringy'', generalizations of the Yang-Mills equations of motion.

%% file: nbifromstring.tex
\chapter{Generalization to the open string}
In this chapter we will apply the technique of the previous chapter to describe
an open string ending on a stack of coincident D-branes. We find that
kappa-symmetry gives rise to generalizations of the superembedding conditions
that are known to give the equations of motion in the abelian case. The case of
a stack of D9-branes is then considered in some detail and the non-linear
generalizations of the super Yang-Mills conditions on the super field strength
derived. This chapter describes the work in \cite{Howe:2005jz}.

\section{The open superstring action}
Our starting point is the Green-Schwarz action for an open string in a general supergravity background and ending on a stack of D-branes. The bulk part of the action is the same as for a closed string, and is given by
\begin{equation}
S_{\mr{bulk}}=-\int_\Sigma d^2\xi\,(\sqrt{-g}+\frac{1}{2}\epsilon^{ij}B_{ij})\,.
\end{equation}
In this expression $g\equiv\det g_{ij}$ where $g_{ij}$ is the induced metric on the string worldsheet, $\Sigma$, given by
\begin{equation}
g_{ij}\equiv E_i^{\phantom{i}\u{a}}E_j^{\phantom{i}\u{b}}\eta_{\u{ab}}\,,
\end{equation}
in terms of the pull-back of the vielbein
\begin{equation}
E_i^{\phantom{i}\u{A}}\equiv\partial_i z^{\u M}E_{\u M}^{\phantom{M}\u A}\,.
\end{equation}
$B_{ij}$ is the pull-back of the NS-NS 2-form potential, or Kalb-Ramond field, also present in supergravity theories and is given by the expression
\begin{equation}
B_{ij}\equiv E_j^{\phantom{j}\u{N}}E_i^{\phantom{i}\u{M}}B_{\u{MN}}=E_j^{\phantom{j}\u{B}}E_i^{\phantom{i}\u{A}}B_{\u{AB}}\,.
\end{equation}
The action has a geometrical interpretation: The first term is simply the area of the worldsheet in spacetime and the second term is the natural coupling of the string, described by the two-surface it sweeps out in spacetime, to a two-form potential $B$, given by $\int_\Sigma B$. For the closed string this would be the full action, but for the open string the worldsheet has a boundary which lies in the worldvolume of a D-brane and which can therefore couple to the gauge field on the brane. The endpoints of the string are like point particles and so interact with the gauge field on the brane like a particle would. For an abelian gauge field the interaction would be given simply by the integral of the pull-back of the gauge field to the boundary of the string\footnote{We will consider only one of the boundaries corresponding to one of the endpoints of the string.},
\begin{equation}
\int_{\partial\Sigma}A\,.
\end{equation}
In the non-abelian case the interaction is given by introducing the Wilson loop
into the path-integral for the string, but as we saw in the previous chapter we
can avoid working with the whole path-integral by introducing fermionic degrees
of freedom, $\eta^{\h\mu}$, in this case living on the two boundaries of the
string, and with action (see \cite{Marcus:1986cm} or \cite{Kraus:2000nj} for a
recent application)\footnote{Note that the normalization of the
$\dot\eta\eta$-term differs from that in \cite{Howe:2005jz} by a factor of $\frac{i}{2}$.}
\begin{equation}
\int_{\partial\Sigma}dt\,\left(\frac{i}{4}\dot{\eta}^{\h\mu}\eta^{\h\nu}\delta_{\h\mu\h\nu}+\dot{z}^M A_M(z,\eta)\right)\,.
\label{eq:S_bdry-orig}
\end{equation}
These $\eta$ are sometimes referred to as Chan-Paton degrees of freedom and can be thought of roughly as keeping track of which of the coincident branes the endpoint of the string is attached to.

Here we will generalize this to get a more geometrical formulation. Since the worldvolume of the D-brane(s) is, in a sense, defined by the boundary of the string we can think of the $\eta^{\h\mu}$ as extra fermionic coordinates on the stack of D-branes. We will call this new space $\h M$. The coordinates of $\h M$ are $z^{\h M}=(z^M,\eta^{\h\mu})$ and we can now generalize the boundary piece of the action for the open string to
\begin{equation}
S_{\mr{bdry}}=\int_{\partial\Sigma}dt\,\dot{z}^{\h M}\mc A_{\h M}=\int_{\partial\Sigma}\mc A\,.
\end{equation}
This now looks nice and geometrical and we see that if we choose a gauge so that $\mc A_{\h\mu}=\frac{i}{4}\eta^{\h\nu}\delta_{\h\nu\h\mu}$ we get back to our original expression, (\ref{eq:S_bdry-orig}).

The action must be invariant under gauge transformations of the B-field,
\begin{equation}
\u B\rightarrow\u B+d\u\Lambda\,,
\end{equation}
with $\u\Lambda$ a one-form on the background space $\u M$. In the closed string case the action is indeed invariant but in the open sting case $\delta S_{\mr{bulk}}$ gives a boundary contribution. This is cancelled by requiring that $\mc A$ also transforms as
\begin{equation}
\mc A\rightarrow\mc A+\h f^*\u\Lambda\,,
\end{equation}
where $\h f^*$ denotes the pull-back to $\h M$. This ensures gauge invariance of the open string action.

\section{Variation of the action}
Let's now take a general variation of the action for the open string and look at the boundary contribution in particular. The variation of the boundary piece is easy to evaluate and we get
\begin{equation}
\delta S_{\mr{bdry}}=-\int_{\partial\Sigma}dt\,\delta z^{\h M}\dot{z}^{\h N}\left(d\mc A\right)_{\h N\h M}\,.
\end{equation}
The variation of the bulk part of the action becomes
\begin{eqnarray}
\lefteqn{\delta S_{\mr{bulk}}=\int_\Sigma d^2\xi\,\Big(\delta z^{\u a}\partial_i\left(\sqrt{-g}g^{ij}E_{j\u a}\right)
+\sqrt{-g}g^{ij}\delta z^{\u A}E_i^{\phantom{i}\u B}T_{\u{BA}}^{\u c}E_{j\u c}}
\nonumber\\
&&{}+\frac{1}{2}\epsilon^{ij}\delta z^{\u A}E_i^{\phantom{i}\u B}E_j^{\phantom{j}\u C}H_{\u{CBA}}\Big)
+\int_{\partial\Sigma}dt\,\delta z^{\u M}\left(-E_{\u M}^{\phantom{M}\u a}E_{1\u a}+\dot{z}^{\u N}B_{\u{NM}}\right)\,,
\nonumber\\
\label{eq:deltaS_bulk}
\end{eqnarray}
with $E_{1\u a}$ the piece perpendicular to the boundary. We are only interested in the boundary contribution coming from the variation of the bulk part of the action which is given by the second integral in (\ref{eq:deltaS_bulk}). Let us call this piece $\delta S_{\mr{bulk-bdry}}$. Because the boundary of the string lies in the space $\h M$ we have to take $z^{\u M}|_{\partial\Sigma}=z^{\u M}(z^{\h M})$. Doing this the boundary contribution coming from the bulk action becomes
\begin{eqnarray}
\delta S_{\mr{bulk-bdry}}=\int_{\partial\Sigma}dt\,
\delta z^{\h M}\partial_{\h M}z^{\u M}\left(-E_{\u M}^{\phantom{M}\u a}E_{1\u a}+\dot{z}^{\h N}\partial_{\h N}z^{\u N}B_{\u{NM}}\right)\,.
\end{eqnarray}
The total boundary contribution from a variation of the action is then
\begin{eqnarray}
\label{eq:var-bdry}
\delta S_{\mr{bdry-total}}=-\int_{\partial\Sigma}dt\,\delta z^{\h N}
\left(\partial_{\h N}z^{\u M}E_{\u M}^{\phantom{M}\u a}E_{1\u a}+\dot{z}^{\h M}K_{\h M\h N}\right)\,,
\end{eqnarray}
where we have defined
\begin{equation}
K\equiv d\mc A-\h f^*B\,.
\end{equation}
We will require $K_{\h\mu\h\nu}$ to be invertible, so that $\dot\eta$ is determined by the $\eta$ equations of motion, and define
\begin{equation}
N^{\h\mu\h\nu}\equiv(K_{\h\mu\h\nu})^{-1}\,.
\end{equation}
We can then use $N^{\h\mu\h\nu}$ and $K_{\h\mu\h\nu}$ to raise and lower $\h\mu$-indices.

\section{Kappa-symmetry}
In analogy to the superparticle case of the previous chapter we now want to demand kappa-symmetry for the open string coupled to the non-abelian gauge-field on the brane. This gives us conditions on the background fields which in the case of the superparticle was seen to amount to the fact that the gauge field had to be a solution to the super Yang-Mills equations. The analog for the string case will give super Yang-Mills plus higher order ''stringy'' corrections.

\subsection{Superembedding interpretation of kappa-symmetry}
As was mentioned in the previous chapter kappa-symmetry has a deep geometric interpretation, it is in fact a manifestation of supersymmetry. This can be seen in the geometric approach known as the superembedding formalism \cite{Sorokin:1988nj,Sorokin:1999jx}. The basic idea is to describe strings and branes as supersurfaces embedded in a larger superspace. In embeddings of ordinary surfaces we are interested in isometric embeddings, \emph{i.e.} embeddings that preserve the lengths of vectors. In the case of superembeddings the analog of an isometric embedding is one that satisfies the so-called superembedding condition
\begin{equation}
E_\alpha^{\phantom\alpha\u a}=0\,,
\label{eq:superembeddingcondition}
\end{equation}
where
\begin{equation}
E_A^{\phantom A\u B}\equiv E_A^{\phantom A M}\partial_M z^{\u N}E_{\u N}^{\phantom N\u B}
\end{equation}
is the analog of the ordinary embedding matrix, $\frac{\partial x^{\u m}}{\partial x^n}$, in the case of ordinary surface theory. In the superspace case we need the vielbeins $E_A^{\phantom AM}$ and $E_{\u A}^{\phantom A\u M}$ to be able to write an object which is invariant under supersymmetry. The superembedding condition, (\ref{eq:superembeddingcondition}), says that at every point the odd tangent directions of the worldvolume form a subspace of the odd tangent directions of the embedding (target) space. A remarkable fact is that in many cases the superembedding condition determines the full dynamics of branes, it gives the field content and the equations of motion for the worldvolume theory as well as requiring the background to be a supergravity solution \cite{Howe:1996mx,Howe:1997wf}. In some cases an extra condition on the field strength on the brane is needed as we will see explicitly when we consider the D9-brane. The Green-Schwarz formulation of superbranes and superstrings, where the string/brane is described as a bosonic surface embedded in a target superspace, can be related to the superembedding approach and the kappa-symmetry present in the Green-Schwarz formulation can then be seen to be simply (the leading component of) the odd diffeomorphisms of the super worldvolume, \emph{i.e.}
\begin{equation}
\label{eq:kappa-transf}
\delta_\kappa z^a=0\,,\qquad \delta_\kappa z^\alpha=v^\alpha\,,
\end{equation}
where $\delta z^A\equiv\delta z^M E_M^{\phantom MA}$. For a nice review of the superembedding formalism see \cite{Sorokin:1999jx}.

\subsection{Kappa-symmetry with boundary fermions}
We now turn to the important question of kappa-invariance of the open string coupled to a non-abelian gauge field. It is known that the bulk part of the action, which is the same as that of the ordinary closed string, will be invariant (up to a boundary term) provided that the background is a solution to the supergravity equations. Assuming this to be the case we only have to worry about the boundary contribution given in (\ref{eq:var-bdry}) for the special case that $\delta z$ is a kappa-variation. Since the stack of coincident branes is now described using the extra fermionic $\eta$-coordinates we need to determine how these should transform under kappa-symmetry. As in the superparticle case in the previous chapter we shall choose this variation so that the terms containing $\dot{\eta}$ cancel. It's easy to see using (\ref{eq:kappa-transf}) that this determines the variation to be
\begin{equation}
\delta_\kappa\eta^{\h\mu}=-\delta_\kappa z^M K_M^{\phantom M\h\mu}\,,
\end{equation}
where the $\h\mu$ index on $K$ has been raised with $N^{\h\mu\h\nu}$. The conditions for kappa-symmetry of the action then become
\begin{equation}
\label{eq:gensuperembeddingcond}
\mc E_\alpha^{\phantom\alpha\u a}=0
\end{equation}
and
\begin{equation}
\mc F_{\alpha B}=0\,,
\end{equation}
where we have defined the generalized superembedding matrix
\begin{equation}
\mc E_A^{\phantom A\u B}\equiv E_A^{\phantom A M}\mc D_M z^{\u N}E_{\u N}^{\phantom N\u B}\,,
\end{equation}
with the derivative defined as
\begin{equation}
\mc D_M\equiv\partial_M-K_M^{\phantom M\h\mu}\partial_{\h\mu}
\end{equation}
and
\begin{equation}
\label{eq:mcFdef}
\mc F_{MN}\equiv K_{MN}-K_{M\h\mu}N^{\h\mu\h\nu}K_{\h\nu N}\,.
\end{equation}
We will refer to these conditions as the (generalized) superembedding condition and $\mc F$ constraint. Later we will analyse the content of these conditions for the case of coincident D9-branes.

In the above derivation we did not treat the $\eta$-coordinates on equal footing with the other coordinates $z^M=(x^m,\theta^\mu)$. If we want to take our new geometrical picture seriously this should be possible however. In fact, we could write the variation of $\eta$ in a more general form similar to the variation $\delta z^a$ as
\begin{equation}
\label{eq:delta-zhatalpha}
\delta_\kappa z^{\h\alpha}\equiv\delta_\kappa z^{\h M}E_{\h M}^{\phantom M\h\alpha}=0\,,
\end{equation}
by introducing the vielbeins $E_{\h M}^{\phantom M\h A}$ on $\h M$. The requirement of kappa-symmetry would then give the covariant conditions
\begin{equation}
E_\alpha^{\phantom\alpha\u a}=0
\end{equation}
and
\begin{equation}
K_{\alpha\h B}=0\,,
\end{equation}
where $E_{\h A}^{\phantom A\u B}\equiv E_{\h A}^{\phantom A\h M}\partial_{\h
M}z^{\u M}E_{\u M}^{\phantom M\u B}$. It is easily seen that the
$\alpha\h\beta$-component of the last condition gives $E_M^{\phantom
M\h\alpha}=K_M^{\phantom M\h\alpha}$ so that (\ref{eq:delta-zhatalpha}) is
consistent with what we had before. Using this condition in the remaining
conditions they become of the same form as previously. The problem is that we
have now introduced a lot more degrees of freedom by introducing vielbeins with
hatted components and in general more conditions have to be imposed to constrain
the extra components. A discussion of how to do this and thereby obtain a fully
covariant approach is given in section 5 of \cite{Howe:2005jz}. We will not need this more sophisticated approach here however and instead we refer the reader to that paper. The conditions on $\mc E$ and $\mc F$ can be viewed as the covariant conditions expressed in a particular basis on $\h M$.

\section{The standard gauge}
\label{sec:standardgauge}
The gauge field $\mc A$ transforms under $U(1)$ gauge transformations with parameter $a'$ and diffeomorphisms of $\h M$ with parameter $\varepsilon^{\h M}$ as
\begin{eqnarray}
\delta\mc A_{\h M}&=&\partial_{\h M}a'+\partial_{\h M}\varepsilon^{\h N}\mc A_{\h N}+\varepsilon^{\h N}\partial_{\h N}\mc A_{\h M}
\nonumber\\
&=&\partial_{\h M}(a'+\varepsilon^{\h N}\mc A_{\h N})+\varepsilon^{\h N}(d\mc A)_{\h N\h M}
\nonumber\\
&=&\partial_{\h M}a+\varepsilon^{\h N}(d\mc A)_{\h N\h M}\,.
\end{eqnarray}
Since $(d\mc A)_{\h\mu\h\nu}$ is invertible by assumption we can use $\varepsilon^{\h\mu}$ to bring the gauge field to the standard form
\begin{equation}
\mc A_{\h\mu}=\frac{i}{4}\eta^{\h\nu}\delta_{\h\nu\h\mu}\quad,\quad\mc A_M\equiv A_M\,.
\end{equation}
This gauge fixing removes the components of the gauge field in the $\eta$-directions and leaves the usual (super) Yang-Mills gauge field, $A$, as we will see. In this gauge the remaining $U(1)$ gauge transformations and diffeomorphisms (of $M$) must be accompanied by a compensating $\eta$-diffeomorphism with parameter
\begin{equation}
\varepsilon^{\h\mu}=2i\delta^{\h\mu\h\nu}\left(\partial_{\h\nu}a-\varepsilon^N A_{N\h\nu}\right)=2i\partial^{\h\mu}a-2i\varepsilon^N A_N^{\phantom N\h\mu}\,,
\end{equation}
in order to stay in the gauge, with
\begin{equation}
A_N^{\phantom N\h\mu}\equiv(-1)^N\delta^{\h\mu\h\nu}\partial_{\h\nu}A_N\,.
\end{equation}
Note that, as above, we will use $\delta$ to raise and lower $\h\mu$-indices in this gauge. The gauge transformation of $A_M$ is now
\begin{equation}
\delta A_M=\partial_M a+\varepsilon^{\h\mu}(d\mc A)_{\h\mu M}=\partial_M a-2iA_M^{\phantom M\h\mu}\partial_{\h\mu}a=D_M a\,,
\end{equation}
where $D_M$ is the Yang-Mills covariant derivative that we met in the previous chapter in (\ref{eq:D-components}). As was found there the covariant derivative obeys the relation
\begin{equation}
[D_M,D_N\}=-(-1)^{M+N}2i\partial^{\h\mu}F_{MN}\partial_{\h\mu}\,,
\end{equation}
where the field strength of $A$ is
\begin{equation}
F_{MN}=\partial_M A_N-(-1)^{MN}\partial_N A_M-2iA_M^{\phantom M\h\mu}\partial_{\h\mu}A_N\,.
\end{equation}

We can now relate $\mc F$ in the standard gauge to the Yang-Mills field strength defined above. First of all we have in this gauge
\begin{equation}
K_{\h\mu\h\nu}=\frac{i}{2}\delta_{\h\mu\h\nu}-B_{\h\mu\h\nu}\,.
\end{equation}
The mixed components of $K$ then become
\begin{eqnarray}
K_{M\h\mu}&=&-A_{M\h\mu}-B_{M\h\mu}=2iA_M^{\phantom M\h\nu}K_{\h\nu\h\mu}+2iA_M^{\phantom M\h\nu}B_{\h\nu\h\mu}-B_{M\h\mu}
\nonumber\\
&=&2iA_M^{\phantom M\h\nu}K_{\h\nu\h\mu}-\tilde B_{M\h\mu}\,,
\label{eq:KMhmu}
\end{eqnarray}
where we've defined the Yang-Mills covariant pull-back as
\begin{equation}
\label{eq:covariant-pull-back}
\tilde v_M\equiv D_Mz^{\u M}v_{\u M}=v_M-2iA_M^{\phantom M\h\mu}v_{\h\mu}\quad,\quad\tilde v_{\h\mu}=v_{\h\mu}
\end{equation}
for a vector $v$ on $\u M$ (as usual $v_{\h M}\equiv\partial_{\h M}z^{\u N}v_{\u N}$).
Using (\ref{eq:KMhmu}) the last term in the definition of $\mc F$, (\ref{eq:mcFdef}), becomes
\begin{eqnarray}
K_{M\h\mu}N^{\h\mu\h\nu}K_{\h\nu N}&=&
4A_M^{\phantom M\h\mu}K_{\h\mu\h\nu}\partial^{\h\nu}A_N
-4iA_{[M}^{\phantom M\h\nu}\tilde B_{|\h\nu|N\}}
+\tilde B_{M\h\mu}N^{\h\mu\h\nu}\tilde B_{\h\nu N}
\nonumber\\
&=&2iA_M^{\phantom M\h\mu}\delta_{\h\mu\h\nu}\partial^{\h\nu}A_N
-4A_M^{\phantom M\h\mu}B_{\h\mu\h\nu}\partial^{\h\nu}A_N
\nonumber\\
&&{}-4iA_{[M}^{\phantom M\h\nu}\tilde B_{|\h\nu|N\}}+\tilde B_{M\h\mu}N^{\h\mu\h\nu}\tilde B_{\h\nu N}\,.
\label{eq:KNK}
\end{eqnarray}
From the expression for $\mc F$,
\begin{equation}
\mc F_{MN}=(dA)_{MN}-B_{MN}-K_{M\h\mu}N^{\h\mu\h\nu}K_{\h\nu N}\,,
\end{equation}
we see that the first term in (\ref{eq:KNK}) combines with $dA$ to give the Yang-Mills field strength $F_{MN}$, while the second and third term in (\ref{eq:KNK}) combine with $B_{MN}$ to give the covariant pull-back of $B$. We then get
\begin{equation}
\label{eq:rel-calF-F}
\mc F_{MN}=F_{MN}-\tilde B_{MN}-\tilde B_{M\h\mu}N^{\h\mu\h\nu}\tilde B_{\h\nu N}\,.
\end{equation}
This is, if you will, the non-abelian generalization of the modified field strength of a single D-brane, $F-B$, in this classical approximation. Observe that the B-gauge invariance, although still present, is no longer manifest in the standard gauge.

\section{Coincident D9-branes}
We will illustrate the procedure of solving the generalized superembedding condition for the case of a stack of D9-branes in a flat type IIB supergravity background. We start by describing the background space $\u M$. The background is an $\mc N=2$ superspace with coordinates $z^{\u M}=(x^{\u m},\theta^{\u\mu})$, where $\u m=0,\ldots 9$ and $\u\mu=1,\ldots 32$. In the IIB case the fermionic directions split up into two 16-component spinors of the same chirality, $\theta^{\u\mu}=\theta^{\mu i}$, with $i=1,2$ and $\mu=1,\ldots,16$. The supersymmetric basis one-forms are given by
\begin{eqnarray}
E^{\alpha i}&=&d\theta^{\alpha i}\nonumber\\
E^{\u a}&=&dx^{\u a}-\frac{i}{2}d\theta^i\gamma^{\u a}\theta^i\,.
\end{eqnarray}
The non-zero components of the torsion are
\begin{equation}
T_{\alpha i\beta j}^{\u c}=-i\gamma^{\u c}_{\alpha\beta}\delta_{ij}\,.
\end{equation}
There is also a non-trivial NS-NS three-form field strength, $H\equiv dB$, whose non-zero components are
\begin{equation}
\label{eq:H-constraint}
H_{\alpha1\beta2\u c}=-i(\gamma_{\u c})_{\alpha\beta}\,.
\end{equation}

The worldvolume of the D9-brane has $9+1=10$ bosonic dimensions, so it fills out the whole (bosonic part of) target space. In fact it will be an $\mc N=1$ superspace as the presence of a D-brane breaks half of the supersymmetries, but in addition it also has the $\eta$-directions that we have introduced. We take the coordinates of the brane to be $z^{\h M}=(x^m,\theta^\mu,\eta^{\h\mu})$.

Because of the diffeomorphism invariance of the brane it is always possible to choose a so-called static gauge, where we take the coordinates of the brane worldvolume to coincide with some of the background coordinates, the remaining background coordinates becoming scalar fields on the brane worldvolume. We can therefore make the following choice of coordinates to describe the brane
\begin{eqnarray}
\u x^m&=&x^m\nonumber\\
\u\theta^{1\alpha}&=&\theta^\alpha\nonumber\\
\u\theta^{2\alpha}&=&\Lambda^\alpha(x,\theta,\eta)\,,
\end{eqnarray}
where the target space coordinates have been underlined for clarity. We will now examine the content of the generalized superembedding condition that we derived by requiring kappa-symmetry of open strings attached to the stack of branes.

\subsection{The generalized superembedding condition}
Without loss of generality we can take the basis one-forms of the $\mc N=1$ worldvolume superspace to be
\begin{eqnarray}
E^\alpha&=&e^\alpha\nonumber\\
E^a&=&(e^b-e^\beta\psi_\beta^{\phantom\beta b})A_b^{\phantom ba}
\label{eq:brane-basis}
\end{eqnarray}
with the dual basis
\begin{eqnarray}
E_\alpha&=&d_\alpha+\psi_\alpha^{\phantom\alpha b}\partial_b\nonumber\\
E_a&=&(A^{-1})_a^{\phantom a b}\partial_b\,.
\end{eqnarray}
Here $(e^\alpha,e^a)$ and $(d_\alpha,d_a)$ denote the flat $\mc N=1$ basis given in (\ref{eq:susy-es}) and (\ref{eq:susy-ds}). The generalized superembedding condition, (\ref{eq:gensuperembeddingcond}), then becomes
\begin{eqnarray}
0&=&\mc E_{\alpha}^{\phantom\alpha\u a}
=E_\alpha^{\phantom\alpha M}\mc D_Mz^{\u M}E_{\u M}^{\phantom M\u a}
=\psi_\alpha^{\phantom\alpha a}-\frac{i}{2}E_\alpha^{\phantom\alpha M}\mc D_M\Lambda\gamma^a\Lambda
\nonumber\\
&=&\psi_\alpha^{\phantom\alpha b}\left(\delta_b^a-\frac{i}{2}\mc D_b\Lambda\gamma^a\Lambda\right)
-\frac{i}{2}\mc D_\alpha\Lambda\gamma^a\Lambda\,.
\end{eqnarray}
This equation is easily solved for $\psi$ and we obtain
\begin{equation}
\psi_\alpha^{\phantom\alpha a}
=\frac{i}{2}\mc D_\alpha\Lambda\gamma^b\Lambda\left(\delta_a^b-\frac{i}{2}\mc D_a\Lambda\gamma^b\Lambda\right)^{-1}\,.
\label{eq:def-psi}
\end{equation}
Since there are no transverse (bosonic) directions we can choose the bosonic part of the superembedding matrix to be simply
\begin{equation}
\mc E_a^{\phantom a\u b}=\delta_a^{\u b}\,,
\end{equation}
which then gives
\begin{equation}
A_a^{\phantom ab}=\delta_a^b-\frac{i}{2}\mc D_a\Lambda\gamma^b\Lambda\,.
\label{eq:def-A}
\end{equation}
We now want to write the other condition coming from the requirement of kappa-symmetry, the so-called $\mc F$ constraint, in a more transparent form by expressing it as conditions on the components of the super Yang-Mills field strength in a flat $\mc N=1$ basis on the brane.

\subsection{The $\mc F$ constraint}
First we fix the standard gauge
\begin{equation}
\mc A_{\h\mu}=\frac{i}{4}\eta_{\h\mu}\quad,\quad\mc A_M\equiv A_M\,,
\end{equation}
as described in section \ref{sec:standardgauge}. The relation between $\mc F$ and the super Yang-Mills field strength, $F$, is then given in (\ref{eq:rel-calF-F}), or in shorter notation
\begin{equation}
\label{eq:mcF-Frel}
\mc F=F-\tilde B-\tilde BN\tilde B\,.
\end{equation}
Recall that $\tilde B$ denotes the (Yang-Mills) covariant pull-back of $B$ defined in (\ref{eq:covariant-pull-back}).

We now want to express this in a flat $\mc N=1$ basis on the brane, $(e^\alpha,e^a)$. The $\mc F$ constraint, $\mc F_{\alpha B}=0$, implies, using (\ref{eq:brane-basis}), that
\begin{eqnarray}
\mc F=\frac{1}{2}E^bE^a\mc F_{ab}
&=&\frac{1}{2}e^\beta e^\alpha\psi_\alpha^{\phantom\alpha a}\psi_\beta^{\phantom\beta b}A_a^{\phantom ac}A_b^{\phantom bd}\mc F_{cd}
-e^be^\alpha\psi_\alpha^{\phantom\alpha a}A_a^{\phantom ac}A_b^{\phantom bd}\mc F_{cd}
\nonumber\\
&&{}+\frac{1}{2}e^b e^a A_a^{\phantom ac}A_b^{\phantom bd}\mc F_{cd}\,.
\end{eqnarray}
Using this in (\ref{eq:mcF-Frel}) we get for the components of the super Yang-Mills field strength in the flat basis defined as, $F\equiv\frac{1}{2}e^Be^Af_{AB}$,
\begin{eqnarray}
f_{\alpha\beta}&=&\psi_\alpha^{\phantom\alpha a}\psi_\beta^{\phantom\beta b}A_a^{\phantom ac}A_b^{\phantom bd}\mc F_{cd}+\tilde b_{\alpha\beta}
+(\tilde bN\tilde b)_{\alpha\beta}\label{eq:mcFconstraint-alphabeta}
\\
f_{\alpha b}&=&-\psi_\alpha^{\phantom\alpha a}A_a^{\phantom ac}A_b^{\phantom bd}\mc F_{cd}+\tilde b_{\alpha b}+(\tilde bN\tilde b)_{\alpha b}
\label{eq:mcFconstraint-alphab}\\
f_{ab}&=&A_a^{\phantom ac}A_b^{\phantom bd}\mc F_{cd}+\tilde b_{ab}+(\tilde bN\tilde b)_{ab}\,,
\label{eq:mcFconstraint-ab}
\end{eqnarray}
where $\tilde b$ denotes the components of $\tilde B$ in the flat basis just as $f$ denotes the components of $F$ in the flat basis. We can make these equations even more explicit by choosing a gauge for the $B$-field.

\subsubsection{Choosing a gauge for $B$}
We will take the $B$-field to be given by 
\begin{equation}
\label{eq:B}
B=id\theta^1\gamma_{\u a}\theta^2\left(E^{\u a}+\frac{i}{3}d\theta^2\gamma^{\u a}\theta^2\right)\,.
\end{equation}
To show that this is indeed a permissable choice of gauge we compute
\begin{eqnarray}
dB&=&id\theta^1\gamma_{\u a}\theta^2\left(dE^{\u a}+\frac{i}{3}d\theta^2\gamma^{\u a}d\theta^2\right)
-id\theta^1\gamma_{\u a}d\theta^2\left(E^{\u a}+\frac{i}{3}d\theta^2\gamma^{\u a}\theta^2\right)
\nonumber\\
&=&\frac{1}{2}d\theta^1\gamma_{\u a}\theta^2d\theta^1\gamma^{\u a}d\theta^1
+\frac{1}{6}d\theta^1\gamma_{\u a}\theta^2d\theta^2\gamma^{\u a}d\theta^2 
+\frac{1}{3}d\theta^1\gamma_{\u a}d\theta^2d\theta^2\gamma^{\u a}\theta^2
\nonumber\\
&&{}-id\theta^1\gamma_{\u a}d\theta^2 E^{\u a}\,.
\end{eqnarray}
Because of our beloved gamma matrix identity 
\begin{equation}
\gamma^a_{\alpha(\beta}(\gamma_a)_{\gamma\delta)}=0
\end{equation}
and the symmetry in $d\theta^1$ the first term is zero. The second and third term cancel using the same identity and we are left with
\begin{equation}
dB=-iE^{\u a}d\theta^2\gamma_{\u a}d\theta^1\,,
\end{equation}
which indeed agrees with the superspace constraint on $H=dB$ given in (\ref{eq:H-constraint}).

It is now a straight-forward exercise to read off the components of $B$ from (\ref{eq:B}) and then compute the covariant pull-back. Expressed in the flat basis on the brane the non-zero components become
\begin{eqnarray}
\tilde b_{\alpha\h\mu}&=&\frac{1}{6}\partial_{\h\mu}\Lambda\gamma^a\Lambda(\gamma_a\Lambda)_\alpha\nonumber\\
\tilde b_{\alpha\beta}&=&\frac{1}{3}D_{(\alpha}\Lambda\gamma^a\Lambda(\gamma_a\Lambda)_{\beta)}\nonumber\\
\tilde b_{a\beta}&=&i(\gamma_a\Lambda)_\beta+\frac{1}{6}D_a\Lambda\gamma^b\Lambda(\gamma_b\Lambda)_\beta\,,
\end{eqnarray}
where the super Yang-Mills covariant derivative $D_A$ is defined in (\ref{eq:D-components}). With this choice of $B$ we get
\begin{equation}
N^{\h\mu\h\nu}\equiv\left(\frac{i}{2}\delta_{\h\mu\h\nu}-B_{\h\mu\h\nu}\right)^{-1}=-2i\delta^{\h\mu\h\nu}\,.
\end{equation}
And from the definition $\mc D_M=\partial_M-K_{M\h\mu}N^{\h\mu\h\nu}\partial_{\h\nu}$ we get
\begin{eqnarray}
\mc D_a&=&D_a-2i\tilde b_{a\h\mu}\partial^{\h\mu}=D_a\\
\mc D_\alpha&=&D_\alpha-2i\tilde b_{\alpha\h\mu}\partial^{\h\mu}=D_\alpha+\frac{i}{3}(\gamma_a\Lambda)_\alpha\,\Lambda\gamma^a\partial^{\h\mu}\Lambda\,\partial_{\h\mu}\,.
\end{eqnarray}
Using this result we get from (\ref{eq:def-A})
\begin{equation}
A_a^{\phantom ab}=\delta_a^b-\frac{i}{2}D_a\Lambda\gamma^b\Lambda
\end{equation}
and using (\ref{eq:def-psi})
\begin{equation}
\psi_\alpha^{\phantom\alpha a}
=\frac{i}{2}\left(D_\alpha\Lambda\gamma^b\Lambda
+\frac{i}{3}(\gamma_a\Lambda)_\alpha\,\Lambda\gamma^a\partial^{\h\mu}\Lambda\,\partial_{\h\mu}\Lambda\gamma^b\Lambda
\right)\left(\delta_a^b-\frac{i}{2}D_a\Lambda\gamma^b\Lambda\right)^{-1}\,.
\end{equation}

Finally we can express the content of the $\mc F$ constraint with this choice of $B$. The last equation, (\ref{eq:mcFconstraint-ab}), simply defines
\begin{equation}
f_{ab}=A_a^{\phantom ac}A_b^{\phantom bd}\mc F_{cd}\,.
\end{equation}
And the content of the $\mc F$ constraint is then using (\ref{eq:mcFconstraint-alphabeta}) and (\ref{eq:mcFconstraint-alphab})
\begin{eqnarray}
f_{\alpha\beta}&=&\frac{1}{3}D_{(\alpha}\Lambda\gamma^a\Lambda(\gamma_a\Lambda)_{\beta)}
+\psi_\alpha^{\phantom\alpha a}\psi_\beta^{\phantom\beta b}f_{ab}\label{eq:falphabeta2}
\nonumber\\
&&{}+\frac{i}{18}\partial_{\h\mu}\Lambda\gamma^a\Lambda\,\partial^{\h\mu}\Lambda\gamma^b\Lambda\,(\gamma_a\Lambda)_\alpha(\gamma_b\Lambda)_\beta
\\
f_{a\beta }&=&i(\gamma_a\Lambda)_\beta+\frac{1}{6}D_a\Lambda\gamma^c\Lambda(\gamma_c\Lambda)_\beta+\psi_\beta^{\phantom\alpha b}f_{ba}\,.
\label{eq:falphab2}
\end{eqnarray}
These two conditions give the non-linear extension appropriate to describing the dynamics of the non-abelian gauge field on a stack of D9-branes of the corresponding super Yang-Mills constraints, (\ref{eq:Falphabeta=0}) and (\ref{eq:Fbgamma}), derived in the previous chapter by requiring kappa-symmetry for the superparticle 
\begin{eqnarray}
f_{\alpha\beta}&=&0\\
f_{a\beta}&=&i(\gamma_a\psi)_\beta\,.\label{eq:f_abeta}
\end{eqnarray}
We see that two lowest order the theory on the stack of D-branes is super Yang-Mills theory with the transverse fermionic coordinate, $\Lambda$, playing the role of the spinorial field strength $\psi$.

The equations (\ref{eq:falphabeta2}) and (\ref{eq:falphab2}) were first derived in a paper by Berkovits and Pershin, \cite{Berkovits:2002ag}, using the pure spinor formulation of the superstring and requiring classical BRST invariance although they use a different type of boundary fermions and therefore only have terms with zero or two $\eta$. In the case of a single D9-brane the equations, first given in \cite{Kerstan:2002au}, take the same form except that the last term in (\ref{eq:falphabeta2}) is missing (and covariant derivatives get replaced by ordinary derivatives). This term is proportional to a commutator of the transverse directions, something which is typical in the case of coincident D-branes as we will see in the next chapter. Indeed, when we quantize the $\eta$, \emph{i.e.} replace them by gamma matrices, this term contains the Poisson-bracket, $-2i\delta^{\h\mu\h\nu}\partial_{\h\mu}\cdot\partial_{\h\nu}\cdot$, which should then be replaced by $-i$ times the \hbox{(anti-)}commutator, so that
\begin{eqnarray}
\lefteqn{\frac{i}{18}\partial_{\h\mu}\Lambda\gamma^a\Lambda\,\partial^{\h\mu}\Lambda\gamma^b\Lambda\,(\gamma_a\Lambda)_\alpha(\gamma_b\Lambda)_\beta}
\nonumber\\
&&\longrightarrow
\frac{i}{36}\{\Lambda^\gamma,\Lambda^\delta\}(\gamma^a\Lambda)_\gamma(\gamma_a\Lambda)_\alpha(\gamma^b\Lambda)_\delta(\gamma_b\Lambda)_\beta\,.
\end{eqnarray}

%% file: action.tex
\chapter{Actions for coincident D-branes}
In this chapter we want to consider the bosonic part of the action for a stack
of coincident D-branes. This action is known in a gauge fixed form only and we
give a proposal for a covariant version using the boundary fermions of the
superstring considered in the last chapter. We demonstrate that the action
suggested from these considerations is indeed gauge-invariant and diffeomorphism
invariant on the new space with extra fermionic directions. It is related to an
action in terms of matrices by quantization of the fermions and we give a short
discussion of this. This chapter presents the results in \cite{Howe:2006rv}. The presentation given here differs a little bit from that in the paper in that the proof of covariance is done in more detail and is a bit more general, and gauge invariance and the equivalence to Myers' action follow instead essentially from this proof.

We will begin, however, with a review of the action for a single D-brane and then discuss its non-abelian extension.

\section{The action for a single D-brane}
In string theory a D$p$-brane is a $(p+1)$-dimensional surface in space-time on which open strings can end. The massless excitations of the open string give rise to a supersymmetric $U(1)$ gauge field, $A_m$, together with scalar fields $\Phi^{m'}$, where $m=0,\ldots,p$ denotes the directions along the worldvolume of the brane and $m'=p+1,\ldots,9$ denotes the directions transverse to the brane. These fields also have their corresponding fermionic superpartners, which we will ignore in this chapter. The gauge field and scalars are confined to the brane worldvolume and to leading order their low-energy effective action is given by the dimensional reduction to $p+1$ dimensions of ten-dimensional supersymmetric Maxwell theory (or, equivalently, super Yang-Mills theory, (\ref{eq:SYM}), with gauge group $U(1)$). There are higher order ''stringy'' corrections to this action, however, in the parameter $\alpha'=l_s^2$, where $l_s$ is the string length scale. It was shown by Leigh in 1989, \cite{Leigh:1989jq}, that including these corrections the action takes the Dirac-Born-Infeld form\footnote{The flat space version of this, \emph{i.e.} $\phi=B=0$ and $g_{mn}=\eta_{mn}$, called simply the Born-Infeld action, was derived from string theory in 1985 by Fradkin and Tseytlin, \cite{Fradkin:1985qd}. For a review of Born-Infeld theory in this context see \cite{Tseytlin:1999dj}.}
\begin{equation}
\label{eq:S_DBI}
S_{\mr{DBI}}=-T_p\int_M d^{p+1}x\,e^{-\phi}\sqrt{-\det\left(g_{mn}+F_{mn}-B_{mn}\right)}\,,
\end{equation}
where $\phi$ is the dilaton, $g_{mn}\equiv\partial_m x^{\u m}\partial_n x^{\u n}G_{\u{mn}}$ is the induced metric on the D-brane worldvolume, $B_{mn}$ is the pull-back of the background Kalb-Ramond field to the D-brane worldvolume and $F_{mn}$ is the field strength of the gauge field $A_m$. The constant in front of the action has dimension $[length]^{-p-1}$ and is known as the tension of the brane. The $\alpha'$ dependence resides in $F$ and for $F$ to have the usual normalization and dimension we should take $F\rightarrow2\pi\alpha'F$. An action of this kind (in flat four-dimensional spacetime and without a $B$-field) had been proposed already in 1934, long before the advent of string theory, by Born and Infeld, \cite{Born:1934gh}, as a possible non-linear extension of electrodynamics in order to render the self-energy of a point charge finite. This effect is a consequence of the fact that for an action of the form (\ref{eq:S_DBI}) there is a maximum value of the field strength, as can be seen in the analogy to the relativistic particle action $\int dt\,\sqrt{1-v^2/c^2}$, where $c$, the speed of light, gives the maximum value of the velocity of the particle. This analogy was in fact the original motivation for their work. Later, in 1962, Dirac considered a similar action for an extended object in his membrane theory of the electron, \cite{Dirac:1962iy}.

In the derivation of this low-energy effective action an important assumption made is that the derivatives of the field strength are small compared to $F$ itself, so that they may be neglected. This means that this action can only be expected to give a good description of the low-energy physics when $F$ is slowly varying and in general it must be supplemented with derivative corrections in $F$. In fact it is known that the effective action contains not only terms with derivatives of the field strength but also of the background fields which are not included in this expression for the action. In this chapter we will therefore assume that all fields are slowly varying so that we can safely neglect these derivative corrections.

The D-brane also interacts with the massless Ramond-Ramond fields of the supergravity background. This interaction is incorporated in a second part of the action, the Wess-Zumino term, which can be written as (\cite{Douglas:1995bn})
\begin{equation}
\label{eq:S_WZ}
S_{\mr{WZ}}=T_p\int_M\,e^{F-B}\sum_n C^{(n)}\,.
\end{equation}
Here $C^{(n)}$ is the pull-back of the $n$-form RR potential of the background and the sum on $n$ runs over odd or even values for a type IIA or type IIB supergravity background respectively. It is understood that in the integrand the $(p+1)$-form terms in the expansion, which can be integrated over the worldvolume of the D$p$-brane, are picked out. The expression for $S_{\mr{WZ}}$ shows that a D$p$-brane is naturally charged under the $(p+1)$-form RR potential.

The fact that branes can support a flux of $F-B$ means that they can also act as charge sources for RR potentials of lower degree than $p+1$, \cite{Douglas:1995bn}. This has an interpretation as bound states of D-branes of different dimension. 

\subsection{Symmetries of the action}
The D-brane action
\begin{equation}
S=S_{\mr{DBI}}+S_{\mr{WZ}}
\end{equation}
is invariant under diffeomorphisms, \emph{i.e.} reparametrizations of the worldvolume, 
\begin{equation}
x^n\rightarrow \tilde x^n(x)\,,
\end{equation}
and also under gauge transformations of the form fields, $A$, $B$ and $C^{(n)}$. The diffeomorphism invariance is easy to see; the Wess-Zumino part is manifestly invariant since it is written in terms of forms only, for the DBI part the integrand transforms with the Jacobian of the transformation and this factor cancels the inverse factor of the Jacobian coming from the transformation of the measure. The gauge transformations of the form fields take the following form
\begin{eqnarray}
A&\rightarrow& A+d\Lambda^{(0)}\\
B&\rightarrow& B+d\Lambda^{(1)}\quad\mbox{and}\quad A\rightarrow A+\Lambda^{(1)}\label{eq:B-transf}\\
C^{(n)}&\rightarrow&C^{(n)}+d\Lambda^{(n-1)}-\Lambda^{(n-3)}H\label{eq:RR-transf}\,,
\end{eqnarray}
where the $\Lambda$ are arbitrary forms of the indicated degree and $H\equiv dB$. The invariance under gauge transformations of $A$ is manifest since the action only involves the field strength, $F\equiv dA$, and this is invariant. The invariance under $B$ gauge transformations follows from the fact that the action only involves $B$ in the combination $F-B$ and under the transformation (\ref{eq:B-transf}) we get
\begin{equation}
\delta(F-B)=d(\delta A)-\delta B=d\Lambda-d\Lambda=0\,,
\end{equation}
with $\Lambda$ a one-form. Finally, under gauge transformations of the RR fields, (\ref{eq:RR-transf}), we have
\begin{eqnarray}
\lefteqn{\delta\Big(e^{F-B}\sum_n C^{(n)}\Big)=e^{F-B}\sum_n\delta C^{(n)}= e^{F-B}\sum_n\left(d\Lambda^{(n-1)}-\Lambda^{(n-3)}H\right)}
\nonumber\\
&=&d\Big(e^{F-B}\sum_n\Lambda^{(n-1)}\Big)+e^{F-B}\sum_n\Lambda^{(n-1)}dB-e^{F-B}\sum_n\Lambda^{(n-3)}H
\nonumber\\
&=&d\Big(e^{F-B}\sum_n\Lambda^{(n-1)}\Big)\,,
\end{eqnarray}
where we have used the fact that $dF=d^2A=0$ and that any form of degree greater than $p+1$ is zero. The action therefore changes by the integral of a total derivative, which is zero (if we ignore the possibility of a boundary).

\subsection{Static gauge}
In order to exhibit the physical degrees of freedom of the system more clearly we can fix a specific gauge. A gauge that is often used is the so-called static gauge (or, more appropriately, Monge gauge) which we met in the case of the D9-brane in the last chapter. The idea is to use diffeomorphism invariance to fix the $p+1$ coordinates of the brane, $x^n$, to coincide with $p+1$ of the spacetime coordinates, $\u x^{\u n}$, 
\begin{eqnarray}
x^n&=&\u x^n\qquad n=0,\ldots,p\,.
\end{eqnarray}
The remaining, transverse, coordinates of spacetime then become scalar fields from the point of view of the brane (these are the scalars referred to in the beginning of this section). They describe the position of the brane in the transverse space. If we denote them by $\Phi$ and denote the directions transverse to the brane by a prime we have
\begin{eqnarray}
\Phi^{n'}(x)&=&\u x^{n'}\qquad n'=p+1,\ldots,9\,.
\end{eqnarray}

The action, (\ref{eq:S_DBI}) and (\ref{eq:S_WZ}), look superficially the same but we can now write the induced metric and the various pull-backs of the background fields in a more explicit form. The induced metric becomes
\begin{eqnarray}
g_{mn}&\equiv&\partial_m x^{\u m}\partial_n x^{\u n}G_{\u{mn}}
\nonumber\\
&=&G_{mn}+\partial_m\Phi^{m'}G_{m'n}+\partial_n\Phi^{m'}G_{m'm}+\partial_m\Phi^{m'}\partial_n\Phi^{n'}G_{n'm'}
\nonumber\\
\end{eqnarray}
and similarly for the gauge fields $B$ and $C^{(n)}$, \emph{e.g.}
\begin{equation}
C^{(1)}=dx^n\partial_n x^{\u m} C_{\u m}=dx^n(C_n+\partial_n\Phi^{m'} C_{m'})\,.
\end{equation}

\section{Myers' action for coincident D-branes}
\label{sec:Myers}
We know that when a number of D-branes become coincident the worldvolume gauge field becomes non-abelian. In fact, when $N$ D-branes are brought together the $U(1)^N$ gauge group of the system of separated branes becomes enhanced to a $U(N)$ gauge group in the limit that the branes coincide. This means that the D-brane action described in the previous section should be replaced by a non-abelian version. To lowest order the Dirac-Born-Infeld part of the action, (\ref{eq:S_DBI}), reduces to Maxwell theory, or $U(1)$ Yang-Mills, whose non-abelian generalization is just $U(N)$ Yang-Mills theory. But if we want to include also the stringy $\alpha'$ corrections and get a non-abelian generalization of the whole non-linear DBI action things are more complicated. To understand this generalization of the DBI action (and also the WZ term) has been an important unsolved problem in string theory for a long time and although, as we shall see, some progress has been made it still remains to a large extent mysterious.

\subsection{The symmetrized trace}
An important step towards a non-abelian version of the DBI action was taken by Tseytlin in \cite{Tseytlin:1997cs}. He argued that the assumption that we made in the abelian case that derivatives of the field strength are small should be replaced, in the non-abelian case, by the assumption that covariant derivatives of the field strength, $DF=\partial F-i[A,F]$, \emph{and} commutators, $[F,F]$ are small. The reason for the assumption that also the commutators of field strengths are small is that
\begin{equation}
[D_m,D_n]F_{kl}=-i[F_{mn},F_{kl}]\,,
\end{equation}
so that the two statements are linked. He showed that under this assumption the non-abelian BI action (the DBI action in flat space) in ten dimensions, corresponding to a D9-brane, becomes
\begin{equation}
\int d^{10}x\,\mr{SymTr}\left(\sqrt{-\det\left(\eta_{mn}+F_{mn}\right)}\right)\,,
\end{equation}
where SymTr denotes the symmetrized trace defined as
\begin{equation}
\mr{SymTr}\left(T^{a_1}\cdots T^{a_n}\right)\equiv\frac{1}{n!}\mr{Tr}\left(T^{a_1}\cdots T^{a_n}+\mbox{permutations}\right)\,,
\end{equation}
with $T^a$ generators of the gauge group and the non-abelian field strength is
\begin{equation}
F_{mn}\equiv \partial_m A_n-\partial_n A_m-i[A_m,A_n]\,.
\end{equation}
Observe that the commutator in the definition of $F$ is treated as one object in the symmetrization.

In \cite{Myers:1999ps} this was extended by Myers to a non-abelian action for the D9-brane in static gauge in a general background (similar results were also derived in \cite{Taylor:1999pr}). The DBI part of the action is\footnote{Our commutators and $B$ field differ by a sign to those of Myers.}
\begin{equation}
-T_9\int d^{10}x\,\mr{SymTr}\left(e^{-\phi}\sqrt{-\det\left(g_{mn}-B_{mn}+F_{mn}\right)}\right)
\end{equation}
and the WZ part
\begin{equation}
T_9\int_{M_{10}}\mr{SymTr}\left(e^{F-B}\sum_n C^{(n)}\right)\,.
\end{equation}
Because there are no transverse directions for the D9-brane and we are in static gauge the pull-backs are trivial and the fields appearing in the action are the same as the background ones.

It has been shown that this symmetrized trace prescription gives the correct effective action up to $F^4$-terms but fails to capture all terms in the effective action at order $F^6$, \cite{Hashimoto:1997gm}. Nevertheless it is interesting to consider it as an approximation and worry about corrections later, which is the point of view we will take here.

\subsection{T-duality}
\begin{figure}[h]
\begin{center}
\setlength{\unitlength}{1cm}
\begin{picture}(10,5)(0,0.5)
\put(2,2){\line(0,1){3}}
\put(2,2){\line(1,0){2.5}}
\put(2,5){\line(1,0){2.5}}
\put(4.5,2){\line(0,1){3}}
\put(2.7,4){$A_1, A_2$}
\put(3,3.5){$\Phi^3$}
\put(2.5,1.5){D2-brane}
\put(9,2){\line(0,1){3}}
\put(9.2,4){$A_2$}
\put(9.2,3.5){$\Phi^1, \Phi^3$}
\put(8.2,1.5){D$1$-brane}
\put(1,1){\vector(1,0){1}}
\put(2.2,1){$x^1$}
\put(1,1){\vector(0,1){1}}
\put(1,2.2){$x^2$}
\put(1,1){\vector(-4,-3){0.4}}
\put(0.2,0.8){$x^3$}
\put(6.5,3.5){\vector(1,0){0.5}}
\put(6.5,3.5){\vector(-1,0){0.5}}
\put(5,4){T-duality along $x^1$}
\end{picture}
\end{center}
\caption{T-duality along one of the worldvolume directions ($x^1$) of a D2-brane turns it into a D1-brane. Conversely, T-duality along a direction transverse to a D1-brane turns it into a D2-brane. Under this duality the roles of the corresponding component of the gauge field and transverse scalar exchange roles, $A_1\leftrightarrow\Phi^1$.}
\label{fig:T-duality}
\end{figure}
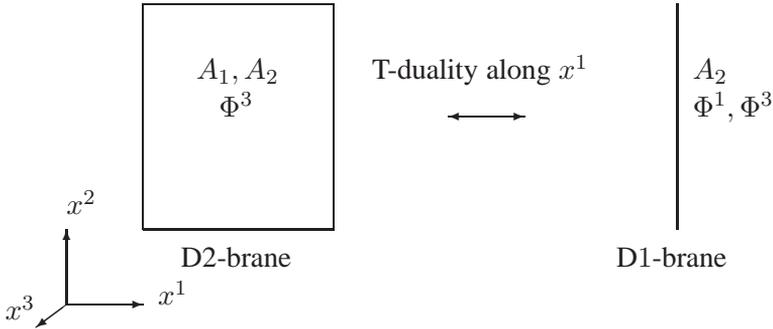

Once we know the action for a D$p$-brane for some value of $p$ we can obtain the action for D$p$-branes for all other $p$ by using an important symmetry in string theory known as T-duality. T-duality acts on a D$p$-brane in the following way: If a coordinate transverse to the D$p$-brane, \emph{e.g.} $y=x^{p+1}$ is T-dualized it becomes a D$(p+1)$-brane with $y$ the extra worldvolume direction and the transverse scalar corresponding to the $y$-direction, $\Phi^{p+1}$, becomes the extra component of the gauge field, $A_{p+1}$. If instead a worldvolume coordinate of the D$p$-brane, \emph{e.g.} $y=x^p$ is T-dualized it becomes a D$(p-1)$-brane with $y$ now an extra transverse direction and the component of the gauge field corresponding to the $y$-direction, $A_p$ becomes the extra transverse scalar, $\Phi^p$. Figure \ref{fig:T-duality} illustrates this for the case of T-duality along one of the worldvolume directions of a D2-brane, giving rise to a D1-brane, and, in the other direction, T-duality along a direction transverse to a D1-brane resulting in a D2-brane.

In a general supergravity background T-duality also affects the background fields and mixes them in a somewhat complicated way. In order to simplify the expression for the action we shall let $E$ denote the combination of the background metric and Kalb-Ramond field 
\begin{equation}
E_{\u{mn}}\equiv G_{\u{mn}}-B_{\u{mn}}\,.
\end{equation}

\subsection{The action}
Repeated T-duality of the D9-brane action then gives the action for a general D$p$-brane. As shown by Myers the Dirac-Born-Infeld part of the action becomes
\begin{eqnarray}
\lefteqn{S_{\mr{DBI}}=-T_p\int d^{p+1}x\,\mr{SymTr}\bigg(e^{-\phi}\sqrt{\det Q}}
\nonumber\\
&&\times\sqrt{-\det\left(\tilde E_{mn}+F_{mn}+\tilde E_{mp'}\left(\left(Q^{-1}-\delta\right)E^{-1}\right)^{p'q'}\tilde E_{q'n}\right)}\,\bigg)\,,
\nonumber\\
\label{S_NDBI}
\end{eqnarray}
where we have defined
\begin{equation}
Q^{m'}_{\phantom{m'}n'}\equiv\delta^{m'}_{n'}-i[\Phi^{m'},\Phi^{p'}]E_{p'n'}\,.
\end{equation}
A tilde denotes covariant pull-back, \emph{i.e.} pull-back using the gauge-covariant derivative on the D-brane instead of the ordinary derivative, \emph{e.g.}
\begin{equation}
\tilde v_n\equiv D_n x^{\u m}v_{\u m}=v_n+D_n\Phi^{m'}v_{m'}=v_n+\partial_n\Phi^{m'}v_{m'}-i[A_n,\Phi^{m'}]v_{m'}
\end{equation}
for a one-form $v$. The Wess-Zumino term takes the form
\begin{equation}
S_{\mr{WZ}}=T_p\int\mr{SymTr}\Big(e^F\tilde P\Big[e^{-ii_\Phi i_\Phi}e^{-B}\sum_n C^{(n)}\Big]\Big)\,,
\label{eq:S_NWZ}
\end{equation}
where $\tilde P[\ldots]$ denotes covariant pull-back of the quantity inside square brackets and $i_{\Phi}$ denotes the inner product of a form with the (in this case matrix valued) vector $\Phi^{n'}$. The inner product is defined as
\begin{equation}
i_\Phi A^{(n)}=\frac{1}{(n-1)!}dx^{\u m_1}\cdots dx^{\u m_{n-1}}\Phi^{n'}A_{n'\u m_{n-1}\cdots\u m_1}
\end{equation}
for $A^{(n)}$ an $n$-form defined on the background. Because of anti-symmetry we have
\begin{equation}
i_\Phi i_\Phi A^{(n)}=\frac{1}{2(n-2)!}dx^{\u m_1}\cdots dx^{\u m_{n-2}}[\Phi^{n'},\Phi^{m'}]A_{m'n'\u m_{n-2}\cdots\u m_1}\,,
\end{equation}
so that it vanishes if the componets of $\Phi$ commute.

The background fields become matrix valued through their dependence on the transverse coordinates and they have to be defined by Taylor expansion in these.

\subsection{Physical consequences}
The fact that, in the non-abelian case, two transverse indices on $B$ or $C^{(n)}$ can be contracted with the commutator of the transverse scalars, implying that coincident Dp-branes can also couple to RR forms of higher degree than $p+1$ contrary to the abelian case, was shown by Myers to give rise to an interesting new physical effect called the dielectric or Myers effect. The effect is an analog of the dielectric effect in ordinary electrodynamics, namely a stack of D-branes can be polarized by a background field into a higher-dimensional (non-commutative) configuration. A specific case is that of $N$ coincident D0-branes in a RR field strength corresponding to a D2-brane. This system can expand into a non-commutative two-sphere which can be interpreted as a bound state of a spherical D2-brane and $N$ D0-branes. This effect is related to many interesting string theory phenomena, \emph{e.g.} so-called giant gravitons and fuzzy funnels (for a review of some of these aspects see \cite{Myers:2003bw}).

\section{Boundary fermion inspired action}
In spite of its usefulness the action of Myers has some shortcomings that we would like to remedy. Firstly it is written in static gauge, which means that it doesn't possess diffeomorphism invariance. Second, the gauge invariance under gauge transformations of the background fields ($B$ and $C^{(n)}$) is far from obvious and very difficult to check because of the complexity of the expressions. Third, we would like to have the full supersymmetric action that reduces to Myers' action when the fermionic fields are set to zero. All of these problems seem very hard to address starting from the action of Myers. On the other hand, in the last chapter we developed a formalism based on the open superstring interacting with the non-abelian gauge field on the brane through boundary fermions. In this formalism supersymmetry is naturally built in and some of the complexities of dealing with matrices can be avoided as these simply become ordinary functions of the boundary fermions.

In \cite{Howe:2006rv} we write down an action for the non-abelian bosonic fields
on a stack of D-branes inspired by the ideas in \cite{Howe:2005jz} described in the previous
chapter. As we have seen from our considerations of the superstring with
boundary fermions we can think of the stack of D-branes as a space with extra
fermionic directions, arising from these boundary fermions, with a $U(1)$ gauge
field $\mc A$ on this space instead of as an ordinary worldvolume with a
non-abelian gauge field. This picture is strictly speaking only valid in some
classical approximation but it is interesting to try to use it to get some
insight into the effective action for coincident D-branes. To really derive the
action in this setting we should solve the generalized superembedding condition
and the $\mc F$ constraint derived from kappa-symmetry. This is however not very
simple and usually has to be done case by case, instead we will here follow the
approach in \cite{Howe:2006rv} and take a short-cut and simply guess what the action (for the bosonic fields) should be. We then show that this action has all the symmetries that it should have and that by gauge-fixing we can recover the action of Myers if we quantize the fermions in the way suggested before.

As we will only consider the bosonic part of the action the background space is simply ordinary ten-dimensional space with coordinates $x^{\u m}$. The stack of branes is then a ''superspace'', $\h M$, with coordinates $y^M=(x^m,\eta^\mu)$ with $m=0,\ldots,p$ and $\mu=1,\ldots,2q$ (since we are only considering the bosonic fields there are no other fermions and we can drop the hats on the indices). Since the object $F-B$ plays an important role in the ordinary DBI action it is natural that its generalization to $\h M$ should be important here. This is simply what we called $K$ in the previous chapter,
\begin{equation}
K_{MN}=(d\mc A)_{MN}-B_{MN}\,,
\end{equation}
where $B_{MN}$ is the pull-back of the background $B$-field to $\h M$. In fact, the simplest possibility would be to just generalize the ordinary DBI part of the action by writing the corresponding thing for the space $\h M$. We replace the $U(1)$ gauge field $A_m$ by the $U(1)$ gauge field on $\h M$, $\mc A_M$, the integral becomes an integral over $x$ and $\eta$ and the determinant must be replaced with a superdeterminant to take into account the fact that some directions are fermionic. We are thus led to consider the straight-forward generalization of the abelian DBI action, (\ref{eq:S_DBI}), to $\h M$
\begin{equation}
\label{eq:DBI-Mhat}
S_{\mr{DBI}}=-T_p\int_{\h M}d^{2q}\eta\,d^{p+1}x\,e^{-\phi}\sqrt{-\mr{sdet}\left(g_{MN}+K_{MN}\right)}\,.
\end{equation}
The definition of the superdeterminant is
\begin{equation}
\label{eq:sdet}
\mr{sdet}\,M\equiv \det(M_{mn}-M_{m\mu}(M_{\mu\nu})^{-1}M_{\nu n})\det(M_{\mu\nu})^{-1}\,.
\end{equation}
For the same reasons as in the ordinary DBI action the action (\ref{eq:DBI-Mhat}) is invariant under diffeomorphisms of $\h M$. We should of course make sure also that the superdeterminant is well-defined and non-zero. This follows from the fact that the induced metric for the bosonic directions, $g_{mn}$, should be non-singular and the fact that we had to assume that $K$ in the fermionic directions, $K_{\mu\nu}$, was also non-singular. In fact, we can already see that this has a chance of reproducing Myers' DBI action from the fact that the superdeterminant is a product of two ordinary determinants, which is precisely the structure that appears in (\ref{S_NDBI}).

The Wess-Zumino term can not be generalized is such a simple way, unfortunately. It is clear however that it should involve the factor $e^K$ and the sum of pull-backs of RR potentials. The observation that in Myers' WZ term there are contractions with commutators of the scalars suggests that some contractions should also be present in our case. We essentially only have one object that we can use to contract indices of forms apart from the metric, namely $N^{\mu\nu}\equiv (K_{\mu\nu})^{-1}$. This gives some motivation for the choice
\begin{equation}
\label{eq:WZ-Mhat}
S_{\mr{WZ}}=T_p\int d^{2q}\eta\,\int_M\sqrt{\det N}\Big(e^{-\frac{1}{2}i_N}e^{K-K_{0,2}}\sum_n C^{(n)}\Big)_{p+1,0}\,.
\end{equation}
The subscript $p+1,0$ means that we should pick out the forms of bosonic degree $p+1$ and fermionic degree $0$ (\emph{i.e.} the forms that have $p+1$ factors of $dx$ and none of $d\eta$), which are then integrated over the $(p+1)$-dimensional bosonic part of the worldvolume, $M$. Similarly $K_{0,2}$ is the part of the two-form $K$ of bosonic degree zero and fermionic degree two, 
\begin{equation}
K_{0,2}\equiv\frac{1}{2}e^\nu e^\mu K_{\mu\nu}
\end{equation}
in terms of a basis of one-forms $\{e^M\}$ on $\h M$. Finally $i_N$ denotes the inner product with $N$ defined for a $(k,l)$-form
\begin{equation}
\omega=\frac{1}{k!l!}e^{\mu_l}\cdots e^{\mu_1}e^{m_k}\cdots e^{m_1}\omega_{m_1\cdots m_k\mu_1\cdots\mu_l}
\end{equation}
as 
\begin{equation}
\label{eq:i_N}
i_N\omega=\frac{1}{k!(l-2)!}e^{\mu_l}\cdots e^{\mu_3}e^{m_k}\cdots e^{m_1}N^{\mu_2\mu_1}\omega_{m_1\cdots m_k\mu_1\cdots\mu_l}\,,
\end{equation}
for $l\geq 2$ and zero otherwise.

\subsection{Covariance of the WZ term}
Apart from forms like $K$ and $C^{(n)}$, which are, by definition, unaffected by a change of basis of (co)tangent space our expression for the Wess-Zumino part of the action, (\ref{eq:WZ-Mhat}), contains objects like $N$ and $K_{0,2}$ which in general depend on which basis they are referred to. Remarkably, as we shall demonstrate in this section, these objects transform in such a way that the expression for $S_{\mr{WZ}}$ remains invariant. The proof is long but basically straight-forward and will provide us with useful relations that simplify the proof of gauge invariance.

We will divide the proof into three steps. The idea is to start with the expression for $S_{\mr{WZ}}$ in a coordinate basis with basis one-forms
$dx^m$ and $d\eta^\mu$ and then change the basis of (co)tangent space in three steps to arrive at a general basis. The steps are as follows

\begin{itemize-indent}
\item{Step 1:}
\begin{eqnarray}
E'^{m}&=&dx^{m}\nonumber\\
E'^{\mu}&=&d\eta^{\mu}+dx^m\phi_m^{\phantom m \mu}.
\label{eq:step1}
\end{eqnarray}
\item{Step 2:}
\begin{eqnarray}
E''^{m}&=&E'^m+E'^\mu\chi_\mu^{\phantom\mu m}\nonumber\\
E''^{\mu}&=&E'^\mu.
\label{eq:step2}
\end{eqnarray}
\item{Step 3:}
\begin{eqnarray}
E'''^{m}&=&E''^{n}\psi_n^{\phantom n m}\nonumber\\
E'''^{\mu}&=&E''^\nu E_\nu^{\phantom\nu\mu}.
\label{eq:step3}
\end{eqnarray}
\end{itemize-indent}
The components of the vielbein at the end can then be related to $\phi$, $\chi$ and $\psi$ as
\begin{eqnarray}
E_m^{\phantom m\mu}&=&\phi_m^{\phantom m\nu}E_\nu^{\phantom\nu\mu}\\
E_\mu^{\phantom\mu m}&=&\chi_\mu^{\phantom\mu n}\psi_n^{\phantom nm}\\
E_m^{\phantom m n}&=&\left(\delta_m^k+\phi_m^{\phantom m\mu}\chi_\mu^{\phantom\mu k}\right)\psi_k^{\phantom kn}\,.
\label{eq:vielbein}
\end{eqnarray}

It will be convenient to have a realization of the operator, call it $\Pi_0$, that projects on purely bosonic forms, \emph{i.e.} for $\omega$ a $(k,l)$-form
\begin{equation}
\Pi_0\omega=\left\{
\begin{array}{cc}
\omega & \mbox{ if $l=0$}\\
0 & \mbox{ if $l>0$}
\end{array}\right.\,.
\end{equation}
Let $\{d_M\}$ be a basis of tangent space of $\h M$ with dual basis $\{e^M\}$. To give an explicit expression for this operator we define the inner product with the basis vectors, $I_{d_M}$, such that for for $k>0$
\begin{equation}
I_{d_n}\omega=\frac{(-1)^l}{(k-1)!l!}e^{\mu_l}\cdots e^{\mu_1}e^{m_{k-1}}\cdots e^{m_1}\omega_{m_1\cdots m_{k-1}n\mu_1\cdots\mu_l}
\end{equation}
and zero otherwise, and for $l>0$
\begin{equation}
\label{eq:Idmu}
I_{d_\nu}\omega=\frac{1}{k!(l-1)!}e^{\mu_{l-1}}\cdots e^{\mu_1}e^{m_k}\cdots e^{m_1}\omega_{m_1\cdots m_k\mu_1\cdots\mu_{l-1}\nu}
\end{equation}
and zero otherwise. Note that this operation is taken to act from the left as this will be more convenient for us. It is then easy to see that
\begin{equation}
e^\mu I_{d_\mu}\omega=l\omega\,,
\end{equation}
and by repeating this we get
\begin{equation}
\frac{1}{n!}e^{\mu_n}\cdots e^{\mu_1} I_{d_{\mu_1}}\cdots I_{d_{\mu_n}}\omega=\left\{
\begin{array}{cc}
\binom{l}{n}\omega&\mbox{ if $n\leq l$}\\
0 & \mbox{ if $n>l$}
\end{array}\right.\,.
\label{eq:Pi0-term}
\end{equation}
Then defining 
\begin{equation}
\Pi_0\equiv\,:e^{-e^\mu I_{d_\mu}}:\,\equiv\sum_{n=0}^\infty\frac{(-1)^n}{n!}e^{\mu_n}\cdots e^{\mu_1} I_{d_{\mu_1}}\cdots I_{d_{\mu_n}}
\end{equation}
gives, using (\ref{eq:Pi0-term}),
\begin{equation}
\Pi_0\omega=\sum_{n=0}^l(-1)^n\binom{l}{n}\omega=\left\{
\begin{array}{cc}
\omega&\mbox{ if $l=0$}\\
0&\mbox{ if $l>0$}
\end{array}\right.\,.
\end{equation}
This is therefore the operator that projects on bosonic forms. From (\ref{eq:WZ-Mhat}) we see that the object of relevance to the WZ term is the $(p+1)$-form piece of the bosonic form
\begin{equation}
\label{eq:Omega}
\Omega\equiv\sqrt{\det N}\Pi_0e^{-\frac{1}{2}i_N}e^{K-K_{0,2}}\sum_n C^{(n)}\,,
\end{equation}
where it is understood that $\Pi_0$ and $i_N$ act on everything to their right. It will also be useful to write the inner product with $N$ defined in (\ref{eq:i_N}) in terms of the inner product with the basis vectors. Using (\ref{eq:Idmu}) it's easy to see that
\begin{equation}
i_N=N^{\mu\nu}I_{d_\mu}I_{d_\nu}\,.
\end{equation}

We are now ready to see how the bosonic form $\Omega$ transforms under the changes of basis in (\ref{eq:step1}), (\ref{eq:step2}) and (\ref{eq:step3}).

\subsubsection{Step 1}
We start from the coordinate basis $\{dy^M\}$ and take as a new basis of cotangent space the one-forms
\begin{eqnarray}
E'^m&=&dx^m\nonumber\\
E'^\mu&=&d\eta^\mu+dx^m\phi_m^{\phantom m\mu}.
\end{eqnarray}
The dual basis is then given by
\begin{eqnarray}
D'_m&=&\partial_m-\phi_m^{\phantom m\mu}\partial_\mu\nonumber\\
D'_\mu&=&\partial_\mu\,.
\label{eq:prime-basis-dual}
\end{eqnarray}
We shall denote all objects expressed in this basis by a prime. It is easy to see that $K'_{\mu\nu}=K_{\mu\nu}$ and therefore we also have 
${N'}^{\mu\nu}\equiv (K'_{\mu\nu})^{-1}=N^{\mu\nu}$ and $(i_N)'=i'_{N'}=i_N$ so we can omit the primes on these quantities without risk of confusion. Expressing all quantities in this new basis we then have
\begin{equation}
\label{eq:pi-prime}
\Pi_0=\,:e^{-d\eta^\mu I_{\partial_\mu}}:\,=\,:e^{-E'^\mu I_{D'_\mu}+E'^m\phi_m^{\phantom m\mu}I_{D'_\mu}}:\,=\Pi_0'e^{E'^m\phi_m^{\phantom m\mu}I_{D'_\mu}}\,,
\end{equation}
where we have removed the ordering symbol on the last factor since it is irrelevant, and
\begin{equation}
K_{0,2}=\frac{1}{2}d\eta^\nu d\eta^\mu K_{\mu\nu}=K'_{0,2}-(\phi K)_{1,1}-(\phi K\phi^\mr{T})_{2,0}\,,
\end{equation}
where we have defined
\begin{equation}
(\phi K)_{1,1}=E'^\mu E'^m\phi_m^{\phantom m\nu}K_{\nu\mu}\quad\mbox{and}\quad
(\phi K\phi^\mr{T})_{2,0}=\frac{1}{2}E'^nE'^m\phi_m^{\phantom m\nu}\phi_n^{\phantom n\mu}K_{\nu\mu}\,. 
\end{equation}
The expression for $\Omega$, given in (\ref{eq:Omega}), therefore becomes
\begin{equation}
\sqrt{\det N}\Pi_0'e^{E'^m\phi_m^{\phantom m\mu}I_{D'_\mu}}
e^{-\frac{1}{2}i_N}e^{K-K_{0,2}'+(\phi K)_{1,1}+(\phi K\phi^\mr{T})_{2,0}}\sum_n C^{(n)}\,.
\label{eq:Omegaprime}
\end{equation}
Consider first the part
\begin{equation}
e^{-\frac{1}{2}i_N}e^{(\phi K)_{1,1}}\cdots\,,
\end{equation}
where the dots denote the additional factors on which $i_N$ acts. Now split the operation $i_N$ into three parts
\begin{equation}
i_N=(i_N)_{11}+(i_N)_{12}+(i_N)_{22}\,,
\end{equation}
where $(i_N)_{11}$ acts only on the factor $e^{\phi K}$, $(i_N)_{22}$ acts only on the rest ($\cdots$) and for $(i_{N})_{12}$ one $I_{d_\mu}$ (recall the expression $i_N=N^{\mu\nu}I_{d_\mu}I_{d_\nu}$) acts on the factor $e^{\phi K}$ and the other $I_{d_\mu}$ acts on the rest (denoted $\cdots$ above). Let's first consider the $(i_N)_{11}$-piece,
\begin{eqnarray}
e^{-\frac{1}{2}(i_N)_{11}}e^{(\phi K)_{1,1}}
&=&\sum_{n=0}^\infty\sum_{k=0}^{[n/2]}
\frac{1}{n!}\frac{1}{k!}\left(-\frac{1}{2}i_N\right)^k\left((\phi K)_{1,1}^n\right)
\nonumber\\
&=&\sum_{n=0}^\infty\sum_{k=0}^{[n/2]}
\frac{1}{n!k!}\frac{n!}{(n-2k)!}(\phi K\phi^\mr{T})_{2,0}^k(\phi K)_{1,1}^{n-2k}\,.
\nonumber\\
\end{eqnarray}
Changing summation variables from $k$ and $n$ to $k$ and $n'=n-2k$ this becomes
\begin{equation}
\sum_{n'=0}^\infty\sum_{k=0}^\infty
\frac{1}{k!}(\phi K\phi^\mr{T})_{2,0}^k\frac{1}{n'!}(\phi K)_{1,1}^{n'}
=e^{(\phi K\phi^\mr{T})_{2,0}}e^{(\phi K)_{1,1}}\,.
\label{eq:i11-result}
\end{equation}
Next let us compute
\begin{eqnarray}
\lefteqn{e^{-\frac{1}{2}(i_N)_{12}}e^{(\phi K)_{1,1}}}
\nonumber\\
&=&\sum_{n=0}^\infty\sum_{k=0}^n
\frac{1}{n!}\frac{1}{k!}(-1)^k 
N^{\mu_k\nu_k}I_{D'_{\mu_k}}\cdots N^{\mu_1\nu_1}I_{D'_{\mu_1}}\left((\phi K)_{1,1}^n\right)I_{D'_{\nu_1}}\cdots I_{D'_{\nu_k}}
\nonumber\\
&=&\sum_{n=0}^\infty\sum_{k=0}^n
\frac{(-1)^k }{k!(n-k)!}(\phi K)_{1,1}^{n-k}
E'^{m_k}\phi_{m_k}^{\phantom{m_k}\nu_k}\cdots E'^{m_1}\phi_{m_1}^{\phantom{m_1}\nu_1}I_{D'_{\nu_1}}\cdots I_{D'_{\nu_k}}\,.
\nonumber\\
\end{eqnarray}
Changing the summations to run over $k$ and $n-k$ this becomes
\begin{equation}
e^{(\phi K)_{1,1}}e^{-E'^m\phi_m^{\phantom m\nu}I_{D'_\nu}}\,.
\label{eq:i12-result}
\end{equation}
From (\ref{eq:i11-result}) and (\ref{eq:i12-result}) we have
\begin{eqnarray}
\lefteqn{e^{-\frac{1}{2}i_N}e^{(\phi K)_{1,1}}\cdots=e^{-\frac{1}{2}((i_N)_{11}+(i_N)_{12}+(i_N)_{22})}e^{(\phi K)_{1,1}}\cdots}
\nonumber\\
&=&e^{-\frac{1}{2}((i_N)_{11}+(i_N)_{12})}e^{(\phi K)_{1,1}}e^{-\frac{1}{2}i_N}\cdots
\nonumber\\
&=&e^{(\phi K\phi^\mr{T})_{2,0}}e^{-\frac{1}{2}(i_N)_{12}}e^{(\phi K)_{1,1}}e^{-\frac{1}{2}i_N}\cdots
\nonumber\\
&=&e^{(\phi K\phi^\mr{T})_{2,0}}e^{(\phi K)_{1,1}}e^{-E'^m\phi_m^{\phantom m\nu}I_{D'_\nu}}e^{-\frac{1}{2}i_N}\cdots
\end{eqnarray}
Thus our expression for $\Omega$ in (\ref{eq:Omegaprime}) becomes
\begin{eqnarray}
\lefteqn{\Omega=\sqrt{\det N}e^{2(\phi K\phi^\mr{T})_{2,0}}}
\nonumber\\
&&\cdot\Pi_0'e^{E'^m\phi_m^{\phantom m\mu}I_{D'_\mu}}
e^{(\phi K)_{1,1}}e^{-E'^m\phi_m^{\phantom m\nu}I_{D'_\nu}}e^{-\frac{1}{2}i_N}e^{K-K_{0,2}'}\sum_n C^{(n)}\,.
\end{eqnarray}

It's now easy to see that because $\Pi_0$ projects on bosonic forms
\begin{equation}
\Pi_0'e^{E'^m\phi_m^{\phantom m\mu}I_{D'_\mu}}e^{(\phi K)_{1,1}}\cdots
=\Pi_0'e^{-2(\phi K\phi^\mr{T})_{2,0}}e^{E'^m\phi_m^{\phantom m\mu}I_{D'_\mu}}\cdots\,,
\end{equation}
so that the corresponding factors in the equation above are cancelled and we find
\begin{equation}
\Omega=\sqrt{\det N}\Pi_0'e^{-\frac{1}{2}i_N}e^{K-K_{0,2}'}\sum_n C^{(n)}=\Omega'
\end{equation}
and the WZ term is thus left unchanged.

\subsubsection{Step 2}
As the next step we change the basis of cotangent space to
\begin{eqnarray}
E''^m&=&E'^m+E'^\nu\chi_\nu^{\phantom\nu m}\nonumber\\
E''^\mu&=&E'^\mu\,.
\label{eq:double-prime-basis}
\end{eqnarray}
The dual basis becomes
\begin{eqnarray}
D''_m&=&D'_m\nonumber\\
D''_\mu&=&D'_\mu-\chi_\mu^{\phantom\mu m}D'_m\,.
\label{eq:double-prime-basis-dual}
\end{eqnarray}
We will look at the transformation of $\Omega$ in two parts.

\subsubsection{Step 2a}
We start by looking at how the projection operator $\Pi_0$ changes when we go from the primed to the double-primed basis. We find using (\ref{eq:double-prime-basis}) and (\ref{eq:double-prime-basis-dual})
\begin{equation}
\Pi'_0=\,:e^{-E'^\mu I_{D_\mu'}}:\,=\,:e^{-E''^\mu I_{D''_\mu}-E''^\mu\chi_\mu^{\phantom\mu m}I_{D''_m}}:\,
=e^{-E''^\mu\chi_\mu^{\phantom\mu m}I_{D''_m}}\Pi''_0\,.
\end{equation}
A term in the expansion of the exponential is
\begin{equation}
\frac{1}{n!}I_{D''_{m_1}}\cdots I_{D''_{m_n}}E''^{\mu_n}\chi_{\mu_n}^{\phantom{\mu_n}m_n}
\cdots E''^{\mu_1}\chi_{\mu_1}^{\phantom{\mu_1}m_1}\,.
\label{eq:Pidoubleprimeexpansion}
\end{equation}
This will be acting on $\Pi_0''\cdots$, \emph{i.e.} on a sum of $(k,0)$-forms (in the double-prime basis) with $k\leq p+1$ because of anti-symmetry. In the end the only terms that contribute to the Wess-Zumino action will be the ones that have precisely $p+1$ factors of $dx^m$, so that they can be integrated over the $(p+1)$-dimensional bosonic subspace of $\h M$. This means that we can forget about contributions to $\Omega$ of the form $I_{\partial_m}(\cdots)$ as they will always have less than $p+1$ factors of $dx^m$. The idea is then that we can remove the $I_{\partial_m}$ part from $I_{D''_m}$ in the above expression. Using (\ref{eq:double-prime-basis-dual}) and (\ref{eq:prime-basis-dual}) we get
\begin{equation}
D''_m=D'_m=\partial_m-\phi_m^{\phantom m\mu}\partial_\mu\,.
\end{equation}
But we also find
\begin{equation}
\partial_\mu=D'_\mu=D''_\mu+\chi_\mu^{\phantom\mu m}D'_m
=D''_\mu+\chi_\mu^{\phantom\mu m}\partial_m-\chi_\mu^{\phantom\mu m}\phi_m^{\phantom m\nu}\partial_\nu\,,
\end{equation}
which implies that
\begin{equation}
\partial_\mu
=P_\mu^{\phantom\mu\nu}\left(D''_\nu+\chi_\nu^{\phantom\nu m}\partial_m\right)
\end{equation}
with
\begin{equation}
\label{eq:def-P}
P_\mu^{\phantom\mu\nu}\equiv\left(\delta_\nu^\mu+\chi_\nu^{\phantom\nu m}\phi_m^{\phantom m\mu}\right)^{-1}\,.
\end{equation}
Thus we can write
\begin{equation}
D''_m=D'_m=\left(\delta_m^n-(\phi P\chi)_m^{\phantom mn}\right)\partial_n-(\phi P)_m^{\phantom m\mu}D''_\mu\,.
\end{equation}
Up to terms in $\Omega$ of the form $I_{\partial_m}(\cdots)$ which are irrelevant, we can then replace $I_{D''_m}$ by $-(\phi P)_m^{\phantom m\mu}I_{D''_\mu}$ in (\ref{eq:Pidoubleprimeexpansion}) and we get
\begin{equation}
\frac{(-1)^n}{n!}
(\phi P)_{m_1}^{\phantom{m_1}\nu_1}I_{D''_{\nu_1}}
\cdots (\phi P)_{m_n}^{\phantom{m_n}\nu_n}I_{D''_{\nu_n}}
E''^{\mu_n}\chi_{\mu_n}^{\phantom{\mu_n}m_n}\cdots E''^{\mu_1}\chi_{\mu_1}^{\phantom{\mu_1}m_1}\,.
\end{equation}
Since this whole expression is acting on a bosonic form in the double-prime basis all the $I_{D''_\mu}$ will have to act on the $E''^\mu$ in the above expression, and we get
\begin{eqnarray}
\lefteqn{\frac{(-1)^n}{n!}(\phi P)_{m_1}^{\phantom{m_1}\mu_1}
\cdots(\phi P)_{m_n}^{\phantom{m_n}\mu_n}
n!\chi_{(\mu_n}^{\phantom{\mu_n}m_n}\cdots\chi_{\mu_1)}^{\phantom{\mu_1}m_1}}
\nonumber\\
{}&=&(-\phi P\chi)_{m_1}^{\phantom{\mu_1}[m_1}\cdots(-\phi P\chi)_{m_n}^{\phantom{\mu_n}m_n]}
\nonumber\\
{}&=&(-\phi P\chi)_{m_1}^{\phantom{\mu_1}n_1}\cdots(-\phi P\chi)_{m_n}^{\phantom{\mu_n}n_n}
\delta_{n_1}^{[m_1}\cdots\delta_{n_n}^{m_n]}\,.
\end{eqnarray}
Now we use the fact that\footnote{The extra minus sign is due to the Minkowski signature.}
\begin{equation}
\varepsilon^{m_1\cdots m_n m_{n+1}\cdots m_{p+1}}\varepsilon_{n_1\cdots n_n m_{n+1}\cdots m_{p+1}}
=-n!(p+1-n)!\delta_{n_1}^{[m_1}\cdots\delta_{n_n}^{m_n]}
\end{equation}
to write this as
\begin{eqnarray}
\lefteqn{-\frac{\varepsilon^{m_1\cdots m_n m_{n+1}\cdots m_{p+1}}}{n!(p+1-n)!}(-\phi P\chi)_{m_1}^{\phantom{\mu_1}n_1}
\cdots(-\phi P\chi)_{m_n}^{\phantom{\mu_n}n_n}\varepsilon_{n_1\cdots n_n m_{n+1}\cdots m_{p+1}}}
\nonumber\\
&&=-\binom{p+1}{n}\frac{\varepsilon^{m_1\cdots m_{p+1}}}{(p+1)!}(-\phi P\chi)_{m_1}^{\phantom{\mu_1}n_1}\cdots(-\phi P\chi)_{m_n}^{\phantom{\mu_n}n_n}
\nonumber\\
&&\qquad\times\delta_{m_{n+1}}^{n_{n+1}}\cdots\delta_{m_{p+1}}^{n_{p+1}}\varepsilon_{n_1\cdots n_{p+1}}\,.
\end{eqnarray}
This we recognize as a term in the expansion of $\det(\delta-\phi P\chi)$. But this can be rewritten using the expression for $P$ in (\ref{eq:def-P}) which gives
\begin{equation}
\phi P\chi=\phi\chi-\phi\chi\phi\chi+\ldots=\delta-\left(\delta+\phi\chi\right)^{-1}\,. 
\end{equation}
Thus we have shown that
\begin{equation}
\Pi'_0\simeq\det(\delta-\phi P\chi)\Pi''_0=\det(\delta+\phi\chi)^{-1}\Pi''_0\,,
\label{eq:pi-double-prime}
\end{equation}
where $\simeq$ means modulo terms of the form $I_{\partial_m}(\cdots)$ which don't affect the action. Now we move on to the rest of the expression for $\Omega$.

\subsubsection{Step 2b}
We first want to show that, in any basis, shifting $K_{\mu\nu}$ by a symmetric matrix $(\delta K)_{\mu\nu}$ we have the relation
\begin{eqnarray}
\lefteqn{\sqrt{\det(K+\delta K)^{-1}}\,\Pi_0\left(e^{-\frac{1}{2}i_{(K+\delta K)^{-1}}}\omega_{0,2r}\right)}
\nonumber\\
&=&\sqrt{\det N}\,\Pi_0\left(e^{-\frac{1}{2}i_N}e^{(\delta K)_{0,2}}\omega_{0,2r}\right)\,,
\label{eq:shift-rel}
\end{eqnarray}
where $N\equiv K^{-1}$ as usual. This expression of course only holds as long as both sides don't diverge (\emph{e.g.} taking $\omega=1$ and $\delta K=aK$ for some number $a$ it is easy to see that $e^{-\frac{1}{2}i_N}e^{aK_{0,2}}$ diverges if $|a|\geq 1$). The bosonic part of the form $\omega$ doesn't enter in the considerations and we've set it to zero for simplicity.

Thinking of $\omega$ as a $r$-linear map of symmetric matrices and forgetting for the moment about the determinant factors the left-hand-side can be written as
\begin{eqnarray}
\lefteqn{\left(-\frac{1}{2}\right)^r\frac{1}{r!}\omega((K+\delta K)^{-1},\ldots,(K+\delta K)^{-1})}
\nonumber\\
&=&\left(-\frac{1}{2}\right)^r\frac{1}{r!}
\sum_{n=0}^r\binom{r}{n}
\nonumber\\
&&\qquad\cdot\omega((K+\delta K)^{-1}-N,\ldots,(K+\delta K)^{-1}-N,\underbrace{N,\ldots,N}_{r-n})
\nonumber\\
&=&\left(-\frac{1}{2}\right)^r 
\sum_{n=0}^r\frac{1}{n!(r-n)!}
\sum_{l=0}^\infty(-1)^{l+n}
\nonumber\\
&&\qquad\cdot\sum_{\{l_i\}_{i=1}^n}
\omega((N\delta K)^{l_1+1}N,\ldots,(N\delta K)^{l_n+1}N, N,\ldots,N)\,.
\nonumber\\
\label{eq:lhs-shift-rel}
\end{eqnarray}
The last sum runs over partitions of $l$ into $n$ parts, so that $\Sigma_i l_i=l$ with $0\leq l_i\leq l$. For example for $l=2$ and $n=3$
the sum would run over
\begin{equation}
\{l_1,l_2,l_3\}=\{2,0,0\},\{1,1,0\},\{1,0,1\},\{0,2,0\},\{0,1,1\},\{0,0,2\}.
\end{equation}

For the right-hand-side of (\ref{eq:shift-rel}) let's start by looking at the part of $e^{-\frac{1}{2}i_N}$ which acts only on the factor $e^{(\delta K)_{0,2}}$. This is
\begin{eqnarray}
\lefteqn{e^{-\frac{1}{2}(i_N)_{11}}e^{(\delta K)_{0,2}}
=\sum_{n=0}^\infty\sum_{k=0}^n\frac{1}{n!k!}\left(-\frac{1}{2}\right)^k i_N^k\left((\delta K)_{0,2}^n\right)}
\nonumber\\
&=&\sum_{n=0}^\infty\sum_{k=0}^n\frac{(-1)^k}{k!n!}
\sum_{l=0}^k\left(\frac{1}{2}\right)^{k-l}\binom{k}{l}\binom{n}{k-l}i_N^{k-l}\left((\delta K)_{0,2}^{k-l}\right)
\nonumber\\
&&l!(n-k+l)!\sum_{\{l_i\}_{i=1}^{n-k}}\frac{1}{d_{\{l_i\}}}
\left(\delta K(N\delta K)^{l_1}\right)_{0,2}\!\!\!\cdots\left(\delta K(N\delta K)^{l_{n-k}}\right)_{0,2}\,,
\nonumber\\
\end{eqnarray}
where the degeneracy of a given partition is 
\begin{equation}
d_{\{l_i\}}=\prod_{j=0}^l d_j!\,,
\end{equation}
where $d_j$ is the number of $l_i$ equal to $j$ for a given partition. Changing summation variable in the $n$ sum to $n'=n-k$ gives
\begin{eqnarray}
\lefteqn{\sum_{n'=0}^\infty\sum_{k=0}^\infty\sum_{l=0}^k
\left(-\frac{1}{2}\right)^{k-l}\frac{1}{((k-l)!)^2}i_N^{k-l}\left((\delta K)_{0,2}^{k-l}\right)}
\nonumber\\
&&(-1)^l
\sum_{\{l_i\}_{i=1}^{n'}}\frac{1}{d_{\{l_i\}}}
\left(\delta K(N\delta K)^{l_1}\right)_{0,2}\cdots\left(\delta K(N\delta K)^{l_{n'}}\right)_{0,2}
\nonumber\\
&=&c\sum_{n=0}^\infty\sum_{l=0}^\infty(-1)^l
\sum_{\{l_i\}_{i=1}^n}\frac{1}{d_{\{l_i\}}}
\left(\delta K(N\delta K)^{l_1}\right)_{0,2}\cdots\left(\delta K(N\delta K)^{l_n}\right)_{0,2}
\nonumber\\
\end{eqnarray}
where we've defined 
\begin{equation}
c\equiv\sum_{k=0}^\infty\left(-\frac{1}{2}\right)^{k}\frac{1}{(k!)^2}i_N^{k}\left((\delta K)_{0,2}^k\right).
\label{eq:c-def}
\end{equation}

Apart from an overall factor of $c\sqrt{\det N}$ the right-hand-side of (\ref{eq:shift-rel}) is then given by (the zero-form part of) the rest of $e^{-\frac{1}{2}i_N}$ acting on the above expression times $\omega$. It is not hard to show that this gives
\begin{eqnarray}
\lefteqn{\left(-\frac{1}{2}\right)^r\sum_{n=0}^r\sum_{l=0}^\infty\sum_{\{l_i\}_{i=1}^n}
\frac{(-1)^{l+n}}{n!(r-n)!}}
\nonumber\\
&&\qquad\times\omega((N\delta K)^{l_1+1}N,\ldots,(N\delta K)^{l_n+1}N,N,\ldots,N)\,.
\end{eqnarray}
This exactly matches (\ref{eq:lhs-shift-rel}) so it only remains to show that
\begin{equation}
c\sqrt{\det N}=\sqrt{\det(K+\delta K)^{-1}}
\end{equation}
in order to prove (\ref{eq:shift-rel}).

Looking at the expression for $c$ in (\ref{eq:c-def}) we see that it is the expansion of $e^{-(\delta K)_{0,2}}$ with each term completely contracted with $N$s in all possible (inequivalent) ways. This can be thought of as a finite-dimensional ''perturbative expansion'' in $\delta K$, and introducing $2q$ bosonic variables $c^\mu$ we can write it in analogy to the path-integral case as
\begin{equation}
c=\frac{\int\prod_\mu dc^\mu\,e^{-\frac{1}{2}c^\mu K_{\mu\nu}c^\nu-\frac{1}{2}c^\mu(\delta K)_{\mu\nu}c^\nu}}{\int\prod_\mu dc^\mu\,e^{-\frac{1}{2}c^\mu K_{\mu\nu}c^\nu}}
=\frac{\sqrt{\det(K+\delta K)^{-1}}}{\sqrt{\det N}}\,,
\end{equation}
where in the last step we have evaluated the gaussian integrals (and cancelled common factors of $\pi$ \emph{etc} between numerator and denominator). This completes the proof of (\ref{eq:shift-rel}).

Now we wish to apply this general relation to the change of basis in (\ref{eq:double-prime-basis}). Under this change of basis we get
\begin{equation}
K'_{\mu\nu}=K''_{\mu\nu}+2\chi_{(\mu}^{\phantom\mu m}K''_{|m|\nu)}+\chi_\mu^{\phantom\mu m}\chi_\nu^{\phantom\nu n}K''_{nm}\equiv K''_{\mu\nu}+(\delta K)_{\mu\nu}\,.
\end{equation}
This means that
\begin{eqnarray}
\lefteqn{\Omega'=\sqrt{\det N'}\,\Pi'_0 e^{-\frac{1}{2}i'_{N'}}e^{K-K'_{0,2}}\sum_n C^{(n)}}
\nonumber\\
&=&\sqrt{\det(K''+\delta K)^{-1}}\,\Pi'_0 e^{-\frac{1}{2}i'_{(K''+\delta K)^{-1}}}e^{-(\delta K)_{0,2}}e^{K-K''_{0,2}}\sum_n C^{(n)}
\nonumber\\
&=&\sqrt{\det N''}\,\Pi'_0 e^{-\frac{1}{2}i'_{N''}}e^{K-K''_{0,2}}\sum_n C^{(n)},\label{eq:Omega-prime}
\end{eqnarray}
where we have made use of (\ref{eq:shift-rel}) in the second step. It remains to express $\Pi_0'$ and $i'_{N''}=(N'')^{\mu\nu}I_{D'_\nu}I_{D'_\mu}$ in terms of the double-prime basis. The result for $\Pi_0'$ we have already derived and using (\ref{eq:pi-double-prime}) we get
\begin{eqnarray}
\Omega'\simeq\det(\delta+\phi\chi)^{-1}\sqrt{\det N''}\,\Pi''_0e^{-\frac{1}{2}i'_{N''}}e^{K-K''_{0,2}}\sum_n C^{(n)}\,.
\end{eqnarray}
Finally
\begin{equation}
i'_{N''}=(N'')^{\mu\nu}I_{D''_\nu}I_{D''_\mu}
+2(N''\chi)^{\mu m}I_{D''_m}I_{D''_\mu}-(\chi^{\mr T}N''\chi)^{mn}I_{D''_n}I_{D''_m}\,.
\end{equation}
In any term containing $I_{D''_m}$ this operator can be moved to the very left. We can then write it as the sum of an $I_{\partial_m}$ piece and an $I_{D''_\mu}$ piece as we have shown in the discussion of the transformation of $\Pi_0$. The latter gives zero as it is acting on a bosonic form (in the double-prime basis) because of the projection operator $\Pi_0''$ and the former we drop since it always gives rise to a form of bosonic degree $<p+1$ that doesn't contribute to the action. This means that we can replace $i'_{N''}$ by $i''_{N''}=(i_N)''$ and we finally get
\begin{eqnarray}
\Omega'&\simeq&\det(\delta+\phi\chi)^{-1}\sqrt{\det N''}\,\Pi''_0e^{-\frac{1}{2}(i_N)''}e^{K-K''_{0,2}}\sum_n C^{(n)}
\nonumber\\
&=&\det(\delta+\phi\chi)^{-1}\Omega''\,.
\end{eqnarray}

\subsubsection{Step 3}
The last step is to take the new basis one-forms
\begin{eqnarray}
E'''^{m}&=&E''^n\psi_n^{\phantom n m}\nonumber\\
E'''^{\mu}&=&E''^\nu E_\nu^{\phantom\nu\mu}\,.
\end{eqnarray}
Then only $\sqrt{\det N''}$ transforms and we get
\begin{equation}
\sqrt{\det N''}=\det(E_\nu^{\phantom\nu\mu})^{-1}\sqrt{\det N'''}\,.
\end{equation}

In going from a coordinate basis to the triple-primed basis we have thus found that 
\begin{equation}
\Omega\simeq\det(\delta+\phi\chi)^{-1}\det(E_\nu^{\phantom\nu\mu})^{-1}\Omega'''
\end{equation}
where again $\simeq$ means equality up to terms of bosonic degree $<p+1$.

The Hodge-dual of (the $(p+1)$-form part of) both sides is a scalar, and, using (\ref{eq:vielbein}), this gives an extra factor of 
\begin{equation}
\det(E_m^{\phantom mn})=\det(\delta+\phi\chi)\det\psi
\end{equation} 
on the right. The WZ action is just the integral of $\mc L_{\mr{WZ}}\equiv*(\Omega_{p+1})$ over $(x,\eta)$-superspace and we have shown that $\mc L_{\mr{WZ}}$ transforms as
\begin{eqnarray}
\mc L_{\mr{WZ}}&=&
\det(E_m^{\phantom mn})\det(\delta+\phi\chi)^{-1}\det(E_\nu^{\phantom\nu\mu})^{-1}\mc L'''_{\mr{WZ}}
\nonumber\\
&=&\det\psi\det(E_\nu^{\phantom\nu\mu})^{-1}\mc L'''_{\mr{WZ}}
\nonumber\\
&=&\det(\psi+\phi\chi\psi-\phi\chi\psi)\det(E_\nu^{\phantom\nu\mu})^{-1}\mc L'''_{\mr{WZ}}
\nonumber\\
&=&\det(E_m^{\phantom mn}-E_m^{\phantom m\mu}(E_\nu^{\phantom\nu\mu})^{-1}E_\mu^{\phantom\mu n})\det(E_\nu^{\phantom\nu\mu})^{-1}\mc L'''_{\mr{WZ}}
\nonumber\\
&=&\mr{sdet}\,E\,\mc L'''_{\mr{WZ}}\,.
\end{eqnarray}
This is precisely right to guarantee that the Wess-Zumino action is invariant under diffeomorphisms of $\h M$ as the measure transforms with the inverse factor. The integral over the $\eta$s should however not be taken literally as we have argued that the $\eta$s have to be treated as operators, being essentially gamma matrices. This non(anti)commutative nature of these coordinates suggests replacing the integral over them with a trace when we interpret them as gamma matrices. 

\subsection{Gauge invariance}\label{sec:gauge-inv}
We now want to show that the action is also invariant under gauge transformations. We first observe that since the gauge field on the brane, $\mc A$, and the background Kalb-Ramond field, $B$, only appear in the combination $K$ the action is manifestly invariant under gauge transformations of these fields. Therefore we only have to show that the Wess-Zumino part of the action is invariant under gauge transformations of the Ramond-Ramond fields, $C^{(n)}$. The gauge transformation of these take the form\footnote{Pull-backs of background forms such as $B$ and $H$ are implicit here.}
\begin{equation}
\delta C^{(n)}=d\Lambda^{(n-1)}-\Lambda^{(n-3)}H
\end{equation}
with $H\equiv dB$. Using the fact that
\begin{equation}
dK=d(d\mc A-B)=-H
\end{equation}
we find that
\begin{equation}
\delta(e^K\sum_n C^{(n)})=e^K\sum_n(d\Lambda^{(n-1)}-\Lambda^{(n-3)}H)=d(e^K\sum_n\Lambda^{(n)})\equiv d\Lambda\,.
\end{equation}
This means that the transformation of the Wess-Zumino term is
\begin{equation}
\delta S_{\mr{WZ}}=\int\sqrt{\det N}\,\Pi_0 e^{-\frac{1}{2}i_N}e^{-K_{0,2}}d\Lambda\,.
\end{equation}
The trick is to now observe that if we take $(\delta K)_{\mu\nu}=\epsilon^M\partial_M K_{\mu\nu}$, with $\epsilon$ infinitesimal, in the relation (\ref{eq:shift-rel}), it implies that
\begin{equation}
\partial_M\left(\sqrt{\det N}\,\Pi_0 e^{-\frac{1}{2}i_N}e^{-K_{0,2}}\omega\right)
=\sqrt{\det N}\,\Pi_0 e^{-\frac{1}{2}i_N}e^{-K_{0,2}}\partial_M\omega\,,
\end{equation}
and since we can go to a coordinate basis, so that there is no torsion, we get also
\begin{equation}
\sqrt{\det N}\,\Pi_0 e^{-\frac{1}{2}i_N}e^{-K_{0,2}}d\omega
=d_0\left(\sqrt{\det N}\,\Pi_0 e^{-\frac{1}{2}i_N}e^{-K_{0,2}}\omega\right)\,,
\end{equation}
where $d_0$ is the bosonic part of $d$. So we see that the gauge transformation of the RR fields change the Lagrangian by a total derivative, leaving the action invariant.

\subsection{Relation to Myers' action}
We will now show that Myers' action that we described in section \ref{sec:Myers} is essentially a gauge-fixed version of the boundary fermion inspired action. We shall fix the gauge in the following way: First we choose the standard gauge for $\mc A$, where we take
\begin{equation}
\mc A_{\mu}=\frac{i}{4}\eta^{\nu}\delta_{\nu\mu}\,.
\end{equation}
The remaining components of $\mc A$ will be denoted by $A_m$. For the coordinates on the D$p$-brane we will choose the static gauge given by
\begin{equation}
x^{\u m}=(x^m,x^{m'}(x,\eta))\qquad m=0,\ldots,p\quad,\quad m'=p+1,\ldots,9\,.
\end{equation}
Finally we will write the components of the fields referred to the Yang-Mills covariant basis of tangent space
\begin{eqnarray}
D_m&=&\partial_m-2i\partial_\mu A_m\partial^\mu\nonumber\\
D_\mu&=&\partial_\mu\,.\label{eq:tilded-basis}
\end{eqnarray}
We will denote the components of a field in this basis by a tilde. We start by looking at the Dirac-Born-Infeld part of the action.

\subsubsection{The DBI term}
This part of the action is the integral of
\begin{equation}
\mc{L}_{\mr{DBI}}=-e^{-\phi}\sqrt{-\mr{sdet}\,L}\,,
\end{equation}
where we have defined
\begin{equation}
L\equiv g+K=d\mc A+E 
\end{equation}
with $E\equiv g-B$. We now determine the components of $L$ with the gauge-fixing given above. We get
\begin{equation}
\label{eq:Lmunu}
L_{\mu\nu}=\frac{i}{2}\delta_{\mu\nu}+E_{\mu\nu}=\frac{i}{2}\delta_{\mu\nu}
+\partial_{\mu}x^{m'}\partial_{\nu}x^{n'}E_{m'n'}\,.
\end{equation}
The superdeterminant, defined in (\ref{eq:sdet}), contains the inverse of the determinant of this, which becomes
\begin{eqnarray}
\det(L_{\mu\nu})^{-1}&=&e^{-\tr\log L_{\mu\nu}}=(2i)^{2q}e^{\tr\log (\delta^{m'}_{n'}-2i\partial_{\mu}x^{m'}\partial^{\mu}x^{p'}E_{p'n'})}
\nonumber\\
&=&(2i)^{2q}\det\left(\delta^{m'}_{n'}+(ME)^{m'}_{\phantom{m'}n'}\right)\,,
\end{eqnarray}
with
\begin{equation}
\label{eq:def-M}
M^{m'n'}\equiv-2i\partial_{\mu}x^{m'}\partial^{\mu}x^{n'}\,.
\end{equation}

For the mixed components of $L$ we get using (\ref{eq:Lmunu})
\begin{eqnarray}
L_{m\mu}&=&(d\mc A)_{m\mu}+E_{m\mu}=-\partial_\mu A_m+\partial_\mu x^{n'}E_{mn'}
\nonumber\\
&=&-2iL_{\mu\nu}\partial^\nu A_m+\partial_\mu x^{n'}\tilde E_{mn'}\,,
\end{eqnarray}
so that the combination entering the definition of the superdeterminant becomes
\begin{eqnarray}
\lefteqn{L_{mn}-L_{m\mu}(L_{\mu\nu})^{-1}L_{\nu n}=(dA)_{mn}+E_{mn}-4L_{\mu\nu}\partial^\nu A_m\partial^\mu A_n}
\nonumber\\
&&{}+4i\partial_\mu A_{[m}\partial^\mu x^{n'}\tilde E_{|n'|n]}
-\tilde E_{mn'}\partial_\mu x^{n'}(L_{\mu\nu})^{-1}\partial_\nu x^{p'}\tilde E_{p'n}\,.
\end{eqnarray}
The first four terms combine to give just
\begin{equation}
\tilde E_{mn}+F_{mn}\,,
\end{equation}
whereas, when we expand $(L_{\mu\nu})^{-1}$, the last term becomes
\begin{eqnarray}
\lefteqn{2i\tilde E_{mn'}
\left(\partial_\mu x^{n'}\partial^\mu x^{p'}+2i\partial_\mu x^{n'}\partial^{\mu}x^{m'}E_{m'q'}\partial^\nu x^{q'}\partial_\nu x^{p'}-\ldots\right)\tilde E_{p'n}}
\nonumber\\
&=&\tilde E_{mn'}\left(\left(\delta_{n'}^{p'}+(ME)^{p'}_{\phantom{p'}n'}\right)^{-1}-\delta_{p'}^{n'}\right)(E_{p'q'})^{-1}\tilde E_{q'n}\,.
\end{eqnarray}

Putting these results together we get that in this gauge
\begin{eqnarray}
\lefteqn{\sqrt{-\mr{sdet}\,L}=\sqrt{-\det\left(L_{mn}-L_{m\mu}(L_{\mu\nu})^{-1}L_{\nu n}\right)\det(L_{\mu\nu})^{-1}}}
\nonumber\\
&=&(2i)^q\sqrt{-\det\left(\tilde E_{mn}+F_{mn}
+\tilde E_{mp'}\left(\left(Q^{-1}-\delta\right)E^{-1}\right)^{p'q'}\tilde E_{q'n}\right)\det Q}
\nonumber\\
\end{eqnarray}
with
\begin{equation}
Q^{n'}_{\phantom{n'}p'}\equiv\delta_{p'}^{n'}+(ME)^{n'}_{\phantom{n'}p'}\,.
\end{equation}

\subsubsection{The WZ term}
We start by expressing everything in the tilded basis given in (\ref{eq:tilded-basis}). We then have for the integrand of the WZ term
\begin{equation}
\Omega=\sqrt{\det\tilde N}\,\tilde\Pi_0 e^{-\frac{1}{2}i_{\tilde N}}e^{\tilde K-\tilde K_{0,2}}\sum_n\tilde C^{(n)}\,.
\end{equation}
Using the fact that $\tilde K_{\mu\nu}=\frac{i}{2}\delta_{\mu\nu}-\tilde B_{\mu\nu}$ together with the relation (\ref{eq:shift-rel}) with $(\delta K)_{\mu\nu}=-\tilde B_{\mu\nu}$ we get
\begin{eqnarray}
\lefteqn{\sqrt{\det(\frac{i}{2}\delta-\tilde B)^{-1}}\,\tilde\Pi_0 e^{-\frac{1}{2}i_{(\frac{i}{2}\delta-\tilde B)^{-1}}}e^{\tilde K_{2,0}+\tilde K_{1,1}}\sum_n\tilde C^{(n)}}
\nonumber\\
&=&\sqrt{\det(-2i\delta)}\,\tilde\Pi_0 e^{-\frac{1}{2}i_{-2i\delta}}e^{F-\tilde B_{2,0}-\tilde B_{1,1}-\tilde B_{0,2}}\sum_n\tilde C^{(n)}
\nonumber\\
&=&(2i)^qe^F\tilde\Pi_0 e^{\frac{1}{2}i_M}e^{-\tilde B}\sum_n\tilde C^{(n)}\,.
\end{eqnarray}

\subsubsection{Quantizing $\eta$}
The final step to get to Myers' action is to pass from the formulation in terms of the $2q$ fermions, $\eta^\mu$, to a description in terms of matrices. We have seen that when we quantize the fermions they become gamma matrices and all fields depending on the fermions therefore also become matrices. Canonical quantization suggests that we replace Poisson-brackets by $-i$ times the commutator,
\begin{equation}
-2i\delta^{\mu\nu}\partial_\mu X\partial_\nu Y\rightarrow -i[X,Y]\,.
\end{equation}
This means that $M^{m'n'}$ defined in (\ref{eq:def-M}) becomes $-i[x^{m'},x^{n'}]$.

It is also natural that the integral over the phase-space of the fermions should be replaced by a trace when they are quantized. Because we have effectively been treating the fermions as classical variables our fields, which are expanded in even powers of the fermions, have been treated as commuting. This means that in this approximation everything is naturally symmetrized and we therefore take
\begin{equation}
i^q\int d^{2q}\eta\,\cdots\rightarrow \mr{SymTr}(\cdots)\,.
\end{equation}
Using this prescription for quantizing the $\eta$-variables we see that we get precisely Myers' action (times a factor of $2^q$, the number of coincident D-branes) from the gauge-fixed version of the boundary fermion inspired action
\begin{eqnarray}
S&=&-T_p\int d^{2q}\eta\,d^{p+1}x\,e^{-\phi}\sqrt{-\mr{sdet}\,(g+K)}
\nonumber\\
&&{}+T_p\int d^{2q}\eta\,\int_M\left(\sqrt{\det N}e^{-\frac{1}{2}i_N}e^{K-K_{0,2}}\sum_nC^{(n)}\right)_{p+1,0}\,.
\nonumber\\
\end{eqnarray}

\section{Discussion}
We have shown in this chapter that there is an underlying covariant description from which Myers' action can be obtained. This action is formulated on an auxiliary superspace with the extra fermionic coordinates corresponding to boundary fermions in the string picture. This means that the exact relation between diffeomorphism invariance on this space and a possible matrix version of diffeomorphism invariance in an ordinary formulation is somewhat obscure. To try to sort this out would be an interesting program for the future.

In the next chapter we shall go on to show that the supersymmetric version of the action we have presented here in fact takes the same form but is now formulated in a background superspace. This is analogous to the abelian case.

%% file: supersymmetric.tex
\chapter{Proof of kappa-symmetry}
The supersymmetric action for a D$p$-brane was found in 1996 by three different groups \cite{Aganagic:1996pe}, \cite{Cederwall:1996ri} and \cite{Bergshoeff:1996tu}. The actions are written in the so-called Green-Schwarz formulation where the brane is taken to be a $(p+1)$-dimensional bosonic space embedded in $\mc N=2$, $D=10$ superspace. Supersymmetry is guaranteed by the presence of kappa-symmetry which, as we learned in our discussion of the superembedding formalism, can be thought of as the remnant of worldvolume supersymmetry after solving the superembedding constraint. In this chapter we will prove that the action formulated in the previous chapter motivated by our considerations of boundary fermions is supersymmetric, {\em i.e.} possesses kappa-symmetry when formulated on a background superspace. We will restrict ourselves to proving it for a type IIB supergravity background for definiteness, the type IIA case follows exactly the same lines. We will also make a (well motivated) simplifying assumption for the components of the induced metric and modified field strength $K$ of the brane as described below.

\section{The setup}
The worldvolume of the brane will be the space $\h M$ with coordinates $y^{\h m}=(x^m,\eta^{\h\mu})$ where $m=0,\ldots,p$ and $\h\mu=1,\ldots,2q$. The background space will be taken to be ten-dimensional type IIB superspace with coordinates $z^{\u M}=(x^{\u m},\theta^{\u\mu})$ where $\u m=0,\ldots,9$ and the 32-component spinor index $\u\mu$ splits up into two 16-component Majorana-Weyl spinors, $\u\mu=\mu i$ with $i=1,2$.

The embedding of the brane worldvolume into target superspace is given by the generalized superembedding matrix
\begin{equation}
E_{\h a}^{\phantom a\u B}\equiv E_{\h a}^{\phantom a\h m}\partial_{\h m}Z^{\u M}E_{\u M}^{\phantom M\u A}\,.
\end{equation}
In order to simplify the calculations we will make the natural assumption that a basis can be chosen on the brane such that only the transverse coordinates depend on the boundary fermions. This means that in particular we have
\begin{equation}
E_{\h\alpha}^{\phantom\alpha\u b}=h_{\h\alpha}^{\phantom\alpha c'}u_{c'}^{\phantom{c'}\u b}
\end{equation}
and
\begin{equation}
E_a^{\phantom a\u b}=h_a^{\phantom a c}u_c^{\phantom c\u b}\,,
\end{equation}
where the transverse directions are denoted by a prime and $u$ is an element of the target space Lorentz group $SO(9,1)$. For the induced metric on the worldvolume,
\begin{equation}
g_{\h a\h b}\equiv E_{\h a}^{\phantom a\u c}E_{\h b}^{\phantom b\u d}\eta_{\u{cd}}\,,
\end{equation}
we get
\begin{equation}
g_{a\h\beta}=0\,,
\end{equation}
so that the mixed components of $g$ vanish. We shall also assume that the same is true for
\begin{equation}
K\equiv d\mc A-B
\end{equation}
in this basis, so that both $g$ and $K$ take a block-diagonal form. Both of
these conditions in fact hold in the covariant formulation described in section
5 of \cite{Howe:2005jz}.

We will denote the basis one-forms of the background and worldvolume by 
\begin{equation}
E^{\u A}=dz^{\u M}E_{\u M}^{\phantom M\u A}\qquad\mbox{and}\qquad e^{\h a}=dy^{\h m}E_{\h m}^{\phantom m\h a}
\end{equation}
respectively. Bosonic worldvolume indices are raised (lowered) with $g^{ab}$ ($g_{ab}$) and fermionic ones with $N^{\h\alpha\h\beta}$ ($K_{\h\alpha\h\beta}$).

We now turn to the constraints on the background fields that are present in the type IIB supergravity background.

\subsection{The type IIB superspace constraints}
We will follow \cite{Bergshoeff:1996tu} and take the type IIB superspace constraints to be\footnote{Note that the constraint on $H$ differs from that used before. This simply means that the $\theta^{1,2}$ used here are linear combinations of the ones used previously.}
\begin{eqnarray}
T_{\u{\alpha i\beta j}}^{\u a}&=&-i(\gamma^{\u a})_{\alpha\beta}\delta_{ij}\label{eq:Tconstraint}\\
H_{\u{\alpha i\beta j c}}&=&i(\gamma_{\u c})_{\alpha\beta}\sigma^3_{ij}\\
(F^{(n+2)})_{\u{a_1\cdots a_n\alpha1\beta2}}&=&ie^{-\phi}(\gamma_{\u{a_1\cdots a_n}})_{\alpha\beta}\qquad\mbox{($n$ odd).}\label{eq:Fnconstraint}
\end{eqnarray}
$T$ is the torsion, $H\equiv dB$ the NS-NS three-form field strength, 
\begin{equation}
F^{(n)}\equiv dC^{(n-1)}-C^{(n-3)}H
\end{equation}
the $n$-form Ramond-Ramond field strengths and $\phi$ is the dilaton. The other components involving spinorial indices vanish in a bosonic background, which is what we will be considering.

\subsection{Kappa-symmetry}
Defining $\delta E^{\u A}\equiv\delta z^{\u M}E_{\u M}^{\phantom M\u A}$ the characteristic property of kappa-symmetry is that
\begin{equation}
\delta_\kappa E^{\u a}=0\,.
\end{equation}
On general grounds the kappa-transformation of the fermionic background coordinates must take the form
\begin{equation}
\delta_\kappa E^{\u\alpha}=\left(\kappa(1-\Gamma)\right)^{\u\alpha}\,,
\end{equation}
where $\frac{1}{2}(1-\Gamma)$ is a projection matrix. This fact can easily be derived in the superembedding formalism (see \cite{Sorokin:1999jx}) and $\Gamma$ can be related to components of the superembedding matrix.

We will also need the transformation of various form fields. A useful result is that under a transformation of the coordinates $\delta z^{\u M}$ a background $n$-form transforms as
\begin{equation}
\delta A^{(n)}=d(i_{\delta z} A^{(n)})+i_{\delta z} (dA)^{(n+1)}\,,
\label{eq:deltaA}
\end{equation}
where $i$ denotes the inner product with a vector. We now turn to the computation of the kappa-variation of the action. We will start with the Wess-Zumino term.

\section{Kappa-variation of the WZ term}
The Wess-Zumino part of the action was written down in (\ref{eq:WZ-Mhat}) and we repeat it here for convenience\footnote{We set $T_p=1$ in this chapter.}
\begin{equation}
S_{\mr{WZ}}=\int d^{2q}\eta\,\int_M\,\sqrt{\det N}\Pi_0e^{-\frac{1}{2}i_N}e^{K-K_{0,2}}\sum_n C^{(n)}\,,\label{eq:WZ-action7}
\end{equation}
where it is understood that the (bosonic) $(p+1)$-form piece is picked. In the same way as when proving the gauge-invariance of the bosonic action in the last chapter we can show that the kappa-variation simply ''passes through'' the first part, $\sqrt{\det N}\Pi_0e^{-\frac{1}{2}i_N}e^{-K_{0,2}}$, and we get
\begin{equation}
\delta_\kappa S_{\mr{WZ}}=\int d^{2q}\eta\int_M\,\sqrt{\det N}\Pi_0e^{-\frac{1}{2}i_N}e^{-K_{0,2}}\delta_\kappa\Big(e^K\sum_n C^{(n)}\Big)\,.
\end{equation}
Using (\ref{eq:deltaA}) and the fact that $d\big(e^K\sum_n C^{(n)}\big)=e^K\sum_n F^{(n+1)}$ we get
\begin{eqnarray}
\lefteqn{\delta_\kappa\Big(e^K\sum_n C^{(n)}\Big)=i_{\delta_\kappa z}d\Big(e^K\sum_n C^{(n)}\Big)+d(\cdots)}
\nonumber\\
&=&e^K\sum_n\frac{1}{(n-1)!}E^{\u A_n}\cdots E^{\u A_2}\delta_\kappa E^{\u A_1} F^{(n)}_{\u{A_1 A_2\cdots A_n}}+d(\cdots)\,,
\label{eq:variation-with-F}
\end{eqnarray}
where the total derivative gives no contribution to the action (see section \ref{sec:gauge-inv}). Using the superspace constraint (\ref{eq:Fnconstraint}) the term that contributes to the variation of the action becomes
\begin{equation}
-ie^{-\phi}e^K\sum_{n\,\,\mr{odd}}^9 \frac{1}{n!}\left(\delta_\kappa E^1e^{\h a_n}\cdots e^{\h a_1}\gamma_{\h a_1\cdots\h a_n}E^2
-E^1e^{\h a_n}\cdots e^{\h a_1}\gamma_{\h a_1\cdots\h a_n}\delta_\kappa E^2\right)\,.
\label{eq:variation1}
\end{equation}
These two terms give two contributions to the variation of the action that we will call $(\delta_\kappa S_{\mr{WZ}})_1$ and $(\delta_\kappa S_{\mr{WZ}})_2$ respectively. In the above equation we have defined the pull-backs of the $D=10$ gamma matrices
\begin{equation}
\gamma_{\h a}\equiv E_{\h a}^{\phantom a \u b}\gamma_{\u b}\,,
\end{equation}
which satisfy the relations
\begin{equation}
\gamma_a\gamma_b+\gamma_b\gamma_a=2g_{ab}\,,\quad\gamma_a\gamma_{\h\beta}+\gamma_{\h\beta}\gamma_a=0
\,,\quad\gamma_{\h\alpha}\gamma_{\h\beta}-\gamma_{\h\beta}\gamma_{\h\alpha}=2g_{\h\alpha\h\beta}\,.
\end{equation}
The second of these follows from our assumption that $g$ have no mixed components. We will focus on the term proportional to $\delta_\kappa E^1$ in (\ref{eq:variation1}). We can extend the sum on $n$ to infinity since the terms with $n>9$ vanish by anti-symmetry. This term then becomes
\begin{eqnarray}
\lefteqn{ie^{-\phi}e^K\sum_{n\,\,\mr{odd}}^\infty\frac{1}{n!}\delta_\kappa E^1e^{\h a_{n+1}}e^{\h a_n}\cdots e^{\h a_1}\gamma_{\h a_1\cdots\h a_n}E_{\h a_{n+1}}^{\phantom{a_{n+1}}2}}
\nonumber\\
&=&
ie^{-\phi}e^K\sum_{l=1}^\infty\frac{1}{(2l-1)!}\sum_{k=0}^le^{\h\alpha_{2l}}\cdots e^{\h\alpha_{2k+1}}e^{a_{2k}}\cdots e^{a_1}
\nonumber\\
&&\times\bigg(
\binom{2l-1}{2l-2k}\delta_\kappa E^1\gamma_{a_1\cdots a_{2k-1}}\gamma_{\h\alpha_{2k+1}\cdots\h\alpha_{2l}}E_{a_{2k}}^{\phantom{a_{2k}}2}
\nonumber\\
&&\qquad{}+\binom{2l-1}{2k}\delta_\kappa E^1\gamma_{a_1\cdots a_{2k}}\gamma_{\h\alpha_{2k+1}\cdots\h\alpha_{2l-1}}E_{\h\alpha_{2l}}^{\phantom{\alpha_{2l}}2}
\bigg)\,,\label{eq:deltaS_WZ1}
\end{eqnarray}
where we have used the fact that the mixed components of $g$ vanish by assumption and ignored terms with an odd number of $e^{\h\alpha}$ as they don't contribute to the action because of our assuption that the mixed components of $K$ vanish. We will consider the two terms in parenthesis separately. The first term in the above equation can be written
\begin{eqnarray}
\lefteqn{\frac{i}{2}e^{-\phi}e^K\sum_{l=1}^\infty\sum_{k=0}^le^{\h\alpha_{2l}}\cdots e^{\h\alpha_{2k+1}}e^{a_{2k}}\cdots e^{a_1}}
\nonumber\\
&&\times\frac{1}{(2l-2k)!(2k)!}\delta_\kappa E^1[\gamma_{a_1\cdots a_{2k}}\gamma_{\h\alpha_{2k+1}\cdots\h\alpha_{2l}},\gamma^b]E_b^{\phantom b2}\,.
\nonumber 
\end{eqnarray}
Changing summation variable to $l'=l-k$ in the sum on $l$ this becomes
\begin{eqnarray}
\lefteqn{\frac{i}{2}e^{-\phi}e^K\sum_{l'=0}^\infty\sum_{k=0}^\infty\frac{1}{(2l')!} e^{\h\alpha_{2l'}}\cdots e^{\h\alpha_1}\frac{1}{(2k)!}e^{a_{2k}}\cdots e^{a_1}}
\nonumber\\
&&\times\delta_\kappa E^1[\gamma_{a_1\cdots a_{2k}}\gamma_{\h\alpha_1\cdots\h\alpha_{2l'}},\gamma^b]E_b^{\phantom b2}\,.
\end{eqnarray}
This term therefore gives the following contribution to the kappa-variation of the action,
\begin{eqnarray}
(\delta_\kappa S_{\mr{WZ}})_{11}&=&
\frac{i}{2}\int d^{2q}\eta\,\int_M\,\sqrt{\det\left(K_{\h\alpha\h\beta}+g_{\h\alpha\h\beta}\right)^{-1}}e^{-\phi}e^{K_{2,0}}
\nonumber\\
&&\times\sum_{k=0}^\infty\frac{1}{(2k)!}e^{a_{2k}}\cdots e^{a_1}\delta_\kappa E^1[\gamma_{a_1\cdots a_{2k}}\h h_\perp,\gamma^b]E_b^{\phantom b2}\,,
\nonumber\\
\label{eq:deltaS_WZ11}
\end{eqnarray}
where we have defined
\begin{eqnarray}
\h h_\perp&\equiv&\sqrt{\det\left(\delta_{\h\alpha}^{\h\beta}+g_{\h\alpha}^{\phantom\alpha\h\beta}\right)}\,\Pi_0e^{-\frac{1}{2}i_N}
\sum_{l=0}^\infty\frac{1}{(2l)!} e^{\h\alpha_{2l}}\cdots e^{\h\alpha_1}\gamma_{\h\alpha_1\cdots\h\alpha_{2l}}
\nonumber\\
&=&\sqrt{\det\left(\delta_{\h\alpha}^{\h\beta}+g_{\h\alpha}^{\phantom\alpha\h\beta}\right)}
\sum_{l=0}^\infty\frac{(-1)^l}{2^ll!}N^{\h\alpha_1\h\beta_1}\cdots N^{\h\alpha_l\h\beta_l}\gamma_{\h\alpha_1\h\beta_1\cdots\h\alpha_l\h\beta_l}\,.
\nonumber\\
\label{eq:def-h-perp}
\end{eqnarray}

The integral over $M$, the bosonic part of $\h M$, in (\ref{eq:deltaS_WZ11}) picks out the (bosonic) $(p+1)$-form piece. We have
\begin{eqnarray}
\lefteqn{\left(e^{K_{2,0}}\sum_{k=0}^\infty\frac{1}{(2k)!} e^{a_{2k}}\cdots e^{a_1}\gamma_{a_1\cdots a_{2k}}\right)_{p+1}}
\nonumber\\
&=&
\sum_{k=0}^{(p+1)/2}\frac{\left(K_{2,0}\right)^k}{k!(p+1-2k)!}e^{a_{p+1}}\cdots e^{a_{2k+1}}\gamma_{a_{2k+1}\cdots a_{p+1}}
\nonumber\\
&=&
\sum_{k=0}^{(p+1)/2}\frac{(-1)^k K_{a_1a_2}\cdots K_{a_{2k-1}a_{2k}}}{2^k k!(p+1-2k)!}e^{a_{p+1}}\cdots e^{a_1}\gamma_{a_{2k+1}\cdots a_{p+1}}\,.
\nonumber\\
\end{eqnarray}
Using the relations
\begin{equation}
e^{a_{p+1}}\cdots e^{a_1}=d^{p+1}x\,\varepsilon^{a_{p+1}\cdots a_1}
\end{equation}
and
\begin{equation}
\frac{1}{(2k)!}\varepsilon^{a_{p+1}\cdots a_1}\gamma_{a_1\cdots a_{2k}}=\sqrt{-\det g_{ab}}\gamma^{(p+1)}\gamma^{a_{p+1}\cdots a_{2k+1}}\,,
\end{equation}
where we have defined
\begin{equation}
\gamma^{(p+1)}\equiv\frac{1}{\sqrt{-\det g_{ab}}}\frac{1}{(p+1)!}\varepsilon^{a_{p+1}\cdots a_1}\gamma_{a_1\cdots a_{p+1}}\,,
\end{equation}
which is easily seen to obey the relations
\begin{equation}
\label{eq:gammma-ids}
\gamma^{(p+1)}\gamma^{\h a}=-(-1)^{\h a}\gamma^{\h a}\gamma^{(p+1)}\quad\mbox{ and }\quad(\gamma^{(p+1)})^2=(-1)^{(p+1)/2}\,,
\end{equation}
we get
\begin{eqnarray}
\lefteqn{\left(e^{K_{2,0}}\sum_{k=0}^\infty\frac{1}{(2k)!} e^{a_{2k}}\cdots e^{a_1}\gamma_{a_1\cdots a_{2k}}\right)_{p+1}}
\nonumber\\
&=&
d^{p+1}x\sqrt{-\det g_{ab}}\sum_{k=0}^{(p+1)/2}\frac{(-1)^k}{2^k k!}
K_{a_kb_k}\cdots K_{a_1b_1}\gamma^{a_kb_k\cdots a_1b_1}\gamma^{(p+1)}
\nonumber\\
&=&d^{p+1}x\sqrt{-\det(g_{ab}+K_{ab})}\,\h h_\parallel\gamma^{(p+1)}\,,
\label{eq:p+1-form-part}
\end{eqnarray}
where we have defined
\begin{equation}
\label{eq:def-h-parallel}
\h h_\parallel\equiv\frac{1}{\sqrt{\det\left(\delta_a^b+K_a^{\phantom ab}\right)}}\sum_{k=0}^{(p+1)/2}\frac{(-1)^k}{2^kk!}
K_{a_1b_1}\cdots K_{a_kb_k}\gamma^{a_1b_1\cdots a_kb_k}\,.
\end{equation}

Using (\ref{eq:p+1-form-part}) in (\ref{eq:deltaS_WZ11}) we get
\begin{eqnarray}
(\delta_\kappa S_{\mr{WZ}})_{11}&=&
-\frac{i}{2}\int d^{2q}\eta\,d^{p+1}x\,\mc L_{\mr{DBI}}\delta_\kappa E^1[\h h_\parallel\gamma^{(p+1)}\h h_\perp,\gamma^b]E_b^{\phantom b2}
\nonumber\\
&=&
-\frac{i}{2}\int d^{2q}\eta\,d^{p+1}x\,\mc L_{\mr{DBI}}\delta_\kappa E^1\gamma^{(p+1)}\{\h h,\gamma^b\}E_b^{\phantom b2}\,,
\label{eq:deltaS_WZ11-final}
\end{eqnarray}
where we have put $\h h\equiv\h h_\parallel\h h_\perp$ and
\begin{eqnarray}
\mc L_{\mr{DBI}}&\equiv&-e^{-\phi}\sqrt{-\mr{sdet}(g+K)}
\nonumber\\
&=&-e^{-\phi}\sqrt{-\det(g_{ab}+K_{ab})}\sqrt{\det\left(K_{\h\alpha\h\beta}+g_{\h\alpha\h\beta}\right)^{-1}}\,.
\label{eq:def-L_DBI}
\end{eqnarray}

Now we return to the second term in (\ref{eq:deltaS_WZ1}). Changing summation variable as for the first term and using (\ref{eq:p+1-form-part}) this term gives a contribution
\begin{eqnarray}
(\delta_\kappa S_{\mr{WZ}})_{12}&=&-i\int d^{2q}\eta\,d^{p+1}x\,\mc L_{\mr{DBI}}
\sqrt{\det\left(\delta_{\h\alpha}^{\h\beta}+g_{\h\alpha}^{\phantom\alpha\h\beta}\right)}
\Pi_0e^{-\frac{1}{2}i_N}
\nonumber\\
&&\times\sum_{l=1}^\infty e^{\h\alpha_{2l}}\cdots e^{\h\alpha_1}\frac{1}{(2l-1)!}\delta_\kappa E^1
\h h_\parallel\gamma^{(p+1)}\gamma_{\h\alpha_1\cdots\h\alpha_{2l-1}}E_{\h\alpha_{2l}}^{\phantom{\alpha_{2l}}2}
\nonumber
\end{eqnarray}
\begin{eqnarray}
&=&-i\int d^{2q}\eta\,d^{p+1}x\,\mc L_{\mr{DBI}}
\sqrt{\det\left(\delta_{\h\alpha}^{\h\beta}+g_{\h\alpha}^{\phantom\alpha\h\beta}\right)}\sum_{l=1}^\infty 2l\frac{(-1)^l}{2^ll!}
\nonumber\\
&&\times N^{\h\alpha_{2l}\h\alpha_{2l-1}}\cdots N^{\h\alpha_2\h\alpha_1}
\delta_\kappa E^1\h h_\parallel\gamma^{(p+1)}\gamma_{\h\alpha_1\cdots\h\alpha_{2l-1}}E_{\h\alpha_{2l}}^{\phantom{\alpha_{2l}}2}
\nonumber
\end{eqnarray}
\begin{eqnarray}
&=&\frac{i}{2}\int d^{2q}\eta d^{p+1}x\,\mc L_{\mr{DBI}}
\sqrt{\det\left(\delta_{\h\alpha}^{\h\beta}+g_{\h\alpha}^{\phantom\alpha\h\beta}\right)}\sum_{l=1}^\infty\frac{(-1)^{l-1}}{2^{l-1}(l-1)!}
\nonumber\\
&&\times N^{\h\alpha_{2l-2}\h\alpha_{2l-3}}\cdots N^{\h\alpha_2\h\alpha_1}
\delta_\kappa E^1\{\h h_\parallel\gamma^{(p+1)}\gamma_{\h\alpha_1\cdots\h\alpha_{2l-2}},\gamma^{\h\beta}\}E_{\h\beta}^{\phantom\beta2}\,.
\end{eqnarray}
Shifting the summation variable to $l'=l-1$ and using the definition of $\h h_\perp$ in (\ref{eq:def-h-perp}) we get
\begin{eqnarray}
(\delta_\kappa S_{\mr{WZ}})_{12}&=&\frac{i}{2}\int d^{2q}\eta\,d^{p+1}x\,\mc L_{\mr{DBI}}
\delta_\kappa E^1\{\h h_\parallel\gamma^{(p+1)}\h h_\perp,\gamma^{\h\beta}\}E_{\h\beta}^{\phantom\beta2}
\nonumber\\
&=&\frac{i}{2}\int d^{2q}\eta\,d^{p+1}x\,\mc L_{\mr{DBI}}\delta_\kappa E^1\gamma^{(p+1)}\{\h h,\gamma^{\h\beta}\}E_{\h\beta}^{\phantom\beta2}\,.
\end{eqnarray}

Adding this contribution to the first one, (\ref{eq:deltaS_WZ11-final}), we get
\begin{eqnarray}
(\delta_\kappa S_{\mr{WZ}})_1=-\frac{i}{2}\int d^{2q}\eta\,d^{p+1}x\,\mc L_{\mr{DBI}}
\delta_\kappa E^1(-1)^{\h b}\gamma^{(p+1)}\{\h h,\gamma^{\h b}\}E_{\h b}^{\phantom\beta2}\,.
\end{eqnarray}

It is not hard to see that the term proportional to $\delta_\kappa E^2$ in (\ref{eq:variation1}) gives a contribution related to the one above by transposition and an extra factor of $(-1)^{\h b}$ and the full kappa-variation of the WZ term becomes
\begin{eqnarray}
\delta_\kappa S_{\mr{WZ}}&=&-\frac{i}{2}\int d^{2q}\eta\,d^{p+1}x\,\mc L_{\mr{DBI}}
\Big(\delta_\kappa E^1(-1)^{\h b}\gamma^{(p+1)}\{\h h,\gamma^{\h b}\}E_{\h b}^{\phantom b2}
\nonumber\\
&&\qquad\qquad{}+(-1)^{(p+1)/2}\delta_\kappa E^2\gamma^{(p+1)}\{\h h^{\mr T},\gamma^{\h b}\}E_{\h b}^{\phantom b1}\Big)\,.
\nonumber\\
\label{eq:deltaS_WZ-final}
\end{eqnarray}

\subsection{The significance of $\h h_\parallel$ and $\h h_\perp$}\label{sec:significanceofh}
The two expressions $\h h_\parallel$ and $\h h_\perp$ have a special physical meaning as Lorentz transformations in spinor representation. To show this let $A_{mn}$ be an anti-symmetric $10\times10$ matrix. We can use $A$ to parameterize a Lorentz transformation as follows. Define
\begin{equation}
\Lambda(A)\equiv\frac{\delta-A}{\delta+A}\,.
\end{equation}
It is easy to see that indeed $\Lambda(A)\in SO(9,1)$. This gives the Lorentz transformation in so-called Cayley parametrization. What is the corresponding transformation acting on a spinor? The answer is (see \cite{Callan:1988wz})
\begin{equation}
\label{eq:M(A)}
M(A)=\frac{1}{\sqrt{\det(\delta+A})}\mbox{\it\AE}(-\frac{1}{2}A_{mn}\gamma^{mn})\,,
\end{equation}
where the ''anti-symmetrized exponential'' $\mbox{\it\AE}$ is defined as
\begin{equation}
\mbox{\it\AE}(X_{mn}\gamma^{mn})\equiv\sum_{k=0}^5\frac{1}{n!}X_{m_1n_1}\cdots X_{m_kn_k}\gamma^{m_1n_1\cdots m_kn_k}\,.
\end{equation}
By definition $M(A)$ then obeys the relation
\begin{equation}
M(A)^{-1}\gamma_m M(A)=\Lambda(A)_m^{\phantom mn}\gamma_n\,.
\end{equation}
One can also show that $M(A)^{-1}=M(-A)=M(A)^{\mr T}$.

Looking at the definition of $\h h_\parallel$ in (\ref{eq:def-h-parallel}) we see that it is exactly of the form (\ref{eq:M(A)}) with $A=K$. This means that it satisfies
\begin{equation}
\h h_\parallel^{\mr T}\gamma_a\h h_\parallel=\Lambda_a^{\phantom ab}\gamma_b\,,
\label{eq:parallel-spinorrep}
\end{equation}
where
\begin{equation}
\label{eq:Lambda-ab}
\Lambda_a^{\phantom ab}\equiv\left(\frac{\delta-K}{\delta+K}\right)_a^{\phantom ab}\,.
\end{equation}
This is a Lorentz transformation of the worldvolume directions of the brane and $\h h_\parallel$ is its spinor representation.

What about $\h h_\perp$? Using the fact that
\begin{equation}
N^{\h\alpha\h\beta}\gamma_{\h\alpha\h\beta}
=N^{\h\alpha\h\beta}h_{\h\alpha}^{\phantom\alpha a'}h_{\h\beta}^{\phantom\beta b'}\gamma_{a'b'}\equiv M^{a'b'}\gamma_{a'b'}
\end{equation}
and
\begin{eqnarray}
\det\left(\delta_{\h\alpha}^{\h\beta}+g_{\h\alpha}^{\phantom\alpha\h\beta}\right)
&=&\det\left(\delta_{\h\alpha}^{\h\beta}+h_{\h\alpha}^{\phantom\alpha a'}h_{\h\gamma}^{\phantom\beta b'}N^{\h\gamma\h\beta}\eta_{a'b'}\right)
\nonumber\\
&=&\det\left(\delta_{a'}^{b'}+M_{a'}^{\phantom{a'}b'}\right)^{-1}\,,
\end{eqnarray}
which is easy to show using $\det(\cdot)=e^{\mr{tr}\log(\cdot)}$, we see from (\ref{eq:def-h-perp}) that $\h h_\perp$ is of the form $M(A)$ with $A^{a'b'}=M^{a'b'}$. Thus we see that $\h h_\perp$ is the spinor representation of the Lorentz transformation
\begin{equation}
\Lambda_{a'}^{\phantom{a'}b'}\equiv\left(\frac{\delta-M}{\delta+M}\right)_{a'}^{\phantom{a'}b'}
\end{equation}
of the coordinates transverse to the brane. In fact it is not hard to show that
\begin{equation}
\h h_\perp^{\mr T}\gamma_{\h\alpha}\h h_\perp=\Lambda_{\h\alpha}^{\phantom\alpha\h\beta}\gamma_{\h\beta}\,,
\end{equation}
where
\begin{equation}
\Lambda_{\h\alpha}^{\phantom\alpha\h\beta}\equiv\left(\frac{\delta-g}{\delta+g}\right)_{\h\alpha}^{\phantom\alpha\h\beta}\,.
\end{equation}

\section{Kappa-variation of the DBI term}
The Dirac-Born-Infeld part of the action is
\begin{equation}
S_{\mr{DBI}}=\int d^{2q}\eta\,d^{p+1}x\,\mc L_{\mr{DBI}}\,,
\end{equation}
where the Lagrangian is (in this case) given in (\ref{eq:def-L_DBI}). Using the fact that $\delta(\mr{sdet}\,M)=\mr{str}(M^{-1}\delta M)$ the kappa-variation becomes
\begin{eqnarray}
\delta_\kappa\mc L_{\mr{DBI}}&=&-\frac{1}{2}\mc L_{\mr{DBI}}\,\mr{str}\left((g+K)^{-1}(\delta_\kappa g+\delta_\kappa K)\right)
\nonumber\\
&=&-\frac{1}{2}\mc L_{\mr{DBI}}(-1)^{\h b}\left((g+K)^{-1}\right)^{\h b\h a}(\delta_\kappa g_{\h a\h b}+\delta_\kappa K_{\h a\h b})\,.
\label{eq:deltaL_DBI}
\end{eqnarray}
We now need to determine the kappa-transformations of $K$ and $g$. This can be done in analogy to the abelian case described in \cite{Bergshoeff:1996tu} or, perhaps better, by using the superembedding interpretation of kappa-symmetry as a worldvolume $\theta$-diffeomorphism. Either way gives
\begin{eqnarray}
\delta_\kappa K&=&-E^{\u C}E^{\u B}\delta_\kappa E^{\u A}H_{\u{ABC}}=(-1)^{\h a}ie^{\h a}\delta_\kappa E^i\gamma_{\h a}E^j\sigma^3_{ij}\,,
\end{eqnarray}
from which we can read off
\begin{eqnarray}
\delta_\kappa K_{\h a\h b}=-(-1)^{\h a+\h b}2i\delta_\kappa E^i\gamma_{[\h a}E_{\h b]}^{\phantom{b)}j}\sigma^3_{ij}\,.
\end{eqnarray}
Since $g_{\h a\h b}=E_{\h a}^{\phantom a\u a}E_{\h b}^{\phantom b\u b}\eta_{\u{ab}}$ we need the kappa-transformation of the (generalized) superembedding matrix. Again this can be determined by either route and we get
\begin{equation}
\delta_\kappa E_{\h a}^{\phantom a\u a}=-(-1)^{\h a\u B}\delta_\kappa E^{\u B}E_{\h a}^{\phantom a\u A}T_{\u{AB}}^{\u a}
=(-1)^{\h a}i\delta_\kappa E^i\gamma^{\u a}E_{\h a}^{\phantom aj}\delta_{ij}\,,
\end{equation}
which gives
\begin{equation}
\delta_\kappa g_{\h a\h b}=2\delta_\kappa E_{\h a}^{\phantom a\u a}E_{\h b}^{\phantom b\u b}\eta_{\u{ab}}
=(-1)^{\h a+\h b}2i\delta_\kappa E^i\gamma_{(\h a}E_{\h b)}^{\phantom{b)}j}\delta_{ij}\,.
\end{equation}

Using these transformations in (\ref{eq:deltaL_DBI}) we get
\begin{eqnarray}
\delta_\kappa\mc L_{\mr{DBI}}&=&-i\mc L_{\mr{DBI}}\delta_\kappa E^i
\Big((-1)^{\h b}((g+K)^{-1})^{\h b\h a}\gamma_{\h a}(P_-)_{ij}
\nonumber\\
&&\qquad\qquad{}+(-1)^{\h a}((g-K)^{-1})^{\h b\h a}\gamma_{\h a}(P_+)_{ij}\Big)E_{\h b}^{\phantom bj}\,,
\end{eqnarray}
where $P_\pm\equiv\frac{1}{2}(1\pm\sigma^3)$. This gives two pieces
\begin{eqnarray}
\delta_\kappa\mc L_{\mr{DBI}}&=&-i\mc L_{\mr{DBI}}\delta_\kappa E^1\left(
((\delta-K)^{-1})^b_{\phantom ba}\gamma^a
+((\delta-g)^{-1})^{\h\beta}_{\phantom\beta\h\alpha}\gamma^{\h\alpha}
\right)E_{\h\beta}^{\phantom\beta1}
\nonumber\\
&&{}-i\mc L_{\mr{DBI}}\delta_\kappa E^2\left(
((\delta+K)^{-1})^b_{\phantom ba}\gamma^a
-((\delta+g)^{-1})^{\h\beta}_{\phantom\beta\h\alpha}\gamma^{\h\alpha}
\right)E_b^{\phantom b2}\,.
\nonumber\\
\end{eqnarray}

From (\ref{eq:parallel-spinorrep}) and (\ref{eq:Lambda-ab}) we get the relation
\begin{equation}
((\delta+K)^{-1})^b_{\phantom ba}\gamma^a=\frac{1}{2}\left(\delta_a^b+\Lambda^b_{\phantom ba}\right)\gamma^a
=\frac{1}{2}(\gamma^b+\h h_\parallel^{\mr T}\gamma^b\h h_\parallel)
=\frac{1}{2}(\gamma^b+\h h^{\mr T}\gamma^b\h h)\,,
\end{equation}
where we have used the fact that $\h h_\perp$ commutes with $\gamma^a$. Similarly we get
\begin{equation}
((\delta-K)^{-1})^b_{\phantom ba}\gamma^a=\frac{1}{2}(\gamma^b+\h h\gamma^b\h h^{\mr T})\,.
\end{equation}
The corresponding relations with hatted indices are
\begin{equation}
((\delta+g)^{-1})^{\h\beta}_{\phantom\beta\h\alpha}\gamma^{\h\alpha}=\frac{1}{2}(\gamma^{\h\beta}+\h h^{\mr T}\gamma^{\h\beta}\h h)
\end{equation}
and
\begin{equation}
((\delta-g)^{-1})^{\h\beta}_{\phantom\beta\h\alpha}\gamma^{\h\alpha}=\frac{1}{2}(\gamma^{\h\beta}+\h h\gamma^{\h\beta}\h h^{\mr T})\,.
\end{equation}

Using these relations the kappa-variation of the DBI Lagrangian becomes
\begin{eqnarray}
\delta_\kappa\mc L_{\mr{DBI}}&=&-\frac{i}{2}\mc L_{\mr{DBI}}\delta_\kappa E^1\left((\gamma^{\h b}+\h h\gamma^{\h b}\h h^{\mr T})\right)E_{\h b}^{\phantom b1}
\nonumber\\
&&{}-\frac{i}{2}\mc L_{\mr{DBI}}\delta_\kappa E^2\left(
(-1)^{\h b}(\gamma^{\h b}+\h h^{\mr T}\gamma^{\h b}\h h)\right)E_{\h b}^{\phantom b2}\,.
\label{eq:deltaS_DBI-final}
\end{eqnarray}

\section{The kappa-symmetry projection operator}
The final step in the proof of kappa-symmetry is to introduce the $32\times32$ matrix
\begin{equation}
\Gamma\equiv
\left(\begin{array}{cc}
0 & \h h\\
(-1)^{(p+1)/2}\h h^{\mr T} & 0
\end{array}\right)\gamma^{(p+1)}\,.\label{eq:proj-op-def}
\end{equation}
It is easy to see that it is symmetric, traceless, and satisfies $\Gamma^2=1$. From this we can construct projection operators $\frac{1}{2}(1\pm\Gamma)$. In fact, consider the following product,
\begin{eqnarray}
\lefteqn{(1+\Gamma)\gamma^{\h a}(\sigma^3)^{\h a}(1-\Gamma)}
\nonumber\\
&=&\left(\begin{array}{cc}
\gamma^{\h a}+\h h\gamma^{\h a}\h h^{\mr T} & (-1)^{\h a}\gamma^{(p+1)}\{\gamma^{\h a},\h h\}\\
(-1)^{(p+1)/2}\gamma^{(p+1)}\{\h h^{\mr T},\gamma^{\h a}\} & (-1)^{\h a}(\gamma^{\h a}+\h h^{\mr T}\gamma^{\h a}\h h)
\end{array}\right)\,.\label{eq:gamma-expression}
\nonumber\\
\end{eqnarray}
Comparing this to (\ref{eq:deltaS_WZ-final}) and (\ref{eq:deltaS_DBI-final}) we see that the variation of the total action, $S=S_{\mr{DBI}}+S_{\mr{WZ}}$, can be written
\begin{equation}
\delta_\kappa S=-\frac{i}{2}\int d^{2q}\eta\,d^{p+1}x\,\mc L_{\mr{DBI}}\delta_\kappa E^{\mr T}(1+\Gamma)\gamma^{\h a}(\sigma^3)^{\h a}(1-\Gamma)E_{\h a}\,,
\label{eq:total-kappa-variation}
\end{equation}
where we have suppressed the $i,j$-indices of the blocks so that for example $\delta_\kappa E=(\delta_\kappa E^1,\delta_\kappa E^2)$. Taking the kappa-transformation of the background fermions to be
\begin{equation}
\label{eq:delta-kappa-E}
\delta_\kappa E=\kappa(1-\Gamma)
\end{equation}
this gives zero and the kappa-symmetry is thus established.

\section{Discussion}
In this chapter we have shown that, under reasonable assumptions about the geometry, the action with boundary fermions is kappa-symmetric and therefore a candidate for a supersymmetric action for coincident D-branes (in the symmetrized trace approximation). In order to have a matrix interpretation we have seen that we should fix the standard gauge, $\mc A_{\h\mu}\sim\eta_{\h\mu}$, and quantize $\eta^{\h\mu}$ by replacing it by $\gamma^{\h\mu}$ so that functions of $\eta$ become matrices. We have also argued that Poisson-brackets should be replaced by $-i$ times commutators, as is standard practice in canonical quantization, and that the $\eta$-integral should be replaced by a symmetrized trace.

What kind of action this gives rise to and to what extent it possesses some kind of matrix-valued diffeomorphism invariance (see \emph{e.g.} \cite{DeBoer:2001uk}) is an interesting problem that deserves further study. It would also be very interesting to see how this proposal compares to other attempts in the literature to write down kappa-symmetric coincident D-brane actions, \emph{e.g.} \cite{Drummond:2002kg,Panda:2003dj}. The process of going from the picture with fermions to a matrix description needs to be understood better and also what happens when one goes beyond the approximation of treating the fermions classically. Perhaps new insights can be gained by using the pure spinor formulation of the superstring in which one can also address the issue of quantum corrections to this action in a covariant way. Furthermore it should be possible to give a clearer picture of the kappa-symmetry of the action along the lines of the unified description in \cite{Howe:1998ts}. These and other interesting questions we hope to be able to address in the near future.

We will end here with an intriguing observation on the non-abelian nature of the geometry suggested by our considerations in this chapter. In section \ref{sec:significanceofh} we showed that the expressions $\h h_\parallel$ and $\h h_\perp$ that naturally appeared have geometrical interpretations as (spinor representations of) Lorentz transformations in the directions parallel to the D-brane and transverse to it respectively. The Lorentz transformation of the worldvolume directions play a role also in the abelian case but the transformation in the transverse directions is a non-abelian phenomenon present only for coincident D-branes. In the ordinary (vector) representation this Lorentz transformation was
\begin{equation}
\Lambda_{a'}^{\phantom{a'}b'}\equiv\left(\frac{\delta-M}{\delta+M}\right)_{a'}^{\phantom{a'}b'}\qquad\mbox{with}\qquad M^{a'b'}\equiv N^{\h\alpha\h\beta}h_{\h\alpha}^{\phantom\alpha a'}h_{\h\beta}^{\phantom\beta b'}\,.
\end{equation}
Now $h_{\h\alpha}^{\phantom\alpha a'}$ is essentially a covariant $\h\alpha$-derivative of the transverse coordinate $z^{a'}$. Since $M$ is two of these contracted with $N$ it is essentially just the Poisson-bracket of two transverse directions. When we pass to the matrix description we therefore get roughly
\begin{equation}
M^{a'b'}\leadsto-i[z^{a'},z^{b'}]
\end{equation}
and
\begin{equation}
\Lambda_{a'}^{\phantom{a'}b'}\leadsto\mr{Sym}\left(\left(\delta_{a'}^{c'}+i[z_{a'},z^{c'}]\right)\left(\delta^{c'}_{b'}-i[z_{b'},z^{c'}]\right)^{-1}\right)\,,
\end{equation}
where the ordering is (naturally) symmetrized in the factors of $[z,z]$. This is a matrix valued Lorentz transformation, using the symmetrized product, in the sense that
\begin{equation}
\mr{Sym}\left(\Lambda_{a'}^{\phantom{a'}c'}\Lambda_{b'}^{\phantom{b'}d'}\eta_{c'd'}\right)=\eta_{a'b'}\,.
\end{equation}
This is no-longer true if we don't take the symmetrized product however, since we then pick up additional commutator terms.

This gives an interesting hint of non-commutative geometry and understanding what it means and how to make these ideas precise is a very interesting problem for the future.

%% file: acknowledgements.tex
\chapter*{Acknowledgments}
Here I would like to take the opportunity to thank the people who made this thesis possible. 

First of all I want to thank my supervisor Ulf Lindstr\"om for his support during these five years, for many interesting discussions and collaborations and for giving me some interesting problems to work on. Fawad Hassan also deserves thanks for always taking the time to explain a difficult subject until you find it easy and for many useful and stimulating discussions.

I am also grateful to my other collaborators on the papers leading up to this thesis. Paul Howe, who is always very nice and helpful and from whom I've learned a great deal of physics and Andreas Bredthauer and Jonas Persson who have provided a great atmosphere and interesting discussions during my visits to Uppsala.

I wish to thank the people in the theoretical physics group in Stockholm, especially Ioanna Pappa with whom I've shared an office for most of my time as a PhD student, for providing a nice and stimulating atmosphere for doing physics and other activities. It has been great to also have the group in Uppsala close by and be able to visit once a week and I am grateful to the PhD students and postdocs there, past and present, for making my visits so fruitful and enjoyable.

Most of this thesis was written in Paris at the Laboratoire de Physique Th\'eorique de l'Ecole Normale Sup\'erieure which has been a great place to spend my last few months of PhD studies. I wish to thank the people of the group there, and especially Costas Bachas, for their hospitality and for providing a very stimulating and friendly atmosphere to finish this work in.

Finally I want to thank my friends and my family for their support during all these years.

%% file: errata.tex
\renewcommand{\theequation}{E.\arabic{equation}}
\setcounter{equation}{0}

\chapter*{Errata}
\addcontentsline{toc}{chapter}{Errata}
\begin{itemize}
\item{\bf General:} Paper [III] has been accepted by JHEP.
\end{itemize}
\section*{Chapter 2}
\begin{itemize}
\item{\bf Page 14 footnote:} "...by lation ones." $\longrightarrow$ "...by latin ones.".
\end{itemize}
\section*{Chapter 3}
\begin{itemize}
\item{\bf Page 35 eq. (\ref{eq:T0-hamiltonian}):} $\eta_0^{\dagger1}\longrightarrow\eta_0^\dagger$ and $\eta_0^2\longrightarrow\eta_0$.
\end{itemize}
\section*{Chapter 5}
\begin{itemize}
\item{\bf Page 61 below eq. (\ref{eq:f_abeta}):} "We see that two..." $\longrightarrow$ "We see that to...".
\end{itemize}
\section*{Chapter 6}
\begin{itemize}
\item{\bf Page 80 below eq. (\ref{eq:Omega-prime}):} "...in the second step." $\longrightarrow$ "...in the third step.".
\end{itemize}
\section*{Chapter 7}
\begin{itemize}
\item{\bf Page 89 eq. (\ref{eq:variation1}):} $-\longrightarrow+$ for the sign in front of the second term.
\item{\bf Page 91 second eq. in (\ref{eq:gammma-ids}):} $(-1)^{(p+1)/2}\longrightarrow(-1)^{(p-1)/2}$.
\item{\bf Page 93 eq. (\ref{eq:deltaS_WZ-final}):} $(-1)^{(p+1)/2}\longrightarrow(-1)^{(p-1)/2}$.
\item{\bf Page 96 eq. (\ref{eq:proj-op-def}):} $(-1)^{(p+1)/2}\longrightarrow(-1)^{(p-1)/2}$.
\item{\bf Page 96 below eq. (\ref{eq:proj-op-def}):}''...that it is symmetric...'' $\longrightarrow$ ''...that it is anti-symmetric...''.
\item{\bf Page 96 eq. (\ref{eq:gamma-expression}):} $(-1)^{(p+1)/2}\longrightarrow(-1)^{(p-1)/2}$.
\item{\bf General:} In the proof of kappa-symmetry in this chapter an argument, due to \cite{Bergshoeff:1996tu}, is used to drop terms proportional to a spinorial derivative
of the dilaton, $\phi$. The argument is that these are proportional to the dilatino which vanishes in a bosonic background. This argument is wrong for the 
following reason\footnote{I would like to thank Dmitri Sorokin for pointing this out to me.}: $\phi$ is in fact a superfield and though the lowest 
component of $D_{\u\alpha}\phi$ is proportional to the dilatino the higher components in the $\theta$-expansion are not and in fact the odd components in
the expansion are proportional to bosonic fields and are in general non-zero even in a bosonic background. This means that the terms proportional to a spinor
derivative of $\phi$ must also be taken into account in the proof of kappa-symmetry as we will now describe.

The superspace constraints, (\ref{eq:Tconstraint}) -- (\ref{eq:Fnconstraint}), should be amended with the constraint
\begin{eqnarray}
(F^{(n+1)})_{\u{\alpha1a_1\cdots a_n}}&=&e^{-\phi}(\gamma_{\u{a_1\cdots a_n}}\chi^2)_\alpha\nonumber\\
(F^{(n+1)})_{\u{\alpha2a_1\cdots a_n}}&=&-(-1)^{n/2}e^{-\phi}(\gamma_{\u{a_1\cdots a_n}}\chi^1)_\alpha\quad\mbox{($n$ even)}\,,\label{eq:E-Fnconstraint}
\end{eqnarray}
where we have defined $\chi^i_\alpha\equiv D_{\alpha i}\phi$. This gives an extra contribution to the kappa-variation of the WZ term, (\ref{eq:WZ-action7}), and using
(\ref{eq:variation-with-F}) and the constraint (\ref{eq:E-Fnconstraint}) the term in $\delta_\kappa\left(e^K\sum_n C^{(n)}\right)$ that contributes to the variation of the action becomes
\begin{eqnarray}
\lefteqn{e^{-\phi}e^K\sum_{n\,\,\mr{even}}^{10}\frac{1}{n!}\big(\delta_\kappa E^1e^{\h a_n}\cdots e^{\h a_1}\gamma_{\h a_1\cdots\h a_n}\chi^2}
\nonumber\\
&&{}-(-1)^{n/2}\delta_\kappa E^2e^{\h a_n}\cdots e^{\h a_1}\gamma_{\h a_1\cdots\h a_n}\chi^1\big)\,.
\label{eq:E-variation1}
\end{eqnarray}
The calculation follows the same steps as the calculation in the thesis. The first term gives instead of (\ref{eq:deltaS_WZ1})
\begin{eqnarray}
\lefteqn{e^{-\phi}e^K\sum_{l=0}^\infty\frac{1}{(2l)!}\sum_{k=0}^le^{\h\alpha_{2l}}\cdots e^{\h\alpha_{2k+1}}e^{a_{2k}}\cdots e^{a_1}}
\nonumber\\
&&\times\binom{2l}{2k}\delta_\kappa E^1\gamma_{a_1\cdots a_{2k}}\gamma_{\h\alpha_{2k+1}\cdots\h\alpha_{2l}}\chi^2\,.
\end{eqnarray}
Shifting the sum on $l$ by $k$ and using the definitions of $\h h_\perp$ and $\h h_\parallel$, (\ref{eq:def-h-perp}) and (\ref{eq:def-h-parallel}), together with the relation
(\ref{eq:p+1-form-part}) and the expression for the DBI Lagrangian (\ref{eq:def-L_DBI}) the contribution to the variation of the WZ term becomes
\begin{equation}
-\int d^{2q}\eta\,d^{p+1}x\,\mc L_{\mr{DBI}}\delta_\kappa E^1\h h\gamma^{(p+1)}\chi^2\,,
\end{equation}
where $\h h\equiv\h h_\parallel\h h_\perp$. Going through the same steps for the second term in (\ref{eq:E-variation1}) and keeping track of the
sign the extra terms missing in the variation of the WZ term in (\ref{eq:deltaS_WZ-final}) are
\begin{eqnarray}
\delta_\kappa S_{\mr{WZ}}'&=&-\int d^{2q}\eta\,d^{p+1}x\,\mc L_{\mr{DBI}}\Big(\delta_\kappa E^1\h h\gamma^{(p+1)}\chi^2
\nonumber\\
&&\qquad\qquad{}+(-1)^{(p-1)/2}\delta_\kappa E^2\h h^{\mr T}\gamma^{(p+1)}\chi^1\Big)\,.
\end{eqnarray}

The missing piece in the variation of the DBI term comes from varying the dilaton factor $e^{-\phi}$ and using the expression for the DBI
Lagrangian (\ref{eq:def-L_DBI}) we get
\begin{equation}
\delta_\kappa\mc L_{\mr{DBI}}'=-\mc L_{\mr{DBI}}\delta_\kappa\phi=-\mc L_{\mr{DBI}}\delta_\kappa E^i\chi^j\delta_{ij}\,.
\end{equation}
This is the term missing in (\ref{eq:deltaS_DBI-final}). The extra term missing in the variation of the total action, $S=S_{\mr{DBI}}+S_{\mr{WZ}}$, (\ref{eq:total-kappa-variation}) is
then
\begin{equation}
\delta_\kappa S'=-\int d^{2q}\eta\,d^{p+1}x\,\mc L_{\mr{DBI}}\delta_\kappa E^{\mr T}(1+\Gamma)\chi\,,
\end{equation}
where we have used the expression for $\Gamma$ in (\ref{eq:proj-op-def}) (corrected as above). This term also vanishes using the expression for 
$\delta_\kappa E$ in (\ref{eq:delta-kappa-E}) and the proof of kappa-symmetry is complete.
\end{itemize}
\vspace{1cm}
\hfill{\tiny Linus Wulff \,\,\, 2007-02-09}